\documentclass[a4paper,11pt]{article}

\usepackage[utf8]{inputenc}
\usepackage[british]{babel}
\usepackage{csquotes}

\usepackage{manfnt}
\usepackage{braket}

\usepackage{debug}

\newcommand{\COMMENTOOK}[1]{}

\usepackage{graphicx}
\usepackage{subcaption} % subfigure env
\usepackage{color} % gb in text from octave output

\usepackage{authblk}
\usepackage{hyperref}
\usepackage{amsmath}
\usepackage{amssymb}
\usepackage{amsfonts}
\usepackage{mathtools}% underbracket

\usepackage{ytableau}

\usepackage{cleveref} % \cref
\usepackage{enumitem}
\numberwithin{equation}{section}

\usepackage[
backend=biber,
style=numeric-comp, %style of citations %numeric %alphabetic,
sorting=none % as they appear %ynt == year name title
]{biblatex}
\newcommand{\rhs}{{r.h.s.} }

\newcommand{\dofs}{{d.o.f.}s }

\newcommand{\wrt}{{w.r.t.} }
\newcommand{\lc}{{lightcone} }
\newcommand{\lcsh}{{lc} } % light cone short

\newcommand{\oh}{ \frac{1}{2} }
\newcommand{\ap}{ \alpha' }
\newcommand{\dap}{ 2\alpha' }

\newcommand{\sap}{ \sqrt{\alpha'} }
\newcommand{\shap}{ \sqrt{\frac{\alpha'}{2}} }
\newcommand{\sdap}{ \sqrt{2\alpha'} }
\newcommand{\stap}{ \sqrt{2\alpha'} }
\newcommand{\ishap}{ \sqrt{ \frac{ 2 }{ \alpha'} } }

\newcommand{\hap}{ \frac{\alpha'}{2} }
\newcommand{\dpap}{ 2\pi \alpha' }

\newcommand{\sw}{ {\pi} } %string width

\newcommand{\Z}{ \mathbb{Z} }

\newcommand{\C}{ \mathbb{C} }

\newcommand{\cA}{{\cal A} }
\newcommand{\cE}{{\cal E} }

\newcommand{\cL}{{\cal L} }

\newcommand{\cN}{{\cal N} }

\newcommand{\Efm}[2]{E^{\underline #1}_{#2}}

\newcommand{\ualpha}{ {\underline \alpha}}
\newcommand{\uepsilon}{ {\underline \epsilon}}
\newcommand{\uA}{ {\underline A}}

\newcommand{\uk}{ {\underline k}}
\newcommand{\ul}{ {\underline l}}
\newcommand{\uL}{ {\underline L}}
\newcommand{\up}{ {\underline p}}
\newcommand{\upl}{ {\underline +}}
\newcommand{\ux}{ {\underline x}}

\newcommand{\uS}{{\underline S}}

\newcommand{\zbar}{{\bar z}}
\newcommand{\eps}{{\epsilon}}
\newcommand{\azp}{{\ualpha_0^+}}
\newcommand{\sN}[1]{ { [#1] } }  %string width
\newcommand{\eq}[1]{
    \begin{align}
        #1
    \end{align}
    }
\newcommand{\del}[0]{
    \partial
    }

\title{
Framed DDF operators and the general solution to Virasoro constraints
}

\author[1]{Dripto Biswas}
\author[1]{Igor Pesando}

\affil[1]
{%
  Dipartimento di Fisica, Universit\`{a} di Torino \authorcr
  and I.N.F.N. -- Sezione di Torino \authorcr
  Via P.\ Giuria 1, I-10125 Torino, Italy % not working \authorcr
}

\hypersetup
{%
  pdfauthor={Dripto Biswas, Raffaele Marotta, Igor Pesando},
  pdfsubject={String Theory},
  pdfkeywords={string theory, physical states},
}

\begin{document}

\maketitle

\begin{abstract}
We define the framed DDF operators by introducing the concept of local
frames in the usual formulation of DDF operators.
In doing so it is possible to completely decouple the DDF operators from the
associated tachyon and show that they are good zero-dimensional
conformal operators.

This allows for an explicit formulation of the
general solution of the Virasoro constraints both on-shell and off-shell.

We then make precise the realization of the
intuitive idea that DDF operators can be used to embed \lc states in
the covariant formulation.
This embedding is not unique, but depends on a coset.
This coset is the little group of the embedding of the \lc
and is associated with a frame.
The frame allows us to embed the $SO(D-2)$ \lc physical
polarizations into the $SO(1,D-1)$ covariant ones in the most general way.

The solution to the Virasoro constraints
is not in the gauge that is usually used.
This happens since the states obtained from DDF operators are
generically the sum of terms
which are partially transverse due to the presence of a projector
but not traceless
and terms which are partially traceless but not transverse.

To check the identification, we verify the matching of
the expectation value of the second
Casimir of the Poincar\'e group for some \lc  states with
the corresponding covariant states built using the framed DDFs.
\end{abstract}

%COMMENTO
\COMMENTOOK{
\section*{To do}
\begin{itemize}
\item
Discuss the DDF operators using the local frame

\item Show the projector

\item Discuss the conformal properties \cite{Erler:2020beb}

\item
Discuss how the local frame allows for an embedding of \lc
polarizations into covariant formalism

\item
Discuss how DDF operators behave as \lc operators and quote properly
the papers by Brower \cite{Brower:1972wj}
and by Goddard, Rebbi and Thorn \cite{Goddard:1972ky}

\item
For off shell consider \cite{Goddard:1972iy}

\item
For states consider \cite{DelGiudice:1970dr}

\item
Discuss the possible local frames

\end{itemize}

} % COMMENTO

\section{Introduction}

String theory is probably the best candidate for quantum gravity and,
as such, it should be able to tell something about both spacelike and
timelike General Relativity singularities.

Spacelike singularities involve time dependent backgrounds and
therefore are very hard and also less studied.
Already at the QFT level time dependent
backgrounds are difficult, since the concept of a particle is lost
because they are created by the interaction with the background
during the time evolution.
String theory has been studied in toy models capable of reproducing a
space-like (or null) singularity, which appears in
space at a specific value of the time coordinate and then disappears.
The easiest way to do so is by generating singularities by
quotienting Minkowski with a discrete group with fixed points,
i.e. orbifolding Minkowski.
In this way, it is possible to produce both space-like singularities
and supersymmetric null singularities
\cite{Horowitz:1989bv,Horowitz:1990sr,Nekrasov:2002kf,Craps:2002ii,
Liu:2002ft,Liu:2002kb,
Fabinger:2002kr,David:2003vn,Craps:2008bv,Madhu:2009jh,Narayan:2009pu,
Narayan:2010rm,Craps:2011sp,Craps:2013qoa} (see also
\cite{Cornalba:2003kd,Craps:2006yb,Berkooz:2007nm} for some reviews).
Another possible way which is a generalization of the previous
orbifolds with null singularity is to consider gravitational shock wave
backgrounds \cite{deVega:1990kk,deVega:1990ke,deVega:1990gq,deVega:1991nm,Jofre:1993hd,Kiritsis:1993jk,DAppollonio:2003zow}.
It happens that in these orbifolds
the four tachyon closed string amplitude diverges in
some kinematical ranges, more explicitly for the Null Boost orbifold
(which may be made supersymmetric and has a null singularity) we have
\begin{equation}
\cA^{(closed)}_{4 T}
\sim
\int_{q\sim \infty} \frac{d q}{|q|} q^{ 4- \alpha' \vec p^2_{\perp\, t}}
,
\end{equation}
so the amplitude diverges for $\alpha' \vec p^2_{\perp\, t}<4$ where
$\vec p_{\perp\, t}$ is the orbifold transverse momentum in $t$ channel.
Until recently, this pathological behavior has been interpreted in the
literature as ``the result of a large gravitational backreaction of the
incoming matter into the singularity due to the exchange of a single
graviton".
This is not very promising for a theory which should tame quantum
gravity.
However, if we perform an analogous computation
for the four point open string function, we find
\begin{equation}
\cA^{(open)}_{4 T}
\sim
\int_{q\sim \infty} \frac{d q}{|q|} q^{ 1- \alpha' \vec p^2_{\perp\, t}}
tr\left(\{T_1, T_2\} \{T_3, T_4\}\right)
,
\end{equation}
which is also divergent when $\alpha' \vec p^2_{\perp\, t}<1$
\cite{Arduino:2020axy,Hikida:2005ec}.
This casts doubts on the backreaction as the main explanation since we are
dealing with open strings at tree level.
This is further strengthened by the fact that three point amplitudes
with massive states may diverge \cite{Arduino:2020axy}
when appropriate polarizations are chosen.
For example, for the three point function of two tachyons and
the first level massive state, we find that, for an
appropriate massive string polarization
\begin{equation}
\cA^{(open)}_{T T M}
\sim
\int_{u\sim 0} \frac{d u}{|u|^{5/2}}  tr\left(\{T_1, T_2\} T_3\right)
.
\end{equation}
This stresses the importance of massive states
and can be interpreted as a complete breakdown of perturbation
theory, i.e. perturbation theory does not exist.
Although it is somewhat obvious that we do not expect to find a well-behaved
perturbation theory because of the singularity,
one could expect some kind of pathology like the series being
asymptotic,
but we find a much worse behavior since
perturbation theory does not exist while it may exist as a complete
theory \cite{Pesando:2022amk}.

The other main type of GR singularity is the timelike singularity.
The prototypical ones and those most studied are associated with Black
Holes.
From this point of view strings can be used
either to try to describe the Black Hole itself
or to compute in an effective way higher spin massive states
or as perturbative theory whose non-perturbative excitations describe
the Black Hole microstates.

The first and third approaches have a very long story that dates back to the 90's.
The second is much more recent.

If we try to identify the highly excited strings with Black Holes, we
have at least to take into account gravitational self-interaction
\cite{Horowitz:1997jc,Damour:1999aw}
in order to try to match the non free massive strings entropy
with that of a Black Holes.
In a similar way, decay rates may be considered \cite{Manes:2001cs,Manes:2003mw,Iengo:2006gm,Iengo:2007ww}.

Finally, applying amplitudes techniques to the scattering of Kerr Black Holes and
merging requires the knowledge of amplitudes involving massive spinning particles,
which could in principle be derived from string amplitudes.

All of the previous studies require massive string states.
Therefore, a better understanding of the string massive spectrum is
required.
Physical string states can be described either in \lc formalism
or in the covariant formalism.
The \lc formalism yields the full physical spectrum,
but this comes at the expense of losing explicit Lorentz covariance.
On the other hand, the covariant formalism requires to select the
physical states using Virasoro conditions.

It has long been known that DDF \cite{DelGiudice:1970dr}
 and Brower \cite{Brower:1972wj} operators generate the full
physical spectrum.
This comes, however, with the drawback of introducing an on-shell tachyonic momentum $|p_T\rangle$, a null vector $q$
 (such that $2\ap p_T\cdot q =-1$) and $D-2$ polarization vectors
 $\epsilon^i(q)$ (such that $q\cdot \epsilon^i(q) =0$)
 in a quite messy bundle of objects.
 
At the same time it seems clear that DDF operators embeds the \lc ones
into the covariant formalism.

The aim of this paper is to make this correspondence clearer and more
streamlined by introducing a local frame $E^{\underline{\mu}}_\nu$,
i.e. a vielbein which encompasses all the previous elements in a
conceptually clearer picture.
In doing so, we find out that it is possible to obtain the general
solution of the Virasoro constraints also off-shell, i.e. dropping the
$L_0$ constraint.
Moreover, we find that, in the appropriate meaning, the \lc polarizations
are Lorentz invariant and that all Lorentz transformations may
be encapsulated in the global rotations of the local frame $E$.

The paper is organized as follows.

In Section \ref{sec:DDF_def} we introduce the notation, conventions and framed DDF operators.

Then in Section \ref{sec:DDF_props} we discuss how framed DDFs differ
from the usual ones and their algebra. We also introduce the Brower
and the ``improved'' Brower operators. These are necessary in order to
discuss the gauge invariance off-shell and how physical states built
using frames which differ by a local rotation are connected.
From this point of view, states obtained using ``improved'' Brower
operators are necessary to rotate the \lc direction despite being null on-shell.

Given this, in Section \ref{sec:DDF_off_shell_and_general_solution} we discuss how to go minimally off-shell while preserving all the other
Virasoro conditions and how to get the general solution to all the Virasoro conditions.

The solution is the general one but not in the simplest gauge,
i.e. the obvious states are a mixture of irreps.
The discussion is performed without technical details
for the first two string levels in order to be able to pinpoint the ideas.
The basic idea is, however, quite simple. Framed DDFs are an explicit way of
embedding in the most general way \lc states into the covariant ones.
%% This is achieved by spliting the generic embedding of the \lc as
%% a chosen embedding  and   the coset element. 

In Section \ref{sec:mapping lc to DDFs} we discuss how framed DDFs
realize the naive idea of embedding \lc states into covariant ones.
This is obtained by comparing the Lorentz algebra restricted to DDFs
operators to the one written on the lightcone.

In Section \ref{sec:DDF_examples} we explicitly examine the details of
the embedding in the simplest cases, i.e., for levels $N=1$ and $N=2$.

Finally, to further strengthen the match in Section \ref{sec:casimirs}
we compare the second order Casimir for the Poincar\'e group for some
states when computed on \lc or in covariant formalism after the
embedding of \lc states into the covariant ones using framed DDFs.

%%%%%%%%%%%%%%%%%%%%%%%%%%%%%%%%%%%%%%%%%%%%%%%%%%%%%%%%%%%%%%%%%%%%%%
%%%%%%%%%%%%%%%%%%%%%%%%%%%%%%%%%%%%%%%%%%%%%%%%%%%%%%%%%%%%%%%%%%%%%%
%%%%%%%%%%%%%%%%%%%%%%%%%%%%%%%%%%%%%%%%%%%%%%%%%%%%%%%%%%%%%%%%%%%%%%

\section{Framed DDF operators and local frames}
\label{sec:DDF_def}
In this section, we present the reformulation of the standard DDF and Brower
operators in terms of a local frame, i.e., a vielbein.
We call them framed DDF.
We start with a brief section on our conventions and then introduce
the framed DDF operators. We then discuss their conformal properties and their
algebra.
It is worth stressing that the original DDF operators are not good
conformal operators for computing off-shell amplitudes, since they have cuts (see \ref{sec:Conformal properties}).

\subsection{The bosonic string}
We write the string action in the conformal gauge
\eq{
  S
  = -\frac{1}{\dpap}
  \int d\tau \int_0^\sw d \sigma
  \oh g_{\mu\nu}
  \left( \dot X^\mu \dot X^\nu - X^{' \mu} X^{' \nu} \right)
  ,
}
using $g_{\mu\nu}$ in place of $\eta_{\mu\nu}$ in order
to make more evident the role of the local frames.
The solution for the open string then reads
\eq{
X^\mu(u, \bar u)
  &=
  L^\mu(u)+ R^\mu(\bar u)
  ,
  \nonumber\\
  L^\mu(u)
  &=
  \oh x^\mu_0 - i \ap p^\mu_0 \ln(u)
  +i \shap \sum_{n\neq 0} \frac{\alpha^\mu_n}{n} u^{-n}
  ,
  \label{eq:L^mu}
  \\
  R^\mu(\bar u)
  &=
  L^\mu(\bar u)
  ,
}
where
\begin{equation}
  u=e^{\tau_E+i \sigma} = e^{i (\tau+\sigma)}
  ~~\mbox{ and }~~
  \arg(\ln(z)) \in (-\pi, \pi]
    ,
\end{equation}
along with the canonical commutation relations,
\eq{
 [x^\mu_0, p_0^\nu]
  &=
  i\, g^ {\mu\nu},
  \nonumber\\
  [\alpha^\mu_m, \alpha_n^\nu]
  &=
  m\, \delta_{m+n,0}\, g^ {\mu\nu}
  .
}
As usual we also define
$\alpha_0^\mu = \sdap p_0^\mu$.

One can introduce the radial ordering $R$ on the complex plane and compute the OPE,
\eq{
   R\left[ L^\mu(u) L^\nu(v) \right]
  =~
  :L^\mu(u)& L^\nu(v):
  -
  \hap g^{\mu\nu}
  \Big[
    \theta(|u|-|v|) \ln(u-v) \nonumber \\
    &+ \theta(-|u|+|v|) \ln(-u+v)
    \Big].
\label{eq:radLL}
}
It follows from \eqref{eq:radLL} the usual useful relations
\eq{
\underbracket{L^{\mu (+)}(u)\, \partial L}{}^{\nu (-)}(v)
  =
  - \partial \underbracket{ L^{\mu (+)}(u)\, L}{}^{\nu (-)}(v)
  =&
  \hap \frac{g^{\mu\nu} }{ u-v },
  \nonumber\\
  \partial \underbracket{L^{\mu (+)}(u)\, \partial L}{}^{\nu (-)}(v)
  =&
  -
  \hap \frac{g^{\mu\nu} }{ (u-v)^2 }
  .
  \label{eq:basic_wick_contractions}
}
For writing down the integrated DDF operator, we consider the tachyon and photon vertices in the $-1$ ghost sector as,
\begin{align}
V_T(x; k_T)
  &=
  c(x) : e^{i k_{T \mu } X^\mu(x, \bar x) }:
  ~~~~\mbox{with } \ap g^{\mu\nu} k_{T \mu } k_{T \nu } = 1
  \nonumber\\
  &=
  c(x) : e^{2 i k_{T \mu } L^\mu(x)} : ~~~~\mbox{when }x>0
  ,
  \nonumber\\
  V_A(x; k, \epsilon)
  &=
  c(x)
  : \epsilon_\mu \partial_x X^\mu(x, \bar x) e^{i k_{\mu } X^\mu(x, \bar x) }:
  ~~~~\mbox{with }
  g^{\mu\nu} k_{\mu } k_{\nu } = g^{\mu\nu} k_{\mu } \epsilon_\nu = 0  
  \nonumber\\
  &=
  c(x)
  : 2 \epsilon_\mu \partial_x L^\mu e^{2 i k_{ \mu } L^\mu(x)} :
  ~~~~\mbox{when }{x>0}
  ,
  \label{eq:tacpho-1}
\end{align}
where we made explicit the dependence on the chiral part of the string
  coordinates $L(u)$.

\subsection{Framed DDF operators}

To reformulate the standard DDF and Brower operators,
leading to the framed definitions
in \cref{eq:DDFflatexpr,eq:A- flatexpr} we introduce a local frame,
i.e., a vielbein.
While for the flat space we are working in, this may seem like overkill, it
is actually useful to clearly distinguish the symmetries of the DDF
and Brower constructions and the action of Lorentz transformations on
the physical states. 

We introduce the vielbein, i.e. the local frame by 
$E_\mu^{\underline{\mu}}$ 
and its inverse $E^\mu_{\underline{\mu}}$
%% As usual we raise the flat indexes using the flat metric as
%% $E^{\mu \underline{\mu}}
%% =
%% E^\mu_{\underline{\nu}} \eta^{\underline{\mu} \underline{\nu}}
%% $.
This  dual local frame is such that
\begin{align}
E_\mu^{\underline{\mu}} E_\nu^{\underline{\nu}} \eta_{\underline{\mu} \underline{\nu}}
= g_{\mu\nu}
,
\end{align}
and has two symmetries: the local and global Lorentz transformations.
As we discuss later in detail
in \ref{sec:N=1 global coords rotations} for the level $N=1$ states,
the global Lorentz transformations associated with
the curved index $\mu$ act in a very
simple way on the DDF states,
while the local Lorentz transformations
discussed in detail in \ref{sec:N=1 local coords rotations}
act on $\underline \mu$ in a complicated way.

The framed DDF and Brower operators are then expressed
using the flat left moving string operators
in the ghost number sector $0$
by
\begin{align}
\uA^i_n(E)
=&
i \ishap
  \oint_{z=0} \frac{d z}{ 2\pi i}
  : \partial_z \uL^i(z) e^{i n \frac{\uL^+(z)}{\ap \up^+_0}} :
  .
  \label{eq:DDFflatexpr}
  \\
   \uA^-_n(E)
  =&
  i \ishap
  \oint_{z=0} \frac{d z}{ 2\pi i}
  :
  \left[
    \partial_z \uL^-(z)
    -
    i \frac{n}{4 \up_0^+}
        \frac{\partial^2_z \uL^+}{\partial_z \uL^+}
    \right]
    e^{i n \frac{ \uL^+(z) }{\ap \up^+_0} } :
  \label{eq:A- flatexpr}   
.
\end{align}
For later convenience we introduce also the improved Brower operators
as\footnote{
Notice that the normal ordering is taken \wrt $\uA$ but this is the
same normal ordering \wrt $\ualpha$ since both have the same vacuum.
}
\begin{align}
   {\tilde \uA}^-_n(E)
  =&
\uA^-_n(E)
-       
  \frac{1}{\ualpha_0^+} \cL_n(E)
  %% \frac{1}{2}           
  %% \sum_{j=2}^{D-1} \sum_{l\in \Z}: \uA^j_l(E)\, \uA^j_{m-l}(E) :
   +
   \frac{D-2}{24}
   \frac{1}{\ualpha_0^+}\,
   \delta_{n, 0}
   ,          
   \\
\cL_m(E)
&=
  \frac{1}{2}           
  \sum_{j=2}^{D-1} \sum_{l\in \Z}: \uA^j_l(E)\, \uA^j_{m-l}(E) :
\label{eq:cL virasoro transverse}
\end{align}     
where the last term  is present only for $n=0$
and it is fundamental to
ensure that the states built from the improved Brower operators are
null on shell and in the critical dimension $D=26$ as shown in
eq. \eqref{eq: vev tAm0 is zero} for the $N=1$ level state and
in eq. \eqref{eq: vev N=2 tAs are zero} for the $N=2$ level states.
In the previous expressions we have introduced the notation
\begin{equation}
\uL^{{\mu}}(z)
=
L^{\underline{\mu}}(z)
=
E^{\underline{\mu}}_\mu\, L^\mu(z)
,
\end{equation}
for the flat chiral string coordinates extended to the whole complex
plane (with a cut on the negative real axis).
Henceforth we usually denote the transformation of any quantity onto
the local frame defined by $E^\mu_{\underline{\mu}}$ by a line under their
respective symbol or indexes.
If necessary we explicitly denote the local frame.

Notice the explicit appearance of the operator $1/\up_0^+$ in the
exponent \cite{Polchinski:1998rq}.
This allows to derive and discuss many properties independently of the
associated tachyonic operator.
In particular only this form of DDF operators are true zero dimensional
conformal operators since they have no cuts when inserted in correlators.

The expression for
$\uA^i_n(E)$
using the curved coordinates is
\begin{align}
 \uA^i_n(E)
  &=
  \ishap
    \oint_{z=0} \frac{d z}{ 2\pi i}
    g_{\mu\nu}\, E^{\mu \underline{i} }\, : \partial_z L^\nu(z)
    \exp
    \left\{
      i n
      \frac{ g_{\rho\sigma}\, E^{\rho \upl}\, L^\sigma(z)
      }{
        \ap g_{\mu\nu}\, E^{\mu \upl}\, p^\nu_0
        }
      \right\}
      :
      .
    \label{eq:DDF_localflat_explicit}
\end{align}

The connection between the usual expression
and the one with explicit frame can be obtained by choosing a null vector
$n^\mu$,
an auxiliary null vector $\bar{n}^\mu$
and the usual transverse polarization $\epsilon_\mu^{(i)}$, such
that 
\begin{equation}
 E^{\upl}_\mu = n_\mu
  ,~~~~
  E^{\underline{-}}_\mu = \bar n_\mu
  ,~~~~
  E^{\underline{i}}_\mu = \epsilon_\mu^{(i)}
  ,
  \label{eq:DDF_usual_fromlocalframe}
\end{equation}
and it is discussed in more details in section
\ref{sec: Differences with the usual formulation}.

\subsection{A better expression for the framed DDF operators}
The previous  framed DDF operators admit a representation which makes the
transversality property of DDF states more clear.
In particular, there is a null vector $E^\mu_{\upl}$ in the flat basis, such that,
\eq{
  \ux^+_0 = g_{\mu \nu} x^\mu_0 E^{\nu \underline{+}}
  ,~~~
  \up^+_0 = - \up_{0 -} = g_{\mu \nu} p^\mu_0 E^{\nu \underline{+}}
  .
}
If we define the part of the chiral coordinate without the momentum
$  \uL^\mu_{(\ne)}(z)$ as
\begin{equation}
  \uL^\mu(z)
  =
  \uL^\mu_{(\ne)}(z)
  -i \ap \up^\mu_0 \ln(z)
  ,
\end{equation}
then we can write the DDF {\sl creator} as
\begin{align}
  \uA^i_{-m}(E)
  &=
  i \sqrt{\frac{2}{\ap}}
  \oint_{z=0} \frac{d z}{ 2\pi i}
  \frac{1}{z^m}
  :
  \left(
  \partial_z \uL_{(\ne)}^i(z)
  -
  \frac{1}{z}
  i \ap \up^i_0
  \right)
  e^{-i m \frac{ \uL_{(\ne)}^+(z) }{\ap \up^+_0} }
  : ,
  \nonumber\\
\end{align}
which allows for a better form upon integration by parts of the
$\frac{1}{z^{m+1}}$ term
\begin{align}
  \uA^i_{-m}(E)
  &=
  i \sqrt{\frac{2}{\ap}}
  \oint_{z=0} \frac{d z}{ 2\pi i}
  \frac{1}{z^m}
  :
  \left(
  \partial_z \uL_{(\ne)}^i(z)
  -
  \frac{\up^i_0 }{ \up^+_0}\,
  \partial_z \uL_{(\ne)}^+(z)
  \right)
  e^{-i m \frac{ \uL_{(\ne)}^+(z) }{\ap \up^+_0} }
  :
  \nonumber\\
  &=
  i \sqrt{\frac{2}{\ap}}
  \oint_{z=0} \frac{d z}{ 2\pi i}
  \frac{1}{z^m}
  : \Pi^{\underline{i}}_\mu
  \partial_z L_{(\ne)}^\mu(z)
  e^{-i m
    \frac{
      E^{\underline{+}}_\mu L_{(\ne)}^\mu(z)
    }{
      \ap E^{\underline{+}}_\mu p^\mu_0}
  }
  :
  \nonumber\\
  &=
  i \sqrt{\frac{2}{\ap}}
\frac{1}{(m-1)!}
\frac{ d^{m-1} }{d z^{m-1} }\Bigg|_{z=0}
  : \Pi^{\underline{i}}_\mu
  \partial_z L_{(\ne)}^\mu(z)
  e^{-i m
    \frac{
      E^{\underline{+}}_\mu L_{(\ne)}^\mu(z)
    }{
      \ap E^{\underline{+}}_\mu p^\mu_0}
  }
  :
  ,
%  \label{eq:DDF_proj}
  \label{eq:DDF creator with projector}
  \end{align}
where we have defined transverse projector
\begin{align}
\Pi^{\underline{i}}_\mu
=
E^{\underline{i}} _\mu
-
\frac{ \up_0^i }{ \up_0^+}\,
E^{\underline{+}}_\mu
=
E^{\underline{i}} _\mu
-
\frac{   E^{\underline{i}}_\nu\, p_0^\nu }{ E^{\underline{+}}_\gamma\,p_0^\gamma}
\,
E^{\underline{+}}_\mu
,~~~~~
p_0^\mu\, \Pi^{\underline{i}}_\mu =0
.
\label{eq:proj} 
\end{align}
This shows clearly that the ``leading order'' of a DDF operator is
\begin{equation}
  \uA^i_{-m}(E)
  \sim
  e^{-i \frac{m}{2\ap \up_0^+} \ux_0^+} (\Pi\, \ualpha)^i_{-m}
  ,
\end{equation}
where the $e^{-i \frac{m}{2\ap \up_0^+} \ux_0^+}$ plays a fundamental
role in the fact that $\uA^i_m$ has zero conformal dimension.

This expression could give the wrong impression that states obtained by
using DDF are transverse; unfortunately, it is not so because the
contour integration becomes a multiple derivative
as in the last line of eq. \eqref{eq:DDF creator with projector} and for $m \ge 2$ derivatives can act on the exponent part which
gives a non-transverse contribution as explicitly shown in eq.
\eqref{eq:T and S Ai-2} for the simplest case of a level $N=2$ state.

%%%%%%%%%%%%%%%%%%%%%%%%%%%%%%%%%%%%%%%%%%%%%%%%%%%%%%%%%%%%%%%%%%%%%%
%%%%%%%%%%%%%%%%%%%%%%%%%%%%%%%%%%%%%%%%%%%%%%%%%%%%%%%%%%%%%%%%%%%%%%
%%%%%%%%%%%%%%%%%%%%%%%%%%%%%%%%%%%%%%%%%%%%%%%%%%%%%%%%%%%%%%%%%%%%%%

\section{Important properties of the framed DDF operators}
\label{sec:DDF_props}
In this section we first discuss the differences between the
framed DDF operators and the usual formulation.
Then we discuss the algebra and the hermiticity properties, and finally
we derive and discuss the fact that only the framed DDF operators are
really zero conformal dimension operators (see also \cite{Erler:2020beb}).
The derivation of the algebra and the hermiticity properties is made in the Appendices
\ref{app:Derivation of algebra}
and
\ref{app:Derivation of hermiticity properties}.

\subsection{Differences with the usual formulation}
\label{sec: Differences with the usual formulation}
The framed DDF construction described above has two significant differences
compared to the standard DDF formulation
- decoupling from the associated tachyonic momentum and being
conformal operators.

Here we discuss the former, and in Section \ref{sec:Conformal properties}
we discuss the conformal properties.

In particular, the decoupling means that we could even act on a
physical state different from the tachyon!

Usually,
starting from the on-shell tachyon vertex operator
$V_T(x; p_T)
= c(x)\, :e^{i p_T\cdot X(x, \bar x)}:
= c(x)\, :e^{2 i p_T\cdot L(x)}:
$ with
\eq{
p_T^2 = 1/\alpha'
}
and imposing
\eq{
2\alpha' p_T \cdot q = 1,
}
for some null vector $q$
(eventually also choosing 
$p_T^+ = -p_T^- = \frac{1}{\sap}$, $p_T^i = 0$
and
$q^+ = q^i = q^2 = 0,~ q^- \neq 0$ ),
one can successively construct not exact BRST invariant states
with momenta given by,
\eq{
p_N = p_T - N\, q,
}
such that the mass is given by,
\eq{
-\alpha' p_N^2 = \alpha' M_N^2 = N-1
,
}
by applying a string of DDF operators $A^i_n$ ($n<0$) on the tachyonic
state as
\begin{equation}
\prod_{n=1}^\infty \prod_{i=2}^{D-2} \left( A^i_{-n}(q, p_T) \right)^{N^i_n}
| p_T \rangle
,
\end{equation}  
with
\begin{equation}
N= \sum_{n=1}^\infty \sum_{i=2}^{D-2} N^i_n
.
\end{equation}

The bosonic DDF operators are defined by:
\begin{eqnarray}
A^i_n(q, p_T)
=
i
\ishap
\oint \, \frac{d z}{2\pi i}\,
\epsilon^{(i)}(q, p_T)\cdot\partial_z L \,
e^{i 2\, n \,q\cdot L(z)}
,
\end{eqnarray}
where $\epsilon_\mu^{(i)}(q, p_T)$ is a polarization vector
with
\begin{equation}
\epsilon^{(i)}(q, p_T)\cdot q = 0
,
\end{equation}  
and  is transforming as a vector under $SO(D-2)$ the little group of massless
states and it is implicitly dependent on $p_T$ through $q$.

Therefore, the associated tachyonic momentum $p_T$ plays a direct role
in ensuring that the usual DDF construction satisfies the harmonic
oscillator algebra 
\eq{
[A^i_m(q, p_T),\, A^j_n(q, p_T)] = m\, \delta^{i j}\, \delta_{m+n,0}.
\label{eq:DDFalg_comparison}
}

In contrast the flat space DDF operators $\uA^i_m(E)$ as constructed in
the previous section automatically satisfy the algebra
(\ref{eq:DDFalg_comparison}) without any reference to the tachyonic
momentum associated to the photon vertex, provided that, $\up^+_0 \neq
0$ in (\ref{eq:DDFflatexpr}), i.e. we have decoupled the tachyonic
momentum from the definition of the (flat space) DDF operators.

Formally one can recover the usual formulation from the framed one by
letting
\begin{equation}
 \frac{ E^{\upl}_\mu }{ \ap \up_0^+} 
 =
 \frac{ n_\mu }{ \ap \up_0^+} 
 \rightarrow
 2 {q_\mu}
   ,
\label{eq: mapping framed usual}
\end{equation}  
which amounts to replacing an operator with a $\C$-number.

\subsection{Algebra of the operators}
The algebra of the framed operators read
\begin{align}
  [\uA^i_m(E),\, \uA^j_n(E)]
  &= m\, \delta_{m+n,0} \delta^{i j}
  ,
  \label{eq:DDFalgebra}
  \\
  [\uA^i_m(E),\, \ualpha_0^+\uA^-_n(E)]
  &=
   m
  \,
  \uA^i_{m+n}(E)
  \label{eq:Ai A- algebra}
  \\
   [\ualpha_0^+ \uA^-_m(E),\, \ualpha_0^+ \uA^-_n(E)]
   &=
   (m-n)\, \ualpha_0^+ \uA^-_{m+n}(E)
   +
   2 m^3\,
   \delta_{m+n, 0}
   ,
  \label{eq:A- A- algebra}
\end{align}
and
\begin{align}
  [\uA^i_m(E),\, \ualpha_0^+ \tilde \uA^-_n(E)]
  &=
0 
  \nonumber\\
   [\ualpha_0^+ \tilde \uA^-_m(E),\, \ualpha_0^+ \tilde \uA^-_n(E)]
   &=
   (m-n)\, \ualpha_0^+ \tilde \uA^-_{m+n}(E)
   +
   \frac{26 - D }{12}
   m^3\,
   \delta_{m+n, 0}
   .
  \label{eq:DDFalgebra_tilde}
\end{align}
Again we observe that this  algebra is independent of the tachyon
momentum as well as the choice of the local frame by construction.
The derivation is fairly straightforward and is given in Appendix \ref{app:Derivation of algebra}.
The only subtlety is in computing the $\tilde \uA^-_m(E)$ algebra,
where it is necessary to start with the explicitly normal ordered
expression for $\sum_{l\in \Z}: \uA^j_l(E)\, \uA^j_{m-l}(E) :$ in
order to get a well-defined expression for the $\C$-number.

\subsection{Hermiticity properties}
The hermiticity properties of the framed operators are the expected ones
\begin{equation}
\left[ \uA^i_m(E) \right]^\dagger
=
\uA^i_{-m}(E)
~~,~~
\left[ \uA^-_m(E) \right]^\dagger
=
\uA^-_{-m}(E)
~~,~~
\left[ \tilde \uA^-_m(E) \right]^\dagger
=
\tilde \uA^-_{-m}(E)
.
\end{equation}  
However in the case of $\uA^-_{-m}(E)$, although straightforward, it is not
trivial because of the normal ordering of
$e^{i \frac{m}{2 \ap \up_0^+} \ux_0^+} \up_0^-$
as discussed in Appendix \ref{app:Derivation of hermiticity properties}.

\subsection{Conformal properties}
\label{sec:Conformal properties}
As already stated, only the framed DDF operators are true conformal
operators of dimension zero.
The usual ones are not when a rigorous approach is adopted.
To understand this statement we consider the basic commutator
\begin{align}
  \left[
    \left.
    \oint_{z=0}
    \frac{d z}{2 \pi i}
    z^{n+1}
 \frac{-2}{\ap} e^{\delta\cdot \partial \uL(z)}
    \right|_{\delta^2}
    ,
    i \sqrt{\frac{2}{\ap}}
    \left.
    \oint_{w=0}
    \frac{d w}{2 \pi i}
    e^{\underline\epsilon \cdot \partial \uL(w) + i \uk \cdot \uL(w)}
    \right|_{\underline\epsilon_{(i)}, \uk_+}
  \right]
    ,
    \label{eq:comm_expLA}
\end{align}
where we restrict ourselves to $\mathcal{O}(\delta^2)$ and
$\mathcal{O}(\underline\epsilon)$ above, to extract the forms of $L_n$ and
$\uA^i_m$ respectively, from the exponentials in
(\ref{eq:comm_expLA}).

This commutator does not depend on the fact that $\uk_+$ is proportional to
$\frac{1}{\ap \up_0^+}$ since there is no $\ux_0^-$ in $L_n$
and therefore this computation also applies to the usual DDF, when
the flat quantities are replaced by the ``curved'' ones.

Then we get,
\begin{align}
  [L_n,\uA^i_m]=&
    i \sqrt{\frac{2}{\ap}}
  \frac{-2}{\ap}
  \oint_{w=0}  
  \oint_{z=w}
  z^{n+1}
  :
  e^{\delta\cdot \partial \uL(z)}
  e^{\underline\epsilon \cdot \partial \uL(w) + i \uk \cdot \uL(w)}
  :
  e^{
    -
    \hap \frac{\delta\cdot \epsilon}{(z-w)^2}
    -
    \hap
    \frac{i \delta\cdot \uk}{z-w}
  }
  \nonumber\\
  =&
    i \sqrt{\frac{2}{\ap}}
  \oint_{w=0}  
  \oint_{z=w}
  z^{n+1}
  :
  \Bigg[
    -
    \hap
    \frac{\delta\cdot \underline\epsilon}{(z-w)^2}
    \frac{i \delta\cdot \uk}{z-w}
    +
    \frac{\delta\cdot \underline\epsilon}{(z-w)^2}
    \delta\cdot \partial \uL(z)
     \nonumber \\
    &+\underline\epsilon \cdot \partial \uL(w)
    \delta \cdot \partial L(z)
    \frac{i \delta\cdot \uk}{z-w}
    \Bigg]
  e^{ i \uk \cdot \uL(w)}
  :
  ,
\end{align}
using the substitution $\delta_\mu \delta_\nu\rightarrow
\eta_{\mu\nu}$
and
$\underline\epsilon\cdot \uk=0$ we get
\begin{align}
  [L_n,\uA^i_m]=&
  i \sqrt{\frac{2}{\ap}}
  \oint_{w=0}  
  :
  \left[
    \epsilon \cdot \partial( w^{n+1} \partial \uL(w) )
    +
    w^{n+1}
    \underline\epsilon \cdot \partial \uL(w)
    i \uk \cdot \partial \uL(w)
    \right]
  e^{ i \uk \cdot \uL(w)}
  :  
  \nonumber\\
  =&
  i \sqrt{\frac{2}{\ap}}
  \oint_{w=0}  
  :
\partial
\left[
  w^{n+1}
  \underline\epsilon \cdot \partial \uL(w)
  e^{ i \uk \cdot \uL(w)}
  \right]
  :
  \nonumber\\
  =&
  i \sqrt{\frac{2}{\ap}}
  \oint_{w=0}  
  :
\partial
  \left[
    w^{ \ap \uk \cdot \up_0 + n + 1}
    \dots
    \right]
  .
  \label{eq:LA_comm}
\end{align}

We observe that in the framed DDF definition we have
$\ap \uk \cdot \up_0 \in\Z$ {\sl by construction}
hence the integrand in (\ref{eq:LA_comm} has no cuts and the integral
is zero proving (the operator relation) that,
\begin{equation}
    [ L_n, \uA^i_m]=0.
    \label{eq:vir_A=0}
\end{equation}

On the other hand, in the usual DDF formulation, using the
mapping in eq. \eqref{eq: mapping framed usual}
we get a result which is generically different from zero and depends on the initial and
final point $x=-r$ on the negative real axis where we must place the
branch cut, explicitly
\begin{align}
    [ L_n, A^i_m(q, p_T)]
  =&
~i \sqrt{\frac{2}{\ap}}
  \,
  \frac{1}{\pi}
  \,
  \sin \left( 2 \pi \ap m q \cdot p_0 \right)
  \,
  (-r)^{n+1} \nonumber \\
  &\times
  \epsilon \cdot \partial L( -r )
  \,
  r^{ \ap m q \cdot p_0 }
  e^{ i m q \cdot L_{(\ne)}(-r) }
  %% \,
  %% \sin \left( \pi \ap \uk \cdot \up_0 \right)
  %% \,
  %% (-r)^{n+1}
  %% \epsilon \cdot \partial L( -r )
  %% \,
  %% r^{ \ap \uk \cdot \up_0 }
  %% e^{ i \uk \cdot \uL_{(\ne)}(-r) }
  .
\end{align}

As far as the $\uA^-$ operators are concerned, we need to be more careful, since the
previous computation fails if we only consider the term with $\partial \uL^-$ since it has a cubic pole.
This issue is discussed in Appendix \ref{app:uA}.
However, the result is still that the framed $\uA^-(E)$  and $\tilde \uA^-(E)$ 
are zero-dimensional conformal operators in the $\up_0^+$ sector, i.e.
\begin{equation}
    [ L_n, \tilde \uA^-_m]=0.
    \label{eq:vir_A-=0}
\end{equation}

%%%%%%%%%%%%%%%%%%%%%%%%%%%%%%%%%%%%%%%%%%%%%%%%%%%%%%%%%%%%%%%%%%%%%%
%%%%%%%%%%%%%%%%%%%%%%%%%%%%%%%%%%%%%%%%%%%%%%%%%%%%%%%%%%%%%%%%%%%%%%
%%%%%%%%%%%%%%%%%%%%%%%%%%%%%%%%%%%%%%%%%%%%%%%%%%%%%%%%%%%%%%%%%%%%%%

%___________________________________________%
\section{The general solution to Virasoro constraints
and on shell and off shell physical states}
\label{sec:DDF_off_shell_and_general_solution}
We are now ready to discuss
how the framed DDF operators offer a way of finding the
general {\sl covariant} solution of the Virasoro constraints.
This is obtained by explicitly implementing, in an appropriate way, the naive idea that DDF operators correspond to the
physical \lc  operators.
In order to obtain the most general covariant state we use the local frame
$E$ to add to the game the dependence on the little group of the
embedding of the lightcone.

In the next section, we examine in more detail how the \lc states can be
embedded into the covariant ones, here we simply take inspiration from
this idea.

In this section we lay down the general ideas but for a better
understanding of the details,
we explicitly carry out the computations for the first
two levels in section \ref{sec:DDF_examples}.

We start with a discussion on how the framed DDFs allow us to define on-shell and (minimally) off-shell states, i.e. states which satisfy all Virasoro
constraints except the mass shell condition.

\subsection{On-shell BRST exact and not exact states}
The most general physical state with ghost number $-1$
which is not BRST exact is given by\footnote{
We have not written $k_+$ in the ket vector since we are on-shell and
therefore it is not independent of the other momentum components.} 
\eq{
    &
    |k_-, k_i, \{N_{i\, n}, N_{c_{-1}}=1 \}_{n\ge 1}; E \rangle
     \nonumber \\
    =&\left[
\prod_{i=2}^{D-1}
\prod_{m=1}^\infty
\frac{ \left( A^i_m{}^\dagger(E) \right)^{N_{i\,m}}}{ \sqrt{ m!\, N_{i\,m}! }}
\right]
c_{-1}\,
%% |k_{T +}, k_{T -}, k_{T i}, 0_a\rangle
|k_{T \rho} = E_\rho^{\underline{\mu}} \uk_{T \mu}, 0_a\rangle
,
\label{eq:generic_DDF_state}
}
with
\begin{equation}
  k_\rho
  =
  E_\rho^{\underline{\mu}} \uk_{T+N\, \mu}
  =
  k_{T \rho}
  -N
  \frac{
       E_\rho^{ \upl}
  }
  {
      2 \ap  g^{\mu\nu} E_\nu^{\upl}\, k_{T \mu}
  }
%%   ~~\leftrightarrow~~
%%   \uk^-= \uk^-_{T+N}
%% ,~~
%% \uk^+=\uk_T^+
%% ,~~
%% \uk^i=\uk_T^i
  ,
  \label{eq:DDF_k_kT_E}
\end{equation}
where we have defined the ``shifted'' $\uk_{T+N}$ momentum as
\begin{equation}
  \uk^-_{T+N}= \uk_T^-
  + \frac{
  N
}{ 2\ap \uk^+}
,~~
N=\sum_{i=2}^{D-1}\sum_{m=1}^\infty m N_{i\, m}
,
\end{equation}  
and where we have  shown the explicit frame dependence.

Notice that the tachyonic momentum  $\uk_T$ is on-shell and
{\sl fixed}.
All possible physical momenta $k_\rho$ ( $k^2<0$ for $N\ge 2$ )
of a particle at level $N$
can be obtained from the fixed momentum $\uk_{T+N}$ by means of the
global Lorentz transformation, i.e., the Lorentz transformation on
curved indexes associated with $E$.
This means that it is not a restriction to take the tachyonic rest
frame $\uk^-_T = \uk^+_T = - \frac{1}{\sdap}$ and $\uk^i_T=0$.

On the other hand the states
\begin{equation}
\prod_{m=1}^\infty (A^i_{-m}(E))^{N^i_m}
(\tilde A^-_{-m}(E))^{N_m}
c_{-1} |k_T; 0_a\rangle
,
\end{equation}
with at least one $N_m \ne 0$ satisfy the Virasoro constraints but are
null, i.e. are BRST exact since in $D=26$ the $\tilde A^-(E)$
satisfy a $c=0$ Virasoro algebra
so that their norm is proportional to 
\begin{align}
\langle k_T | \tilde A^-_{0} | k_T \rangle
\propto
\sdap \uk_T^- + \frac{D-2}{24} \frac{ 1}{ \sdap \uk_T^+}
=
\frac{ 1}{\sdap \uk_T^+}
\left(-\ap \uk_T^2 + \frac{D-2}{24} \right) 
,
\end{align}
which vanishes when the tachyon
$\ap (2 \uk^+_T \uk^-_T - \vec \uk_T^2 ) = -1$ is on-shell.
They play, however, a role when going minimally off-shell as we discuss now.

\subsection{Going minimally off-shell}
\label{sec:going minimally offshell}
As elaborated in Section \ref{sec:DDF_props}
an advantage of this formulation is that the
$\uA^i_m(E)$  and $\tilde A^-_m(E)$ do not refer to the
tachyon momentum and always satisfy the same algebra
\cref{eq:DDFalgebra,eq:DDFalgebra_tilde}
independently of it.
This allows us to easily go minimally off-shell by setting the tachyon
momentum to be off-shell\footnote{
It should be also possible to do the same with the usual DDF but in a more
complicated way}.
In doing so, all the Virasoro conditions are satisfied but the mass shell
one - in fact using \cref{eq:vir_A=0,eq:vir_A-=0}, we have for $n\ge 1$
\begin{equation}
  L_n \prod_{m=1}^\infty
  (A^i_{-m}(E))^{N^i_m} (\tilde A^-_{-m}(E))^{N_m}|k_T\rangle
  =
  \prod_{m=1}^\infty (A^i_{-m}(E))^{N^i_m} (\tilde A^-_{-m}(E))^{N_m} L_n |k_T\rangle
  =
  0
  ,
  \label{eq:L_Astate_comm}
\end{equation}
since for a tachyonic off-shell state we still have
$ L_n |k_T\rangle = 0 $ for $n\ge 1$.
Notice, however, that these off-shell states may have a negative norm
since they are off-shell.
This is better discussed in Section \ref{sec:DDF_examples} with the
use of examples.

\subsection{The general solution to Virasoro constraints}

Let us turn to the main point: how to obtain the general solution to the
Virasoro constraints using the framed DDF.
Instead of writing the more general expression we discuss the first
non-trivial levels.

We start with the $N=1$ level where the generic on-shell framed DDF state is
\begin{align}
  |G_1(E)\rangle
  =
  \underline{\epsilon}_i\, \uA^i_{-1}(E) |\uk_T\rangle
  +
  \underline{\epsilon}_-\, \tilde\uA^-_{-1}(E) |\uk_T\rangle
  ,
  \label{eq:DDFcov_lvl1}
\end{align}
which shows the true independent \dofs associated to the \lc
polarizations $\underline{\epsilon}_i$
since the Brower state
$\tilde\uA^-_{-1}(E) |\uk_T\rangle$
is null on shell and decouples from everything and the associated
polarization $\underline{\epsilon}_{-}$ is irrelevant.
The number of independent $\underline{\epsilon}_i$
is obviously independent of the frame $E$
and can be chosen at will.
In principle, they depend on the frame $E$ but the
two polarizations $\underline{\epsilon}_i(E)$
and $\underline{\epsilon}_i(\hat E)$ chosen in two different reference
frames related by a {\sl global} Lorentz rotation are the same,
since $\underline{\epsilon}_i$ do not transform.
%%but the state  $  |G_1(E)\rangle$ .
The sets
$\{\underline{\epsilon}_i(E),\, \underline{\epsilon}_-(E)\}$
and
$\{\underline{\epsilon}_i(\hat E),\, \underline{\epsilon}_-(\hat E)\}$
which are
related by a {\sl local} Lorentz transformation
can be mapped into each other.
Their relation is non-trivial and requires the use
of gauge transformations generated by a Brower state as detailed in
\ref{sec:N=1 gauge equiv with Brower}, i.e. a change
in $\underline{\epsilon}_-(E)$.

Therefore, to obtain the general solution we can keep the polarizations
$\underline{\epsilon}_i$ fixed and change the frame $E$ by a {\sl
global} Lorentz rotation which leaves $\underline{\epsilon}_i$ invariant.

Having in mind the goal of obtaining the general solution by changing the frame
$E$ while keeping the polarization fixed, we have not made the dependence on $E$ explicit. 

%% While $\underline{\epsilon}_i$ is the \dofs of the state it is not the
%% general solution of the Virasoro constraint.
The general solution to the Virasoro constraints %up to null states
is obtained by expressing
the previous non-null state in a covariant basis as
(see eq. \eqref{eq:level 1 most general gauge transf} for details)
\begin{align}
  |G_1(E)\rangle
  =&
  \left[
    \underline{\epsilon}_i\,
    %%              [\underline{\Pi}(E)]^i_\mu
    \left(
    E_{\mu}^{\underline{i}}
    -
    \frac{\uk^i_T}{ \uk^+_T}
    E_{\mu}^{\underline{+}}
    \right)
    +
    \underline{\epsilon}_-\,
    \left(
    E_{\mu}^{\underline{-}}
    +
    \frac{ -2 \uk^-_{T+1} \uk^+_T + 2 \vec \uk^2_T}{ 2 (\uk^+_T)^2}
    E_{\mu}^{\underline{+}}
    -
    \frac{\uk^j_T}{ \uk^+_T}
    E_{\mu}^{\underline{j}}
\right)
\right]\, \alpha^\mu_{-1} |k\rangle
  \nonumber\\
  =&
  \left[
  \underline{\epsilon}_i\,
  \left(
  E^{\underline{i}}_\mu
  -
  \frac{E^{\underline{i}}_\rho k^\rho}{ E^{\underline{+}}_\rho k^\rho}
  E_{\mu}^{\underline{+}}
  \right)
  +
  \underline{\epsilon}_-\,
  \left(
  E_{\mu}^{\underline{-}}
  +
  \frac{
    ( -2 E_{\rho}^{\underline{-}} E_{\sigma}^{\underline{+}}
    + E_{\rho}^{\underline{j}} E_{\sigma}^{\underline{j}}
    ) k^\rho\, k^\sigma
  }{ 2 ( E_{\rho}^{\underline{+}} k^\rho)^2 }
    E_{\mu}^{\underline{+}}
    -
    \frac{\uk^j}{ \uk^+}
    E_{\mu}^{\underline{j}}
\right)
  \right]
  \alpha^\mu_{-1} |k\rangle
,
  \nonumber\\
\mbox{with  }  
&
k_\mu
=
E_\mu^{\underline{\nu}}\, \uk_{T+1\, \nu}
=
E_\mu^{\underline{\nu}}\,
\left( \uk_{T\, \nu} - \frac{\delta^-_\nu}{ 2 \ap \uk_T^+} \right)   
,
  \label{eq:DDF_cov_embedding_lvl1}
\end{align}
where $|\uk_{T+1}\rangle$ is the level $N=1$ physical state momentum
obtained from the tachyon momentum $|\uk_T\rangle$.

From the previous expression
we read off the polarizations of
the general solution to the Virasoro constraints for
a $N=1$ state as a function of the physical momentum $k$ as
\begin{align}
\epsilon_\mu(k)
&=
\left(
\underline{\epsilon}_i
-
\underline{\epsilon}_-\,
\frac{\uk^i}{ \uk^+}
\right)
E^{\underline{i}}_\mu
+
\underline{\epsilon}_-\,
E_{\mu}^{\underline{-}}
+
\left(
  -
  \underline{\epsilon}_i
  \frac{E^{\underline{i}}_\rho k^\rho}{ E^{\underline{+}}_\rho k^\rho}
  +
  \underline{\epsilon}_-
  \frac{
    ( -2 E_{\rho}^{\underline{-}} E_{\sigma}^{\underline{+}}
    + E_{\rho}^{\underline{j}} E_{\sigma}^{\underline{j}}
    ) k^\rho\, k^\sigma
  }{ 2 ( E_{\rho}^{\underline{+}} k^\rho)^2 }
\right)
E_{\mu}^{\underline{+}}
  ,
  \nonumber\\
\mbox{with  }  
&
k_\mu
=
E_\mu^{\underline{\nu}}\, \uk_{T+1\, \nu}
,~~~~
k^2=0
,
\end{align}
in which it is clear that
the momentum $k$ can point in any direction due to the action
of the local frame $E$ on the fixed $\uk_{T+1}$.
The polarization is rotated
accordingly while still depending on the right number of \dofs.
This is the reason why this is the general solution.
Actually, only the little group of $\uk_{T+1}$ matters.

Obviously, it is by far simpler to consider the general solution up to null
states which reads
\begin{align}
  \epsilon_\mu(k)
  =
  \underline{\epsilon}_i\,
  %%              [\underline{\Pi}(E)]^i_\mu
  \left(
  E_{\mu}^{\underline{i}}
  -
  \frac{E^{\underline{i}}_\rho k^\rho}{ E^{\underline{+}}_\rho k^\rho}
  E_{\mu}^{\underline{+}}
  \right)
  ,~~~~
  k_\mu
=
E_\mu^{\underline{\nu}}\, \uk_{T+1\, \nu}
.
\end{align}

Let us now consider the general on shell-level $N=2$ DDF state.
This can be written using the framed DDFs as
\begin{align}
  |G_2(E)\rangle
  =&
  \underline{T}_i\, \uA^i_{-2}(E)\, |\uk_T\rangle
  +
  \underline{S}_{i j}\, \uA^i_{-1}(E)\, \uA^j_{-1}(E)\, |\uk_T\rangle
\nonumber\\
&
+
  \underline{g}_{i -}\, \uA^i_{-1}(E)\, \tilde \uA^-_{-1}(E)\, |\uk_T\rangle
+
  \underline{g}_{- -}\, \left( \tilde \uA^-_{-1}(E) \right)^2\, |\uk_T\rangle
  +
  \underline{g}_-\, \tilde \uA^-_{-2}(E)\, |\uk_T\rangle
,
  \label{eq:DDFcov_lvl2}
\end{align}
which shows the true independent \dofs associated to the \lc
polarizations $\underline{T}_i$ and $\underline{S}_{i j}$
since the other polarizations
$\underline{g}_{i -}$, $\underline{g}_{- -}$ and
$\underline{g}_{-}$ are associated to null states (when on-shell).
The polarizations $\underline{T}_i$ and $\underline{S}_{i j}$
are independent and can be chosen at will in a given frame $E$.
The number of polarizations is obviously independent of the frame $E$.
These polarizations are invariant under a global Lorentz rotation,
but the polarizations in different frames related by a tangent Lorentz are not. They can be mapped into each other by a
transformation which is non-trivial and requires the use
of gauge transformations generated by Brower null states.

We can then rewrite the previous state using the usual ``curved
index'' operators.
From this expression we can read the general covariant solution to
the Virasoro constraints for level $N=2$.
As in the $N=1$ case this happens because the proper number of \dofs
is embedded into the covariant polarization in a momentum-dependent
way.
The arbitrary direction of the physical momentum $k$
is achieved by the local frame $E$ which allows the
momentum to point in any direction.

We first write the generic non-null part of the previous state
in an oscillator flat basis $\ualpha^\mu_{-n}$.
We could write this expression using the projector
$\underline{\Pi}(E)$ but it is also instructive to write the full
expanded expression down as
\begin{align}
  |G_2\rangle
  =&
  \left(
  \underline{T}_i \underline{T}^{(i)}_l
%%  +
%%  \underline{S}_{i j} \underline{T}^{(i j)}_l
  \right)
  \ualpha^l_{-2} |\uk_{T+2}\rangle
  +
  \left(
  \underline{T}_i \underline{T}^{(i)}_+
  +
  \underline{S}_{i j} \underline{T}^{(i j)}_+
  \right)
  \ualpha^+_{-2} |\uk_{T+2}\rangle
  \nonumber\\
  &
  +
  \left(
%  \underline{T}_i \underline{S}^{(i)}_{l m}
%  +
  \underline{S}_{i j} \underline{S}^{(i j)}_{l m}
  \right)
  \ualpha^l_{-1} \ualpha^m_{-1} |\uk_{T+2}\rangle
  +
  \left(
  \underline{T}_i \underline{S}^{(i)}_{+ m}
  +
  \underline{S}_{i j} \underline{S}^{(i j)}_{+ m}
  \right)
  \ualpha^+_{-1} \ualpha^m_{-1} |\uk_{T+2}\rangle
  \nonumber\\
  &+
  \left(
  \underline{T}_i \underline{S}^{(i)}_{+ +}
  +
  \underline{S}_{i j} \underline{S}^{(i j)}_{+ +}
  \right)
  \ualpha^+_{-1} \ualpha^+_{-1} |\uk_{T+2}\rangle
  ,
\end{align}
where we used
 \begin{align}
\uA^i_{-2} |\uk_T\rangle
=&         
  \left(
  \underline{T}^{(i)}_l   \ualpha^l_{-2} 
  +
  \underline{T}^{(i)}_+ \ualpha^+_{-2}
    \right)
|\uk_{T+2}\rangle
+
\left(
\underline{S}^{(i)}_{+ m} \ualpha^+_{-1} \ualpha^m_{-1}
+
\underline{S}^{(i)}_{+ +} \ualpha^+_{-1} \ualpha^+_{-1}
\right)
|\uk_{T+2}\rangle
,
\nonumber\\
\uA^i_{-1} \uA^j_{-1} |\uk_T\rangle
=&
\left(   
%%  \underline{T}^{(i j)}_l \ualpha^l_{-2}
%%  +                   
  \underline{T}^{(i j)}_+ \ualpha^+_{-2}
  \right)
   |\uk_{T+2}\rangle
+
\left(
\underline{S}^{(i j)}_{l m} \ualpha^l_{-1} \ualpha^m_{-1}
+
\underline{S}^{(i j)}_{+ m} \ualpha^+_{-1} \ualpha^m_{-1}
+
\underline{S}^{(i j)}_{+ +} \ualpha^+_{-1} \ualpha^+_{-1}
\right)
|\uk_{T+2}\rangle
,
\end{align}
and $|\uk_{T+2}\rangle$ is the level $N=2$
physical state momentum obtained from the
tachyon momentum $|\uk_T\rangle$.
There is no $\underline{S}^{(i)}_{l m}$ due to the little group
transformation property of $\uA^i_{-2} |\uk_T\rangle$.
Similarly for $\underline{T}^{(i j)}_l$.
The expressions for
$\underline{T}^{(i)}_l$, $\underline{T}^{(i j)}_l$,
$\underline{S}^{(i)}_{l m}$, $\underline{S}^{(i j)}_{l m}$ 
can be obtained in an algorithmically straightforward way and in the
present case are given in \cref{eq:Ai-2,eq:ij11state}.
%%\igor{add eqs}

We then change to the usual ``curved'' coordinates and obtain the
general covariant solution for the $N=2$ on-shell states
up to null states 
(see \cref{eq:T and S Ai-2,eq:curved TSvals11} for the explicit expressions)
\begin{align}
  T_\mu
  =&
  \left(
  \underline{T}_i \underline{T}^{(i)}_l
%%  +
%%  \underline{S}_{i j} \underline{T}^{(i j)}_l
  \right)
  E^{\underline{l}}_\mu
  +
  \left(
  \underline{T}_i \underline{T}^{(i)}_+
  +
  \underline{S}_{i j} \underline{T}^{(i j)}_+
  \right)
  E^{\underline{+}}_\mu
  ,
  \nonumber\\
  S_{\mu\nu}
  =&
    \left(
%%  \underline{T}_i \underline{S}^{(i)}_{l m}
%%  +
  \underline{S}_{i j} \underline{S}^{(i j)}_{l m}
  \right)
  E^{\underline{l}}_\mu   E^{\underline{m}}_\nu
  +
    \left(
  \underline{T}_i \underline{S}^{(i)}_{+ m}
  +
  \underline{S}_{i j} \underline{S}^{(i j)}_{+ m}
  \right)
  E^{\underline{+}}_\mu   E^{\underline{m}}_\nu
  \nonumber\\
  &+
  \left(
  \underline{T}_i \underline{S}^{(i)}_{+ +}
  +
  \underline{S}_{i j} \underline{S}^{(i j)}_{+ +}
  \right)
  E^{\underline{+}}_\mu   E^{\underline{+}}_\nu
.
\label{eq:DDF emebdding in covariant}
\end{align}
This is the general solution of the level $N=2$ Virasoro constraints up to null states.
It is given by the linear superposition of the proper number of
independent solutions, for example associated to the polarization
$\underline{T}_i$ we have the covariant solution
\begin{align}
  T_\mu^{(i)}
  =&
  \underline{T}^{(i)}_l\,  E^{\underline{l}}_\mu
  +
  \underline{T}^{(i)}_+\,  E^{\underline{+}}_\mu
  ,
  \nonumber\\
  S_{\mu\nu}^{(i)}
  =&
  \underline{S}^{(i)}_{+ m}
  E^{\underline{+}}_\mu   E^{\underline{m}}_\nu
  +
  \underline{S}^{(i)}_{+ +}
  E^{\underline{+}}_\mu   E^{\underline{+}}_\nu
,
\label{eq:one DDF emebdding in covariant}
\end{align}
which depends on the frame $E$ or said differently on the on-shell
momentum $k = E \uk_{T+2}$ which is related to the fixed $\uk_{T+2}$
but it is anyhow arbitrary because of the action of $E$.

Notice how this solution is not in the usual gauge, where the
polarizations are transverse and traceless
(see \eqref{eq:Ai-2} for the explicit form).

Some steps along this line of thought (but using the covariant formalism) were
taken in \cite{Labastida:1988wi} where a projector was introduced.

\subsection{The \lc string amplitudes point of view}
Keeping $\uk_T$ and therefore $\uk_{T+N}$ fixed and varying $E$
for discussing the spectrum is perfectly fine.
It also implies that polarizations have an ``absolute'' meaning since
they are the polarizations seen in a fixed reference frame, for
example the rest frame of massive particles.
For example $\uepsilon_i$ is the polarization of a photon as seen in
a well-defined frame.

However, it is not the best choice for dealing with amplitudes.
In fact, in the presence of $M$ particles with momenta
$k_{\sN r}= E_{\sN r}\, \uk_{T+N_{\sN r}}$ ($r=1,2, \dots M$)
we should be obliged to consider different frames $E_{\sN r}$.
This implies the use of DDF operators
$\uA^i_n( E_{\sN r})$
each of which
contains different flat coordinates
left-moving string coordinates $\uL_{\sN r}^\mu(z)$ then
\begin{equation}
  %% e^{i k_{\sN r \mu}\,     L^\mu(z) }
  %% =
  %% e^{i \uk_{T+N_{\sN r} \nu}\,    E_{\sN r \mu}^{\underline{\nu} }\, L^\mu(z)}
  %% =
  %% e^{i \uk_{T+N_{\sN r} \nu} \uL_{\sN r}(z)}
  e^{i n \frac{ \uL_{\sN r}(z) } { \ap \up^+_{0 \sN r}} }
    =
    e^{i n
      \frac{ E_{\sN r \mu}^{\underline{+}} L^\mu(z) }
           { \ap E_{\sN r \mu}^{\underline{+}} p^\mu_{0}}
    }
  .
\end{equation}
This complicates the evaluation of DDF amplitudes.
The simplest point of view for evaluating amplitudes is to keep $E=E_0$
fixed and vary $\uk_T$.
This means we have many $  k_{\sN r}$,
explicitly we write
\begin{equation}
  k_{\sN r}
  =
  E_{\sN r}\, \uk_{T+N_{\sN r}}
  =
  E_0\, \uk_{T+N_{\sN r} \sN r}
  .
\end{equation}
This is always possible since for any $k_{\sN r}$ we can determine
the corresponding momentum at fixed $E_0$ as
$
\uk_{T+N_{\sN r} \sN r}
= E_0^{-1}\,k_{\sN r}
= E_0^{-1}\,   E_{\sN r}\, \uk_{T+N_{\sN r}}
$.

However, this implies that polarizations are not anymore
``absolute'' but they represent quantities measured in
different frames, so they cannot be compared directly.

These two points of view can be seen explicitly in the case of $M$
photons where we write
\begin{align}
  \epsilon_{\sN r \mu}(k_{\sN r})
  =&
  \underline{\epsilon}_i\,
  \left(
  E_{\sN r \mu}^{\underline{i}}
  -
  \frac{E^{\underline{i}}_{ \sN r \rho} k^\rho_{\sN r}}
       { E^{\underline{+}}_{\sN r \rho} k^\rho_{\sN r}}
  E_{\sN r\mu}^{\underline{+}}
  \right)
  ,~~~~
  k_{\sN r \mu}
=
E_{\mu \sN r}^{\underline{\nu}}\, \uk_{T+1\, \nu}
\nonumber\\
=&
  \underline{\epsilon}_{ \sN r i}\,
  %%              [\underline{\Pi}(E)]^i_\mu
  \left(
  E_{0 \mu}^{\underline{i}}
  -
  \frac{E^{\underline{i}}_{0 \rho} k^\rho_{\sN r}}
       { E^{\underline{+}}_{0 \rho} k^\rho_{\sN r }}
  E_{0 \mu}^{\underline{+}}
  \right)
  ,~~~~
  k_{ \sN r \mu}
=
E_{0 \mu}^{\underline{\nu}}\, \uk_{\sN r T+1\, \nu}
,
\end{align}
where the transversality condition reads in the first case
\begin{equation}
  \epsilon_{\sN r \mu}(k_{\sN r})\,
  k_{ \sN r \mu}
  =
   \underline{\epsilon}_i\,
   \uk_{T+1}^i
   ,
\end{equation}
so that it is always satisfied because of  the same flat coordinate
condition
and in the second case
\begin{equation}
  \epsilon_{\sN r \mu}(k_{\sN r})\,
  k_{ \sN r \mu}
  =
   \underline{\epsilon}_{\sN r i}\,
   \uk_{\sN r  T+1}^i
   ,
\end{equation}
so that it is satisfied because we have adapted the flat polarization
to the different flat frames.

%%%%%%%%%%%%%%%%%%%%%%%%%%%%%%%%%%%%%%%%%%%%%%%%%%%%%%%%%%%%%%%%%%%%%%
%%%%%%%%%%%%%%%%%%%%%%%%%%%%%%%%%%%%%%%%%%%%%%%%%%%%%%%%%%%%%%%%%%%%%%
%%%%%%%%%%%%%%%%%%%%%%%%%%%%%%%%%%%%%%%%%%%%%%%%%%%%%%%%%%%%%%%%%%%%%%

\section{Mapping \lc operators into framed DDFs}
\label{sec:mapping lc to DDFs}
The results of the previous section are naturally understood as the
embedding of \lc Fock space into covariant Fock space.
The local frame $E$ allows the original \lc to be rotated
to point in any possible \lc direction and
therefore yields the general solution to the Virasoro constraints.

Let us make this more formal.
We have a family of injective algebra homomorphisms $i(E)$
parameterized by the local frame $E$ of the \lc operators into the covariant
ones given by
\begin{align}
  \alpha^i_{n (\lcsh)}\,
  \stackrel{i(E)}{  \rightarrow}
  \uA^i_n(E)
  ,
\end{align}
with inverse
\begin{align}
   \uA^i_n(E),\,
\tilde \uA^-_n(E)
  \stackrel{i^{-1}(E)}{  \rightarrow}
  \alpha^i_{n (\lcsh)},\,
  0
  ,
\end{align}
which shows that $\tilde \uA^-_n(E)$ are the proper operators to
consider since they are in the kernel
and are the fundamental operators to consider in this map,
since otherwise we would have the mapping between \lc composite
operators and the Brower operators as
$
\hat \alpha^-_{n (\lc)} \leftrightarrow \uA^-_n(E)
$
.
The fact that the improved Brower operators are the fundamental ones
is also seen in the fact that they are the correct operators needed to get the
off-shell gauge invariance in the space of off-shell DDF  states as
shown in section \ref{sec:N=1 gauge equiv with Brower}.

We can also reformulate the previous paragraph by saying that when we
also consider the null states generated by improved Brower operators,
we can span the whole momentum space, i.e also $\uk^+=0$ but the point $k_\mu=0$.

One could think of extending the previous homomorphism to the zero
mode sector as
\begin{align}
  \alpha^i_{n (\lcsh)},\,
  x_{0 (\lcsh)}^i,\,
  x_{0 (\lcsh)}^-,\,
  p_{0 (\lcsh)}^i,\,
  p_{0 (\lcsh)}^-\,
  \stackrel{\hat i(E)}{  \rightarrow}
  \uA^i_n(E),\,
  \ux_{0}^i(E),\,
  \ux_{0}^-(E),\,
  \up_{0}^i(E),\,
  \up_{0 }^-(E)\,
  ,
\end{align}
but this is tricky since the zero modes are connected to the Fock
space on which we represent them and so we cannot get a clear map
between algebras.

Instead we can get an homomorphism $\phi$ between the vector spaces as
\begin{equation}
  \phi\left(
  \prod_{i\, n}\,  %\prod_n\,
  \left[\alpha^i_{n (\lcsh)} \right]^{N^i_n}\,
  |k_{- (\lcsh)}, k_{i (\lcsh)}\rangle
  \right)
  =
  \prod_{i\, n}\, %\prod_n\,
  \left[\uA^i_{n} (E) \right]^{N_{i\,n}}\,
  \Bigg|\uk_{-},\, \uk_{i},\, \uk_{+}= \frac{\uk_i^2+\sum N_{i\,n}-1}{2 \uk_-}\Bigg\rangle
.
\end{equation}

Explicitly from the generic level $N=2$ \lc state
(in \lc there is no $k_+$ dependence),
\begin{align}
  |G_{2 (\lcsh)}\rangle
  =
  {T}_i\,  \alpha^i_{-2 (\lcsh)} |k_{- (\lcsh)}, k_{i (\lcsh)}\rangle
  +
  {S}_{i j}\, \alpha^i_{-1 (\lcsh)} \alpha^j_{-1 (\lcsh)} |k_{-(\lcsh)}, k_{i (\lcsh)}\rangle
  ,
\end{align}
we can get the covariant state \eqref{eq:DDFcov_lvl2}
which has the ``queue'' of $\ualpha^+(E)$ operators.
These $\ualpha^+(E)$ operators vanish on the \lc and are therefore not
present in the \lc state.

To complete the mapping, we need to compute the covariant momentum.
The tachyon momentum associated with the DDF construction,
can be reverse engineered and
has a well-defined and computable $\uk_{T +}$.
It can be calculated from $\uk_{T -}$, $\uk_{T i}$, the level
$N$ and the mass shell condition of the tachyon.

Using the previous idea of embedding, we can now connect the physical
covariant polarizations with the ones used in \lc formulation
thus making explicit the obvious naive idea.

%% In this section we show that the usual light cone polarization is
%% actually embedded in the covariant states flat space DDF state.

This identification is supported by the mapping of the Lorentz algebra
expressed using the DDF and Brower operators\footnote{
Actually we write the action of
Lorentz generators on DDF and Brower operators using DDF and Brower
operators.
This is not the same as expressing the Lorentz generators using DDF
and Brower operators since they span a subset of all covariant operators.
}
into
the \lc Lorentz algebra
(this is very similar to what was done in \cite{Goddard:1972ky})
and by the explicit computation of the second Casimir of the Poincar\'e group
in the \lc formalism and in the covariant one.

As detailed in appendix \ref{app:Lorentz_algebra_using_DDF}
the action of Lorentz generators on DDF $\uA^i_n$ and
Brower $\tilde A^-_n$ ($n\ge 1$) operators can be written as
\begin{align}
  \underline{M}^{i j}|_{DDF}
  =&
  i
  \sum_{n\ne 0} \frac{1}{n} \uA_n^{i}(E)\, \uA^j_{-n}(E)
  ,
  \nonumber\\
  \underline{M}^{+ i}|_{DDF}
  =&
  0
  ,
  \nonumber\\
  \underline{M}^{+ -}|_{DDF}
  =&
  \mbox{not possible}
  ,
  \nonumber\\
    \underline{M}^{- i}|_{DDF}
  =&
  i
  \sum_{m\ne 0} \frac{1}{m} \uA_m^-(E) \uA^i_{-m}(E)
  %% -
  %% \frac{1}{ (\ualpha_0^+)^2} \underline{M}^{+ i}(E) \cL_0(E)
  \nonumber\\
  =&
  i
  \sum_{m\ne 0} \frac{1}{m} {\tilde \uA}_m^-(E) \uA^i_{-m}(E)
  +
  \frac{i}{ 2 \ualpha_0^+}
  \sum_{m\ne 0} \frac{1}{m} \cL_m(E) \uA^i_{-m}(E)
  %% -
  %% \frac{1}{ (\ualpha_0^+)^2} \underline{M}^{+ i} \cL_0(E)
.
\label{eq:Lorentz using DDFs}
\end{align}     
Few comments are needed.
The generator $\underline{M}^{+ -}$ acts only on $\uA^-$ and requires
the use of zero modes.
In the same way the generator $  \underline{M}^{- i}|_{DDF}$ also has
a part which depends on zero modes.
These two points are discussed in the Appendix \ref{app:Lorentz_algebra_using_DDF}.

Finally, note that the generator $  \underline{M}^{- i}|_{DDF}$ is not
normal-ordered - the difference from the normal-ordered version is a
term that gives no change on the variations of DDFs.

%%%%%%%%%%%%%%%%%%%%%%%%%%%%%%%%%%%%%%%%%%%%%%%%%%
\subsection{Comparison with the \lc expressions}

If we use the \lc gauge where $\alpha^+_{0 (\lcsh)}$ is the only non vanishing
operator, from the Virasoro conditions we get ($n\ne 0$)
\begin{align}
  L_n
  =&
  \oh
  \sum_m \alpha_{n-m} \cdot \alpha_{m}
  ~\Rightarrow~
  -\alpha^+_{0  (\lcsh)} {\hat \alpha}^-_n
  + \oh \sum_m \alpha_{n-m  (\lcsh)}^i  \alpha_{m (\lcsh)}^i
  =
  0
  \nonumber\\
  &
  \Longrightarrow
  {\hat \alpha}^-_n
  =
  \frac{1}{2 \alpha^+_{0 (\lcsh)} }
  \sum_m \alpha_{n-m (\lcsh)}^i   \alpha_{m (\lcsh)}^i
     .
\end{align}
We can then compute for $n\ne 0$
\begin{align}
  [ \alpha^+_{0 (\lcsh)} \alpha^j_{n  (\lcsh)},\, \alpha^-_{m  (\lcsh)}]
  =
   n\,
   \alpha^j_{n+m  (\lcsh)}
  ,
\end{align}
and then the \lc Lorentz generators
\begin{align}
  M^{+ i}_{ (\lcsh) }
  =&
  x_0^+ p_0^i - x_0^i p_0^+
, 
\nonumber\\
M^{- +}_{ (\lcsh) }     
  =&
  x_0^- p_0^+ - x_0^+ p_0^-
  ,
  \nonumber\\   
  M^{- i}_{ (\lcsh) }
  =&
  x_0^- p_0^i - x_0^i p_0^-
  -
  i \sum_{n=1}^\infty
  \frac{ \hat \alpha^-_{-n (\lcsh)} \alpha^i_{n  (\lcsh)}
    -  \alpha^i_{-n  (\lcsh)}  \hat \alpha^-_{n (\lcsh)}
  }{n}
,
\end{align}
which matches the expression for the DDF in \eqref{eq:Lorentz using DDFs}
when this is restricted to non-zero modes,
%% \subsection{Explicit mapping between DDF and \lc operators}
%%We can summarize
and we make 
the map between DDF and Brower operators and the \lc
ones for $n\ne 0$ as
\begin{align}
  \uA^i_n(E) \leftrightarrow \alpha^i_{n  (\lcsh)}
  ,~~
 \uA^-_n(E) \leftrightarrow  \hat \alpha^-_{n (\lcsh)}
 ,~~
  \tilde \uA^-_n(E) \leftrightarrow  0 %\alpha^-_{n  (l c)}
  ,
\end{align}
where the mapping for the improved Brower operators is zero since they
contain the transverse Virasoro generators
$\cL_n \leftrightarrow L_{n
(\lcsh)}^{(transverse)}\sim \hat \alpha^-_{n (\lcsh)}$
so it cancels, as follows from the first mapping rule.

%%%%%%%%%%%%%%%%%%%%%%%%%%%%%%%%%%%%%%%%%%%%%%%%%%%%%%%%%%%%%%%%%%%%%%
%%%%%%%%%%%%%%%%%%%%%%%%%%%%%%%%%%%%%%%%%%%%%%%%%%%%%%%%%%%%%%%%%%%%%%
%%%%%%%%%%%%%%%%%%%%%%%%%%%%%%%%%%%%%%%%%%%%%%%%%%%%%%%%%%%%%%%%%%%%%%

\section{Examples of framed DDF states}
\label{sec:DDF_examples}
In this section we discuss the lowest levels in order to clarify
the ideas and observations put forward in the previous section.
We start with the massless $N=1$ level and then discuss the $N=2$ level.
A similar discussion was done in the early days of string theory
in \cite{DelGiudice:1970dr} where the physical and null states were
discussed in $D=4$ in the covariant formalism.

In particular, we are interested in discussing the gauge chosen by DDF
states,
the local and global Lorentz transformations,
and the role of the Brower states, also in connection with the off-shell extension of the states.

\subsection{Level 1 states}
We start with the simplest case of the photon where computations can
also be done on the back of an envelope.

\subsubsection{Level 1 DDF states}
The $N=1$ DDF reads
\begin{align}
    \uA^i_{-1} |k_T\rangle
    =
    \left[
      \ualpha^i_{-1}
    -
    \frac{\uk^i }{ \uk^+ } \ualpha^+_{-1}
    \right]
    |\uk_{T + 1}^-, \uk_{T}^+, \uk_{T }^i \rangle   
,
\end{align}
so that the physical momentum is related to the tachyon momentum as
and the only momentum component which differs from $\uk_T$ is
$\uk_{T}^-$, explicitly
\begin{equation}
  \uk^-_{T+1} = - \uk_{T+1\,+} = \uk_{T +}^- + \frac{1}{2\ap \uk^+}
  ,~~~~
  \uk_- = - \uk^+ = \uk_{T -}
  ,~~~~
  \uk_i = \uk^i = \uk_{T i}
  .
\end{equation}

Comparing with the most general covariant state 
\begin{equation}
  |\epsilon, k\rangle
  = \uepsilon_\mu \ualpha^\mu_{-1} | \uk_\nu\rangle
  = \epsilon_\mu \alpha^\mu_{-1} | k_\nu\rangle
  ,
\end{equation}
we get that the components of the flat polarization are
\begin{equation}
  \underline{\epsilon}^{(i)}_j=\delta^i_j
  ,~~~~
  \underline{\epsilon}^{(i)}_+=- \frac{\uk^i }{ \uk^+}
  ,~~~~
  \underline{\epsilon}^{(i)}_-=0
  ~~
  \leftrightarrow
  ~~
  \underline{\epsilon}^{(i)}_\mu
  =
  \delta^i_\mu - \frac{\uk^i }{ \uk^+} \delta^+_\mu 
  ,
\end{equation}
and in curved ones are
\begin{equation}
 {\epsilon}^{(i)}_\mu
  =
  E^{\underline{i}}_\mu
  - \frac{E^{\underline{i}}_\rho k^\rho }{ E^{\underline{+}}_\sigma k^\sigma} E^{\underline{+}}_\mu 
  .
\end{equation}
This form reveals the projector discussed in \eqref{eq:DDF creator with projector}.

The only Virasoro condition reads
\begin{equation}
  L_1  |\epsilon, k\rangle = 0
  ~~\Rightarrow~~
  k^\mu \epsilon_\mu=0
  ,
\end{equation}
and it is trivially satisfied because of the projector.

It can also be verified in flat coordinates as
\begin{equation}
  k^\mu \epsilon_\mu
  =
  \uk^i_T \cdot 1
  +
  \uk^+_T  \cdot \frac{\uk^i_T }{ \uk_{ T -}}
  \equiv
  0
  .
\end{equation}

\subsubsection{Level 1 Brower state}
In the discussion on the gauge transformation below we use the
``Brower state''
\begin{align}
  \uA^-_{-1} | \uk_T \rangle
  &=
  \left\{
  \left[
    e^{-i \frac{\ux_0^+}{\dap \up_0^+} }
    \left(
    \ualpha^-_{-1}
    -
    \frac{\ualpha_0^-}{\ualpha_0^+} \ualpha_{-1}^+
    \right)
    \right]
    -
    \frac{1}{2 \ualpha_0^+}
    \left[
    2\, e^{-i \frac{\ux_0^+}{\dap \up_0^+} }
    \frac{\ualpha_{-1}^+}{\ualpha_0^+}
    \right]
    \right\}
    | \uk_T \rangle
    ,
\end{align}
where the $[ \dots]$ are the contributions from the two terms in
$\uA^-$.
Notice the effect of normal ordering in the first contribution.
Evaluating explicitly the previous expression we get
\begin{align}
  =&
  \left[
    \ualpha^-_{-1}
    -
    \frac{1}{\uk^+}
    \left(
    \uk_T^-
    +
    \frac{1}{\dap \uk^+}
    \right)
    \ualpha_{-1}^+
    \right]
  |\uk^-=   \uk_T^-    +   \frac{1}{\dap \uk^+}\,
  \uk^+ = \uk_T^+
  \rangle
  \nonumber\\
  =&
  \left[
    \ualpha^-_{-1}
    -
    \frac{\uk^-_{T+1}}{\uk^+}
    \ualpha_{-1}^+
    \right]
  |\uk_{T+1}\rangle  
  .
\end{align}
Again the flat polarization is given by
\begin{equation}
\underline{\epsilon}^{(-)}_\mu
  =
  \delta^-_\mu - \frac{\uk^-_{T+1} }{ \uk^+} \delta^+_\mu 
  ,
\end{equation}
and in curved one is
\begin{equation}
 {\epsilon}^{(-)}_\mu
  =
  E^{\underline{-}}_\mu
  - \frac{E^{\underline{-}}_\rho k^\rho }{ E^{\underline{+}}_\sigma k^\sigma} E^{\underline{+}}_\mu 
  .
\end{equation}

This expression satisfies the Virasoro conditions
\begin{align}
  L_1\,  \uA^-_{-1} | \uk_T \rangle
  =&
  (-\ualpha_0^+ \ualpha_{-1}^- -\ualpha_0^- \ualpha_{-1}^+ )
  \uA^-_{-1} | \uk_T \rangle
  =0
  ,
\end{align}
in an obvious way because of the projector $\Pi^-$.

It is also important to compute the norm of this state, which may also be {\sl negative} when off-shell.
This can be achieved in two ways -
either by direct computation on the explicit expression
\begin{align}
  \langle \uk_{T+1} |
  \left[
    \ualpha^-_{1}
    -
    \frac{\uk^-_{T+1}}{\uk^+}
    \ualpha_{1}^+
    \right]
  \,
    \left[
    \ualpha^-_{-1}
    -
    \frac{\ul^-_{T+1}}{\ul^+}
    \ualpha_{-1}^+
    \right]
  | \ul_{T+1}\rangle
  =&
  2
  \frac{\uk^-_{T+1}}{\uk^+}
  \,
  \delta(\uk -\ul)\,
  ,
\end{align}
or  by using the $\uA^-_m$ algebra (see \eqref{eq:A- A- algebra}) as,
\begin{align}
  \langle \uk_T |\uA_{1}^-\, \uA_{-1}^- | \ul_T\rangle
  =&
  \langle \uk_T |
  \left(
  2 \frac{\uA_{0}^-}{\ualpha_0^+} + 2 \frac{(1)^3}{(\ualpha_0^+)^2}
  \right)
  | \ul_T\rangle
  =
  2
  \frac{\uk^-_{T+1}}{\uk^+}
  \,
  \delta(\uk_T -\ul_T)\,
  .
\end{align}
Notice the different bras and kets in the two computations.

Actually we are also interested in the ``improved Brower state''
\begin{align}
  \tilde\uA^-_{-1} | \uk_T \rangle
  =
  \left(
  \uA^-_{-1}
  -
\frac{1}{2\ualpha_0^+} 2 \vec \uA_{-1} \cdot \vec \uA_{0} 
  \right)
  | \uk_T \rangle \nonumber \\
  =
  \left[
    \ualpha_{-1}^-
    +
    \frac{ -2 \uk^-_{T+1} \uk^+ + 2 \vec \uk^2}{ 2 (\uk^+)^2} \ualpha_{-1}^+
    -
    \frac{\uk^j}{ \uk^+} \ualpha_{-1}^j
    \right]
  | \uk_{T+1} \rangle
  ,
  \label{eq:Atilde1}
\end{align}
with norm
\begin{align}
  \langle \uk_T|  \tilde\uA^-_{1}   \tilde\uA^-_{-1} | \ul_T \rangle
  =
  \langle \uk_T|
  \left[
  2 \frac{1}{\ualpha^+_0 } \tilde\uA^-_{0}
  +
  \frac{26-D}{12} \frac{1}{ (\ualpha^+_0)^2 }
  \right]
  | \ul_T \rangle  \nonumber \\
  =
  - \frac{-2 \uk^-_{T+1} \uk^+ +  \vec \uk^2}{ 2 (\uk^+)^2}
  \delta(\uk-\ul)
  \label{eq: vev tAm0 is zero}
  ,
\end{align}
which vanishes on-shell only.

\subsubsection{Global Lorentz rotations
%% : to the very same DDF
%%   polarization corresponds all
%%   physical states related by a Lorentz transformation
}
\label{sec:N=1 global coords rotations}

As we stated at the beginning the local frame $ E^\mu_{\underline{\mu}}$ is acted upon by
two independent groups: the local Lorentz group, i.e. the group of tangent space
rotations, and the global Lorentz group.
Any global Lorentz transformation can be reabsorbed into $E$
without changing $\uA$.
More explicitly we have
\begin{align}
  \uepsilon^{(i)}_j\, \uA^j_{-1}(E) \ket{\uk_T}
  =&
  \epsilon^{(i)}_\mu(E)\, \alpha_{-1}^\mu |k\rangle
%%  ,
  %%  \nonumber\\
  ~~
  \Rightarrow
  ~~
  \epsilon^{(i)}_\mu(E)
  =&
  \uepsilon^{(i)}_j
  \left(
  E^{\underline{j}}_\mu
  -
  \frac{E^{\underline{j}}_\nu k^\nu }{ E^\upl_\nu k^\nu } E^\upl_\mu
  \right)
  ,
\end{align}
so that a global Lorentz transformation acts trivially on framed DDF
polarization, or said in a different way, {\sl framed DDF polarizations are
  Lorentz invariants with respect to global rotations}.
This means that the generic solution of the Virasoro conditions can be
expressed using a set of parameters, the DDF polarizations which are
invariant under global Lorentz transformations at the cost of changing $E$.

However, when two or more states are present we generically need as
many DDF polarizations in a given frame $E$
as states, but these polarizations describe the
states in all possible global coordinates which are related by a
global Lorentz transformation.
In fact, for two vectors the scalar product is invariant  
\begin{align}
  \epsilon_{\sN 1}^{(i_1)}(k_{\sN 1}) \cdot \epsilon_{\sN 2}^{(i_2)}(k_{\sN 2}) 
  =&
    \uepsilon_{\sN 1 j}^{(i_1)}\, \delta^{j l}\, \uepsilon_{\sN 2l}^{(i_2)}
  ,
\end{align}
and can be reproduced using two fixed DDF polarizations.

\subsubsection{Local Lorentz rotations 1: on-shell gauge equivalence
of different frames}
\label{sec:N=1 local coords rotations}

Suppose that we want to describe the very same physical state using two
different frames $E$ and $\hat E$.
Since the physical state is the same
these two frames are related by a tangent space Lorentz rotation.
Doing so, we actually get two different states that describe the same
physical state and must therefore be gauge
equivalent, i.e.
\begin{equation} 
  | k, \epsilon_{\hat E}\rangle
 =
 | k, \epsilon_{E}\rangle
 +
 L_{-1} \xi | k\rangle
 ,
\end{equation}
for some number $\xi$.
When we rewrite the previous equivalence using DDF states we get
\begin{align}
  \uepsilon_{(\hat E) i} \uA^i_{-1}(\hat E) |\uk_{T (\hat E)}\rangle
  =
  \uepsilon_{(E) i} \uA^i_{-1}(E) |\uk_{T (E)}\rangle
 +
 L_{-1}\,  \xi | k\rangle
 ,
\end{align}
and we require that both DDF states have the same physical momentum $|k\rangle$.
This means that
\begin{align}
  E^\mu_{\underline{\mu}} k_\mu
  =
  (\uk_{T (E) +} + \frac{1}{2\ap \uk_{ (E) -}},
  \uk_{ (E) -},
  \uk_{ (E) i}
  )
  ,~~~~
  \hat E^\mu_{\underline{\mu}} k_\mu
  =
  (\uk_{T (\hat E) +} + \frac{1}{2\ap \uk_{(\hat E) -}},
  \uk_{(\hat E) -},
  \uk_{(\hat E) i}
  )
  ,
\end{align}
or
\begin{align}
  \left(\uk_{T (\hat E) +} + \frac{1}{2\ap \uk_{(\hat E) -}},
  \uk_{(\hat E) -},
  \uk_{(\hat E) i}
  \right)
  =
  E^{-1} \hat E\,
  \left(\uk_{T (E) +} + \frac{1}{2\ap \uk_{(E) -}},
  \uk_{(E) -},
  \uk_{(E) i}
  \right)
  .
\end{align}

Then for the polarization we get
\begin{align}
  \uepsilon_{(\hat E) i} \hat E^{\underline{i}}_\mu
  =
  \uepsilon_{(E) i} E^{\underline{i}}_\mu
  +
  \xi k_\mu
  .
\end{align}
This is a set of $D$ equations in $D-1$ unknown, i.e.
$\uepsilon_{(E) i}$ and $\xi$.
The system is soluble on-shell where $k^2=0$ and $k\cdot \epsilon=0$.
In fact contracting with $k$ and (an auxiliary null vector) $\bar k$ ($k\cdot \bar k =-1$) we get
\begin{align}
  \uepsilon_{(\hat E) i} \uk^i_{\hat E}
  &=
  \uepsilon_{(E) i} \uk^i_{E}
  +
  \xi
  k^2
  ~~\Rightarrow~~
  0= \xi k^2
  ,
  \nonumber\\
  \uepsilon_{(\hat E) i} {\bar \uk}^i_{\hat E}
  &=
  \uepsilon_{(E) i} {\bar \uk}^i_{E}
  -
  \xi
  ,
\end{align}
where the first equation is identically zero on shell and the second
yields $\xi$.

\subsubsection{Local Lorentz rotations 2: off-shell gauge equivalence of different frames}
\label{sec:N=1 gauge equiv with Brower}
The result of the previous section is annoying since it is valid only
when the states are on-shell, while we would like to be able to go off-shell.

There is a way out using the Brower states generated by ${\tilde \uA}^-$.
Actually the two different (off-shell) states must be gauge equivalent as
\begin{equation} 
  | k, \epsilon_{(\hat E)}\rangle
 =
 | k, \epsilon_{(E)}\rangle
 +
 \zeta\, {\tilde \uA}^-_{-1}(E) |\uk_T\rangle
 .
\end{equation}
However, while the off-shell Brower state satisfies all Virasoro
conditions but the $L_0$ one, it may have {\bf negative norm} and 
only when it is on-shell it has zero norm.

Again, it is a system of $D$ equations in $D-1$ unknowns $\uepsilon_i$
and $\zeta$ when $\hat \uepsilon_i$, $\hat E$, $E$ and $k$ are given. 
However, it is soluble because the action of $L_1$ kills both the right and
left sides.
Let us discuss this in more detail.
Equating the states in global operator basis, we get (using \eqref{eq:Atilde1})
\begin{align}
\hat \uepsilon_i\, \Pi^{\underline{i}}_\mu(\hat E)
&=
\uepsilon_i\, \Pi^{\underline{i}}_\mu(E)
+
\zeta
\left(
    E_{\mu}^{\underline{-}}
    +
    \frac{ -2 \uk^-_{T+1} \uk^+ + 2 \vec \uk^2}{ 2 (\uk^+)^2} E_{\mu}^{\underline{+}}
    -
    \frac{\uk^j}{ \uk^+} E_{\mu}^{\underline{j}}
    \right)
    .
\label{eq:level 1 most general gauge transf}    
\end{align}     
Multiplying by $E^\mu_{\underline{-}}$ we can extract $\zeta$ as
\begin{align}
\zeta
=&
\hat \uepsilon_i\, \Pi^{\underline{i}}_\mu(\hat E)\, E^\mu_{\underline{-}}
,
\end{align}
and by $E^\mu_{\underline{i}}$ we can extract $\uepsilon_i$ as
\begin{align}
\uepsilon_i
=&
\hat \uepsilon_j\, \Pi^{\underline{j}}_\mu(\hat E)\, E^\mu_{\underline{i}}
-
\frac{\uk_i}{ \uk^+}\, 
\hat \uepsilon_j\, \Pi^{\underline{j}}_\mu(\hat E)\, E^\mu_{\underline{-}}
.
\end{align}
The equation which is obtained by multiplying with $E^\mu_{\underline{+}}$ is then
identically satisfied.

%__________________________________________________%
\subsection{Level $2$ states}
We now quickly examine the next level since computations are heavier and
not very illuminating.
In this case, we encounter for the first time two kinds of DDF state,
even though they belong to the same $D=26$ irrep.

\subsubsection{Level $2$ state from $\uA^i_{-2}$}
Similar to the previous example using (\ref{eq:DDFflatexpr}) we now compute
\begin{align}
\uA^i_{-2} |k_T\rangle
  =&
  \Biggl[
    \ualpha^i_{-2}
    -
    \frac{2}{ \sdap \uk_{ T}^+ } \ualpha^+_{-1} \ualpha^i_{-1}
\nonumber\\
    &+
    \left(
    \frac{-1}{ \sdap \uk_{ T}^+} \ualpha^+_{-2} 
    +
    \frac{1}{ (\sdap \uk_{ T}^+)^2 } \ualpha^+_{-1} \ualpha^+_{-1}     
    \right)
    \sdap \uk^i_T
    \Biggr]
    \Ket{\uk_{T}^- +  \frac{2 }{\dap \uk_{ T}^+}, \uk_{T}^+, \uk_{T i}} 
    \label{eq:Ai-2}
\nonumber\\
  =&
  \left(
  T^{(i)}_\mu\, \alpha^\mu_{-2}
  +
  S^{(i)}_{\mu\nu}\, \alpha^\mu_{-1} \alpha^\nu_{-1}
  \right)
  |k_\nu\rangle
  ,
\end{align}
so that the covariant polarizations are
\begin{align}
  T^{(i)}_\mu 
  =&
  E^{\underline{i}}_{\mu} 
  -
  \frac{\uk^i_T }{ \uk_{ T}^+} E^{\underline{+}}_{\mu}
  =
  \Pi^{\underline{i}}_\mu(E)
  ,
  \nonumber\\
  S^{(i)}_{\mu \nu} 
  =&
  S^{(i)}_{\nu \mu} = 
  -\frac{2}{\sdap \uk_{T}^+}
  E^{\underline{+}}_{(\mu}~
  E^{\underline{i}}_{\nu)}
  -
  2
  \frac{\uk^i_T }{ \uk_{ T}^+}
  \frac{1}{\sdap \uk_{T -}}
  \Efm{+}{\mu}~\Efm{+}{\nu}
  \nonumber\\
  =&
  \Pi^{\underline{i}}_{( \mu }(E)\, 
    E^{\underline{+}}_{\nu)} \frac{N}{\sdap \uk^+} \frac{1}{1!}
,
\label{eq:T and S Ai-2}
\end{align}
and the momentum is
\begin{equation}
  \uk = \uk_{T + 2} 
  .
\end{equation}
The Virasoro conditions are given by,
\begin{equation}
L_2:~
S^{\mu}_\mu
  +
  2 \sdap k^\mu T_\mu
  =
  0
  ,~~~~
L_1:~
  \sdap k^\nu S_{\nu \mu}
  +
  T_\mu
  =
  0
  ,
\end{equation}
and can be verified using the expressions for the covariant
polarizations above.
In particular using the properties of the projector
\begin{equation}
k^\mu\, \Pi^i_\mu(E)
=
\Pi^i_\mu(E)\, E^+_\nu \eta^{\mu\nu}
=
0
,
\end{equation}  
we see that $T$ is transverse, $S$ is traceless but not transverse
so $L_2$ condition is automatic while $L_1$ must be checked.
%% \begin{align}
%%   &
%%   0 + \sdap k^\mu_T T^{(i)}_\mu
%%  = k^\mu_T\left(\Efm{i}{\mu} + \frac{\uk^i_T}{\uk_{T-}} \Efm{+}{\mu}\right) = 
%%   0
%%   ,
%%   \nonumber\\
%%   \left(\Efm{i}{\mu} + \frac{\uk^i_T}{\uk_{T-}} \Efm{+}{\mu}\right) &+ \sdap k^\nu_T\left(  \frac{2}{\sdap \uk_{T -}}\Efm{+}{(\mu}~\Efm{i}{\nu)}
%%   +
%%   2
%%   \frac{\uk^i_T }{ \uk_{ T -}}
%%   \frac{1}{\sdap \uk_{T -}}\Efm{+}{\mu}~\Efm{+}{\nu}\right) = 0,
%% \end{align}
%% respectively.

\subsubsection{Level 2 from $\uA^i_{-1}\uA^j_{-1}$}
For two DDF operators successively acting on a pure momentum state we
have some {\sl self-interactions}.

The corresponding level $2$ DDF state is given by
\begin{align}
  \uA^i_{-1}\uA^j_{-1}\ket{k_T}
  = &
  \Bigg\{
  \left(
  \ualpha^i_{-1}
  + \frac{\uk^i_T}{\uk_{T-}}\ualpha^+_{-1}
  \right)
  \left(
  \ualpha^j_{-1}
  + \frac{\uk^j_T}{\uk_{T-}}\ualpha^+_{-1}\right)
\nonumber\\
&
+ \delta^{ij}
\left[
\frac{1}{\sdap \uk_{T-}}\frac{\ualpha^+_{-2}}{2}
+
\frac{1}{2}\left(\frac{\ualpha^+_{-1}}{\sdap    \uk_{T-}}\right)^2
  \right]
  \Bigg\}
\ket{\uk_{T}^- + \frac{2}{2\alpha' \uk_{T}^+},\uk_{T}^+,\uk_{T}^i}
\label{eq:ij11state}
\nonumber\\
  =&
  \left(
  T^{(ij)}_\mu\, \alpha^\mu_{-2}
  +
  S^{(ij)}_{\mu\nu}\, \alpha^\mu_{-1} \alpha^\nu_{-1}
  \right)
  |k_\nu\rangle
  ,
\end{align}
with curved polarizations
\begin{align}
&T^{(i j)}_\mu
=
\frac{\delta^{ij}}{2\sdap \uk_{T-}}
E^{\underline{+}}_\mu
,
\\
&
S^{(i j)}_{\mu\nu}
=
\Pi^{\underline{i}}_{ (\mu}(E)\,
\Pi^{\underline{j}}_{ \nu)}(E)
+
\frac{\delta^{ij}}{2(\sdap \uk_{T}^+)^2}
E^{\underline{+}}_\mu\, 
E^{\underline{+}}_\nu
.
\label{eq:curved TSvals11}
\end{align}
The polarizations for $i\ne j$ are transverse and traceless since
$
\Pi^{\underline{i}}_{ (\mu}(E)\,
\Pi^{\underline{j}}_{ \nu)}(E)\, \eta^{\mu \nu}
= \delta^{i j}
$.
On the contrary the ones for $i=j$ are neither transverse nor traceless.

As before the Virasoro conditions read,
\begin{equation}
L_2:~ S^{\mu}_\mu
  +
  2 \sdap k^\mu T_\mu
  =
  0
  ,~~~~
  L_1:~
  \sdap k^\nu S_{\nu \mu}
  +
  T_\mu
  =
  0
  ,
\end{equation}
and can be verified as follows,
\eq{
S^\mu_\mu = &\uS_{ii} = 1; ~ 2\sdap k^\mu T_\mu = - \delta^{ii} = -1 \\
&\implies S^{\mu}_\mu
  +
  2 \sdap k^\mu T_\mu
  =
  0.
  \label{eq:vir1_11}
}
and,
\eq{
&\sdap k^\nu S_{\nu i} = \sdap\left(\uk_T^+\frac{\uk_T^i}{\uk_{T-}} + \uk^i_T\right) = (-\uk^i_T + \uk^i_T) = 0; \\
\sdap k^\nu S_{\nu+} = &\sdap \left(\uk_T^+\left\{\frac{\uk^i_T \uk^j_T}{\uk_{T-}^2} + \frac{\delta^{ij}}{4\alpha' \uk_{T-}^2}\right\}\right) + \frac{\uk^i_T \uk^j_T}{\uk_{T-}} + \frac{\delta^{ij}}{2 \sdap \uk_{T-}} = 0 \\
&\implies \sdap k^\nu S_{\nu \mu}
  +
  T_\mu
  =
  0
  .
\label{eq:vir2_11}
}

%------------------------------------------
\subsubsection{Level 2 Brower states in $\uk^i=0$ frame}
To make computations easier we can perform them in the $\uk^i=0$ frame
where the particle can still move in $x^1$ direction and therefore it
is not the rest frame even if the rest frame belongs to this class.

Despite this ``nice'' kinematics, the results are quite opaque.

In this frame we have the following partial simplification
\begin{equation}
\Pi^{\underline{i}}_\mu
~\rightarrow~
E^{\underline{i}}_\mu
~~\mbox{ but }~~
\Pi^{\underline{-}}_\mu(E, \ualpha_0^-)
=
E^{\underline{-}}_\mu - \frac{\ualpha_0^-}{\uk_T^+} E^{\underline{+}}_\mu
,
\end{equation}
since $\ualpha_0^-\sim \uk^-\ne 0$ is not vanishing.
Moreover, the actual value of $\ualpha_0^-$ in the state does depend on
the order of $\uA^-$ since $\uk^-$ is altered by the action of the DDF and
Brower operators.

We can then evaluate the following improved Brower states
%% \begin{align}
%% \uA^-_{-2} |\uk_T \rangle
%% =&
%% \Bigg[
%% \left(
%% \Pi^{\underline{-}}_\mu(\uk_T^-)\, \alpha^\mu_{-2}
%% -
%% \frac{2}{(\sdap \uk_T^+)}\,
%% \Pi^{\underline{-}}_\mu(\uk_T^-)\, \alpha^\mu_{-1}\, \alpha^+_{-1}
%% \right)
%% \nonumber\\
%% &
%% -
%% \frac{1}{(\sdap \uk_T^+)}
%% \left(
%% 3 \frac{ \alpha_{-2}^+ }{ (\sdap \uk_T^+) }
%% - 5 \left( \frac{ \alpha_{-1}^+ }{ (\sdap \uk_T^+) }\right)^2
%% \right)
%% \Bigg]
%% |\uk_{T+2}\rangle
%% ,
%% %
%% \end{align}
%% and the improved ones
\begin{align}
\tilde \uA^-_{-2} |\uk_T \rangle
=&
\Bigg\{
\left[
%% \Pi^{\underline{-}}_\mu(\uk_T^-)\, \alpha^\mu_{-2}
\ualpha^-_{-2}
-
\frac{2\ap \uk_T^+ \uk_T^-}{(\sdap \uk_T^+)^2}\,
\ualpha^+_{-2}
%% -
%% \frac{2}{(\sdap \uk_T^+)}\,
%% \Pi^{\underline{-}}_\mu(\uk_T^-)\, \alpha^\mu_{-1}\, \ualpha^+_{-1}
-
\frac{2}{(\sdap \uk_T^+)}\,
\ualpha^-_{-1} \ualpha^+_{-1}
+
\frac{4\ap \uk_T^+ \uk_T^-}{(\sdap \uk_T^+)^3}\,
\left( \ualpha^+_{-1} \right)^2
\right]
\nonumber\\
&
+\left[
-
\frac{1}{(\sdap \uk_T^+)}
\left(
3 \frac{ \ualpha_{-2}^+ }{ (\sdap \uk_T^+) }
- 5 \left( \frac{ \ualpha_{-1}^+ }{ (\sdap \uk_T^+) }\right)^2
\right)
\right]
\nonumber\\
&
-
\frac{1}{(\sdap \uk_T^+)}\,
\oh
\left[
\vec \ualpha_{-1}^2
+
\frac{D-2}{2}
\left(
- \frac{ \ualpha_{-2}^+ }{ (\sdap \uk_T^+) }
+\oh \left( \frac{ \ualpha_{-1}^+ }{ (\sdap \uk_T^+) }\right)^2
\right)
\right]
\Bigg]
|\uk_{T+2}\rangle
,
\nonumber\\
=&
\Bigg\{
\left(
E^{\underline{-}}_\mu
+
\frac{D-14 - 8\ap \uk_T^+ \uk_T^-}{4 (\sdap \uk_T^+)^2}
E^{\underline{+}}_\mu
\right)\,
\alpha_{-2}^\mu
\nonumber\\
&+
\left(
-
\frac{2}{ \sdap \uk_T^+ }\,
E^{\underline{-}}_\mu\, E^{\underline{+}}_\nu
+
\frac{42-D}{8 (\sdap \uk_T^+)^3}\,
E^{\underline{+}}_\mu\, E^{\underline{+}}_\nu
-
\frac{1}{2 (\sdap \uk_T^+)}
E^{\underline{j}}_\mu\, E^{\underline{j}}_\nu
\right)\,
\alpha_{-1}^\mu\, \alpha_{-1}^\nu
\Bigg\}
|\uk_{T+2} \rangle
,
\end{align}
and
\begin{align}
(\tilde \uA^-_{-1})^2 |\uk_T \rangle
=&
%% \Pi^{\underline{-}}_\mu(\uk_{T+1}^-)\, \alpha^\mu_{-1}\,
%% \Pi^{\underline{-}}_\nu(\uk_T^-)\, \alpha^\nu_{-1}
%% |\uk_{T+2}\rangle
%% =
\Bigg\{
\left[
-
\frac{1}{\sdap \uk_T^+}\,
\ualpha_{-2}^-
+
\left( \ualpha_{-1}^- \right)^2
-
\frac{4\ap \uk_T^+ \uk_T^- +1 }{(\sdap \uk_T^+)^2}\,
\ualpha_{-1}^- \ualpha_{-1}^+
+
\frac{ (2\ap \uk_T^+ \uk_T^- +1)^2 }{(\sdap \uk_T^+)^4}\,
\left( \ualpha_{-1}^+ \right)^2
\right]
\nonumber\\
&
+
\left[
\frac{1}{(\sdap \uk_T^+)^3}\,
\ualpha_{-2}^+
-
\frac{1 }{(\sdap \uk_T^+)^2}\,
\ualpha_{-1}^- \ualpha_{-1}^+
+
\frac{ 2\ap \uk_T^+ \uk_T^- }{(\sdap \uk_T^+)^4}\,
\left( \ualpha_{-1}^+ \right)^2
\right]
\Bigg\}
|\uk_{T+2}\rangle
\nonumber\\
=&
\Bigg\{
\left(
-
\frac{1}{\sdap \uk_T^+}\,
E^{\underline{-}}_\mu
+
\frac{1}{(\sdap \uk_T^+)^3}\,
E^{\underline{+}}_\mu
\right)
\alpha_{-2}^\mu
\nonumber\\
&+
\Bigg(
E^{\underline{-}}_\mu\, E^{\underline{-}}_\nu
+
\frac{(2 \ap \uk_{T}^+ \uk_{T}^-)^2
      + 6\ap \uk_{T}^+ \uk_{T}^-
      + 1
      }{ (\sdap \uk_{T}^+)^4 }
E^{\underline{+}}_\mu\, E^{\underline{+}}_\nu
 \nonumber\\
 &
 \phantom{\Bigg[}
+
\frac{-4\ap \uk_{T}^+ \uk_{T}^- -2 }{(\sdap \uk_{T}^+)^2}
E^{\underline{+}}_\mu\, E^{\underline{-}}_\nu
\Bigg)
\alpha_{-1}^\mu\, \alpha_{-1}^\nu
\Bigg\}
|\uk_{T+2}\rangle
,
\end{align}
and
\begin{align}
\uA^i_{-1}\,\tilde\uA^-_{-1} |\uk_T \rangle
=&
\Bigg[
-
\frac{ 1 }{ \sdap \uk_T^+ }\,
\ualpha_{-2}^i
+
\ualpha_{-1}^i \ualpha_{-1}^-
-
\frac{2\ap \uk_T^+ \uk^-_{T}}{ (\sdap \uk_T^+ )^2}\,
\ualpha_{-1}^i\, \ualpha_{-1}^+
\Bigg]
|\uk_{T+2}\rangle
\nonumber\\
=&
\Bigg\{
\left(
-
\frac{1}{\sdap \uk_T^+}
E^{\underline{+}}_\mu
\right)\,
\alpha^\mu_{-2}
+
\left(
E^{\underline{i}}_\mu\,
E^{\underline{-}}_\mu
-
\frac{2\ap \uk_T^+\uk^-_{T}}{ (\sdap \uk_T^+)^2 }\,
E^{\underline{i}}_\mu\,
E^{\underline{+}}_\mu
\right)
\alpha_{-1}^\mu\, \alpha_{-1}^\nu
\Bigg\}
|\uk_{T+2} \rangle
,
\end{align}
where  
in the state $\tilde \uA^-_{-2} |\uk_T \rangle$
the first two $[\dots]$ are $\uA^-_{-2} |\uk_T \rangle$
 and the last one is the contribution from
$\cL_{-2}$.
In particular, the first $[\dots]$ is from $\oint \partial \uL^- \dots$
and the second $[\dots]$ is from
$\oint \frac{\partial^2 \uL^+}{\partial \uL^+} \dots$

In the state $(\tilde \uA^-_{-1})^2 |\uk_T \rangle$
the first $[\dots]$ is from $\oint \partial \uL^- \dots$
and the second $[\dots]$ is from
$\oint \frac{\partial^2 \uL^+}{\partial \uL^+} \dots$.
The contributions from $\cL_{-1}$ are absent in the other states
due to the choice of frame.

They have off-shell norms given by,
\begin{align}
\langle \ul_T | \tilde \uA^-_{2}\, \tilde\uA^-_{-2} |\uk_T \rangle 
&=
\frac{(26-D) +8 (1 + 2 \ap \uk_T^+ \uk_T^-)}{2  ( \sdap \uk_T^+)^2 }
\delta(\uk -\ul)
\nonumber\\
\langle \ul_T | \left( \tilde \uA^-_{1}\right)^2\,
\left(\tilde\uA^-_{-1}\right)^2 |\uk_T \rangle 
&=
8
\frac{
\left( 2 \ap \uk_T^+ \uk_T^- +1 \right)
\left( 4 \ap \uk_T^+ \uk_T^- +3 \right)
}{ ( \sdap \uk_T^+)^4 }
\delta(\uk -\ul)
\nonumber\\
\langle \ul_T | \tilde\uA^-_{1}\,  \uA^i_{1}\,
\uA^i_{-1}\,\tilde\uA^-_{-1} |\uk_T \rangle 
&=
\delta^{i j}
\frac{2 \ap \uk_T^+ \uk_T^- +1}{ ( \sdap \uk_T^+)^2 }
\delta(\uk -\ul)
,
\label{eq: vev N=2 tAs are zero}
\end{align}
which vanish in the critical momentum and on-shell where
$1 + 2 \ap \uk_T^+ \uk_T^-=0$.

Notice that these states are not orthogonal off-shell, in fact
\begin{align}
\langle \ul_T | \tilde \uA^-_{2}\, \left(\tilde\uA^-_{-1}\right)^2 |\uk_T \rangle 
&=
3 \frac{2 \ap \uk_T^+ \uk_T^- +1}{ ( \sdap \uk_T^+)^2 }
\delta(\uk -\ul)
.
\end{align}

\subsubsection{Off-shell gauge equivalence of different frames}
We are now in a position to observe that starting from an arbitrary frame, one can always choose the rest frame (or vice versa).

Consider for example the state,
\begin{align}
\uA^i_{-1}(\hat E)\,
\uA^j_{-1}(\hat E)\,|\hat \uk_T\rangle
,
\end{align}
with $i\ne j$.
We want to show that the expansion
\begin{align}
\uA^i_{-1}(\hat E)\,
\uA^j_{-1}(\hat E)\,|\hat \uk_T\rangle
=&
\Bigg[
R^{i j}_l\, \uA^l_{-2}(E)
+
R^{i j}_{l m}\, \uA^l_{-1}(E)\, \uA^m_{-1}(E)
\nonumber\\
+
G^{i j}_2\,\tilde\uA^-_{-2}(E)&
+
G^{i j}_1\,\tilde\uA^-_{-1}(E)\,\tilde\uA^-_{-1}(E)
+
G^{i j}_l\,\uA^l_{-1}(E)\,\tilde\uA^-_{-1}(E)
\Bigg]
\,|\uk_T\rangle
.
\end{align}
is well-defined. To see this, note that the previous relation amounts to $D + \oh D(D+1)= \oh( D^2+ 3 D)$ equations
for $\alpha^\mu_{-2}$ and $\alpha^\mu_{-1}\alpha^\nu_{-1}$, in
$2(D-2) + 2 + \oh (D-2)(D-1) =\oh( D^2 +D -2)$ unknown coefficients
$R_l, G_l$, $G_1, G_2$ and $R_{l m}$. 
The excess of $D+1$ equations is only apparent since both states
satisfy the off-shell $L_2$ and $L_1$ Virasoro conditions which yield
precisely $D+1$ equations associated to $\alpha^\mu_{-1}$ and $1$.

%------------------------------------------
\subsection{A curiosity}
With the formulation based on the local frame and the use of $\up_0^+$, the
DDF operators are true zero-dimensional conformal operators. Therefore, it
is possible to consider physical states built using different local
frames.
For example
\begin{align}
\uA^i(\tilde E)\, \uA^j(E)\, |\uk_T(E)\rangle
,
\end{align}
which is still a physical state.

%%%%%%%%%%%%%%%%%%%%%%%%%%%%%%%%%%%%%%%%%%%%%%%%%%%%%%%%%%%%%%%%%%%%%%
%%%%%%%%%%%%%%%%%%%%%%%%%%%%%%%%%%%%%%%%%%%%%%%%%%%%%%%%%%%%%%%%%%%%%%
%%%%%%%%%%%%%%%%%%%%%%%%%%%%%%%%%%%%%%%%%%%%%%%%%%%%%%%%%%%%%%%%%%%%%%

\section{Mean value of second order Casimir for Poincar\'e group for some physical states}
\label{sec:casimirs}
%% \ytableausetup{smalltableaux}
\ytableausetup{boxsize=0.5em}

In order to substantiate the map between DDF states and \lc ones and
in order to understand the average spin content of the DDF states, we compute the mean second order Casimir for some states.

In particular, at level $N=2$, we know that there is only one Poincar\'e
irrep. despite the existence of two classes of DDF states, therefore,
these must have the same Casimirs and we want to verify this.

The computation can be done in two ways.
The first is to evaluate on the \lc or equivalently using the
Lorentz algebra of DDF operators and using eq. \eqref{eq:C2 lc
  quantization}.
The second way is to express the DDF states in a covariant basis and
then perform the polarization decomposition into Poincar\'e irreps.

We start with the \lc computation and then move to the covariant one.

\subsection{Poincar\'e algebra and Casimirs}
The Poincar\'e algebra $iso(1,D-1)$ reads
\begin{align}
  [ M_{\mu \nu},\, M_{\rho \sigma}]
  =&
  2 i \eta_{\rho [\mu} M_{\nu] \sigma}
  -
  2 i \eta_{\sigma [\mu} M_{\nu] \rho}
  ,
  \nonumber\\
  [M_{\mu \nu},\, P_\rho]
  =&
  2 i \eta_{\rho [\mu} P_{\nu]}
.
\end{align}

We can then define the tensor
\begin{equation}
  W_{[ \mu\nu\rho] }
  =
  M_{[\mu\nu }\, P_{\rho]}
  =
  P_{[\rho}\,  M_{\mu\nu] }
  \equiv
  \frac{1}{3!} \sum_\sigma   W_{ \sigma(\mu) \sigma(\nu) \sigma(\rho) }
  =
  \frac{1}{3}
  \left[
    M_{\mu\nu }\, P_{\rho}
    +
    2 M_{\rho [\mu }\, P_{\nu]}
    \right]
  ,
\end{equation}
which is well defined independently of the order of $M$ and $P$.
It commutes with $P$, i.e.
\begin{equation}
  [  W_{[ \mu\nu\rho] },\, P_\sigma]
  =
  0.
\end{equation}
Then the simplest Casimir invariant is the second order Casimir invariant, given
by the scalar
\begin{align}
-\frac{3^2}{3!}
  C_2\left( iso(1,D-1)\right)
  =&
  W_{[ \mu\nu\rho] } W^{[ \mu\nu\rho] }
  =
  3!
  \sum_{\mu<\nu<\rho}    W_{[ \mu\nu\rho] } W^{[ \mu\nu\rho] }
  \nonumber\\
  &=
  -\frac{1}{3}
  \left(
  P_\rho (M^2)_{\mu \mu} P_\rho
  +
  2 P_\rho (M^2)_{\mu \rho} P_\mu
  \right)
  ,
\end{align}
where the normalization is chosen  so that we get the simple relation
\eqref{eq:C2 ISO vs C2 SO} between the Poincare Casimir and the space
rotation one.

Higher Casimir can be obtained by considering the square of
\begin{equation}
  W_{[ \mu_1\nu_1 \dots \mu_n\nu_n \rho] }
  =
  M_{[\mu_1\nu_1 }\, \dots M_{\mu_n\nu_n }\, P_{\rho]}
  ,
\end{equation}
i.e.
\begin{align}
  C_{2 n}
  =&
  W_{[ \mu_1\nu_1 \dots \rho] } W^{[ \mu_1\nu_1 \dots \rho] }
  .
\end{align}

Notice that the usual Minkowski algebra $so(1,D-1)$ Casimir
is {\bf not} a Casimir for the Poincar\'e algebra since
\begin{align}
  [ M_{\mu \nu} M^{\mu \nu},\, P_\rho]
  =&
  4 i M_{\rho \nu} P^\nu
  +
  4 \frac{1-D}{2} P_\rho
  .
\end{align}

%%%%%%%%%%%%%%%%%%%%%%%%%%%%%%%%%%%%%%%%
\subsubsection{$C_2(iso(1,D-1))$ in the rest frame}
The expression for $W$ can be simplified a lot if we notice that in
string
\begin{align}
  M_{\mu\nu}
  &=
  2 x_{[\mu} p_{0 \nu]}
  +
  J_{\mu\nu}
  ,
  \nonumber\\
  P_\mu
  =&
  p_{0 \mu}
  ,
\end{align}
where 
\eq{
J^{\mu\nu} = i \sum_{n=1}^\infty \frac{2}{n} \alpha^{[\mu}_{-n}\alpha^{\nu]}_{n},
}
is manifestly anti-symmetric and
contains only the non-zero modes of the Lorentz generator $M_{\mu\nu}$.
In fact we get
\begin{equation}
  W_{[ \mu\nu\rho] }
  =
  J_{[\mu\nu }\, p_{0 \rho]}
  ,
\end{equation}
where the only dependence on zero modes is through $p_0$.

The meaning of this expression is that the tachyonic string state
$|k\rangle$ is a scalar, in fact 
\begin{equation}
  W_{[ \mu\nu\rho] } |k \rangle =0
  ~~\Rightarrow~~
  C_2\equiv 0
  .
\end{equation}

In the following, we consider massive states and in order to simplify
the computations we choose the rest
frame where
\begin{equation}
  k_0 =M,~~~~
  k_I=0,~~~~
  I=1,2,\dots D-1
  ,
\end{equation}
then we get that the only non-vanishing $W$ components are
\begin{equation}
  W_{I J 0}
  =
  \frac{1}{3}
  J_{I J} p_{0\, 0}
  =
    \frac{1}{3}
  J_{I J}\, M
  .
\end{equation}
Then the Casimir $C_2( iso(1, D-1) )$ reads,
\begin{equation}
  C_2( iso(1, D-1) )
  =
  - M^2\, \sum_{I<J} J_{I J} J^{I J}
  =
  M^2\, C_2( so(D-1) )
\label{eq:C2 ISO vs C2 SO}
  .
\end{equation}
This means that the states are classified according to $SO(D-1)$,
i.e. according to the ``spin''.

%%%%%%%%%%%%%%%%%%%%%%%%%%%%%%%%%%%%%%%%
\subsection{Casimir of $\alpha^i_{-N} |k\rangle$ physical \lc states}
We want now to compute the second Casimir $C_2$ for the level
$N=2$ \lc $\alpha^i_{-N} |k\rangle$ state.

Since the computation has the same order of difficulty for generic
$N$ we will do this more general one and then restrict to $N=2$.
Moreover it is also interesting ho have an estimate of the average ``spin''
contained in these states.

First we compute
\begin{align}
  i J^{i j} \alpha^l_{-N (\lcsh)} |k\rangle
  =&
  2 \delta^{l [i} \alpha^{j ]}_{-N  (\lcsh)} |k\rangle
,
\end{align}
and its squared modulus
\begin{align}
  \sum_{i<j}
  \parallel   i J^{i j} \alpha^l_{-N  (\lcsh)} |k\rangle \parallel^2
  =&
  N \cdot (D-3)
  .
  \end{align}
We then compute
\begin{align}
  i J^{- j} \alpha^l_{-N  (\lcsh)} |k\rangle
  &=
  \frac{1}{\sdap k^+}
  \sum_{l=1}^{N-1}
  \sum_{r,s=2}^{D-1}
  \left[
    \oh \delta^{j l} \delta^{r s}
    -
    \frac{N}{l} \delta^{j (r} \delta^{s) l}
    \right]
  \alpha^r_{-1  (\lcsh)}   \alpha^s_{-1  (\lcsh)} |k\rangle
  ,
\end{align}
and its squared modulus
\begin{align}
  \sum_{j}
  \parallel   i J^{- j} \alpha^l_{-N  (\lcsh)} |k\rangle \parallel^2
  =&
  \frac{1}{2\ap (k^+)^2}
  \sum_{l=1}^{N-1}
  \sum_{j,r,s=2}^{D-1}
  l (N -l)
  \nonumber\\
  &
  \times \Biggl[
  \left(
  \oh \delta^{j l} \delta^{r s}
  -
  \frac{N}{N-l} \delta^{j (r} \delta^{s) l}
  \right)
  \left(
  \oh \delta^{j l} \delta^{s r}
  -
  \frac{N}{l} \delta^{j (s} \delta^{r) l}
  \right)
\nonumber\\  
&+
  \left(
  \oh \delta^{j l} \delta^{r s}
  -
  \frac{N}{l} \delta^{j (r} \delta^{s) l}
  \right)
  \left(
  \oh \delta^{j l} \delta^{r s}
  -
  \frac{N}{l} \delta^{j (r} \delta^{s) l}
  \right)  
  \Biggr]
  \nonumber\\
  =
  \frac{1}{2\ap (k^+)^2}
  &
  \Biggl[
    \sum_{l=1}^{N-1} l (N -l) \frac{D-2}{4}
    +
N^2 (N-1) \frac{D-2}{2}
  \nonumber\\
  &
  +
  \sum_{l=1}^{N-1} l (N -l) \frac{D-2}{4}
-
  N \sum_{l=1}^{N-1} (N -l)
  +
  N^2    \sum_{l=1}^{N-1} \frac{N -l}{l} \frac{D-1}{2}
  \Biggr]
  .
\end{align}
Using the relations
\begin{align}
  \sum_{l=1}^{m-1} l (m -l)
  =&
  \frac{ m (m^2 -1)}{6}
  ,
  \nonumber\\
  \sum_{l=1}^{m-1} \frac{m -l}{l}
  =&
  m \sum_{l=1}^{m-1} \frac{1}{l}
  - (m-1)
  =
  m H_{m-1}
  - (m-1)
  ,
  \nonumber\\
  \ap M_N^2
  &=
  N-1
  ,
\end{align}
where $H_m$ is the $m$-th harmonic number
we can evaluate the average second Casimir to be
\begin{align}
  \frac{1}{M_N^2} \langle \hat C_2( iso(1, D-1) ) \rangle
  =&
  \langle \hat C_2(so(D-1)) \rangle
  \nonumber\\
  =
  \frac{D-3}{N}
  +
  \frac{1}{24 (N-1)}
  &\left[
    \left( (6 H_{N-1} +1 ) N^2 -1 \right) (D-1)
    - 18 N^2 + 12 N + 1
    \right]
  \nonumber\\
  &\sim
  \frac{1}{4} N \ln(N) (D-1)
  +O(N)
  ,
\end{align}
where $\hat C_2$ is the operator associated with the Casimir.
This value is rather suppressed \wrt the Casimir
we can expect from the symmetric traceless irrep. which should be of
order $N^2$ as suggested by eq. \eqref{eq:C2 (a^i)^n}.

In the case of interest $N=2$ we get
\begin{align}
  \ap \langle \hat C_2( iso(1, D-1) ) \rangle
  =&
  \langle \hat C_2(so(D-1)) \rangle
  =
  (D-3)
  +
  \frac{9}{8} (D-2)
  =
  50
  ,
  \label{eq:C2 N2 a1a1 lc}.
\end{align}
which matches exactly the covariant expressing eq.
\eqref{eq:C2 N2 covariant} only for $D=26$
and the \lc Casimir for the $(\alpha^i_{-1})^N |k\rangle$ state in
\eqref{eq:C2 N2 a2 lc} for all $D$.

%%%%%%%%%%%%%%%%%%%%%%%%%%%%%%%%%%%%%%%%
\subsection{Casimir of $(\alpha^i_{-1})^N |k\rangle$ physical \lc states}
As in the previous case
since the computation has the same order of difficulty for generic
$N$ we will do this more general one and then restrict to $N=2$.

We start from the easiest part
\begin{align}
  -i J^{i j} ( \alpha^l_{-1  (\lcsh)} )^N |k \rangle
  =&
  -N\,
  \delta^{i l}
  (\alpha^l_{-1 (\lcsh)} )^{N-1} \alpha^j_{-1 (\lcsh)} |k \rangle
  ,
\end{align}
then we can compute the square norm
\begin{align}
  \sum_{i<j} \parallel -i J^{i j} ( \alpha^l_{-1 (\lcsh)} )^N |k \rangle \parallel^2
=&
N^2\, (D-3)\, N!
.
\end{align}
For computing the remaining contribution we start from
\begin{align}
[i J^{- j},\, \alpha^l_{-1 (\lcsh)}]
=&
\delta^{j l} \hat \alpha^-_{-1 (\lcsh)}
+
\frac{1}{\alpha^+_{0 (\lcsh)} }
\sum_{n=1}^\infty \frac{1}{n} \alpha^l_{-n-1 (\lcsh)} \alpha^j_{n (\lcsh)}
-
\frac{1}{\alpha^+_{0 (\lcsh)}}
\sum_{n=1}^\infty \frac{1}{n} \alpha^j_{-n (\lcsh)} \alpha^l_{n-1 (\lcsh)}
\nonumber\\
\implies
&
\delta^{j l} \frac{ \vec \alpha_{-2 (\lcsh)} \cdot \vec \alpha_1(\lcsh)}
{\alpha^+_{0 (\lcsh)}}
+
\frac{1}{\alpha^+_{0 (\lcsh)}}
\alpha^l_{-2 (\lcsh)} \alpha^j_{1 (\lcsh)}
-
\frac{1}{\alpha^+_{0 (\lcsh)}}
\frac{1}{2} \alpha^j_{-2 (\lcsh)} \alpha^l_{1 (\lcsh)}
,
\label{eq:J-j a^l-1}
\end{align}
where we have kept the terms which give a non-vanishing contribution
in 
\begin{align}
  i J^{- j} ( \alpha^l_{-1 (\lcsh)})^N |k \rangle
  =&
  \sum_{k=0}^{N-1}
  ( \alpha^l_{-1 (\lcsh)})^k 
[i J^{- j},\, \alpha^l_{-1 (\lcsh)}]
( \alpha^l_{-1 (\lcsh)})^{N-k-1} |k \rangle
,~~~~
j\ne l
,
\end{align}
taking also into account that we have chosen the rest frame for which
$\alpha^i_{0 (\lcsh)} \rightarrow k^i=0$.
It is then easy to show that
\begin{align}
  i J^{- j} ( \alpha^l_{-1 (\lcsh)} )^N |k \rangle
  =&
  -
  \frac{1}{\sdap k^+}
  \frac{N (N-1)}{2}
( \alpha^l_{-1 (\lcsh)} )^{N-1} \alpha^j_{-2 (\lcsh)} |k \rangle
,~~~~
j\ne l
  .
\end{align}
The case $j=l$ is the same but with a different overall coefficient
which  arises because of $\delta^{j l}$ in \eqref{eq:J-j a^l-1}.
We get therefore
\begin{align}
  i J^{- l} ( \alpha^l_{-1 (\lcsh)} )^N |k \rangle
  =&
  + \frac{3}{2}
  \frac{1}{\sdap k^+}
  \frac{N (N-1)}{2}
( \alpha^l_{-1 (\lcsh)} )^{N-1} \alpha^l_{-2 (\lcsh)} |k \rangle
  .
\end{align}
Now we can compute the square norm
\begin{align}
  \sum_{j\ne l} \parallel   i J^{- j} ( \alpha^l_{-1 (\lcsh)} )^N |k \rangle
  \parallel^2
  +
  \parallel   i J^{- l} ( \alpha^l_{-1} )^N |k \rangle \parallel^2 \nonumber \\
  =
  [ 9 + D-3 ]\,
  \frac{1}{ 8 \ap (k^+)^2 }\,
  \frac{N (N-1)}{2}\,
  N!
  .
\end{align}

Now we can evaluate the average second Casimir to be
\begin{align}
    \frac{1}{M_N^2} \langle \hat C_2( iso(1, D-1) ) \rangle
  =&
  \langle \hat C_2(so(D-1)) \rangle
  \nonumber\\
  =
  N\, (D-3)&
  +
  \frac{N (N-1)}{16} (D+6)
  \nonumber\\
  \sim
  \frac{N^2}{16} (D+&6)
  +O(N)
  .
  \label{eq:C2 (a^i)^n}
\end{align}

This value of order $N^2$ is what we can expect from a symmetric
traceless irrep.
The coefficient is very likely, less than the pure irrep because of the
contamination with other irreps.

In the case of interest $N=2$ we get
\begin{align}
  \ap \langle \hat C_2( iso(1, D-1) ) \rangle
  =&
  \langle \hat C_2(so(D-1)) \rangle
  =
  2 (D-3)
  +
  \frac{1}{8} (D+6)
  =
  50
  ,
    \label{eq:C2 N2 a2 lc}.
\end{align}
which matches exactly the covariant expression in eq.
\eqref{eq:C2 N2 covariant}  only for $D=26$
and the \lc state of the previous section \eqref{eq:C2 N2 a1a1 lc} for all $D$.

%%%%%%%%%%%%%%%%%%%%%%%%%%%%%%%%%%%%%%%%
\subsection{Casimir of $N=2$ physical DDF states:
  decomposing tensors and computing the Casimirs}
We know that
$\uA^i_{-2} | \uk_T\rangle$ and
$\uA^i_{-1} \uA^j_{-1}| \uk_T\rangle$
are two different pieces of the same irrep then we expect that their
Casimirs be equal, i.e
\begin{equation}
  C_2\left( \uA^i_{-2} | \uk_T\rangle  \right)
  \equiv
  \frac{
  \langle \ul_T | \uA^i_2\, \hat C_2\, \uA^i_{-2} | \uk_T\rangle
  }{
  \langle \ul_T | \uA^i_2\,\uA^i_{-2} | \uk_T\rangle
  }
  =
  C_2\left( \uA^i_{-1} \uA^j_{-1}| \uk_T\rangle  \right)
  \equiv
  \frac{
  \langle \ul_T | \uA^i_{1} \uA^j_{1}\, \hat C_2\, \uA^i_{-1} \uA^j_{-1} | \uk_T\rangle
  }{
  \langle \ul_T | \uA^i_{1} \uA^j_{1}\, \uA^i_{-1} \uA^j_{-1} | \uk_T\rangle
  }
  ,
\end{equation}
where $\hat C_2$ is the Casimir operator which acts for example as
\begin{equation}
  \hat C_2\,
  s_{ (\ydiagram{2,1})  \mu \nu | \rho }%\YTxxy }
  \,
  \ualpha^\mu_{m_1} \ualpha^\nu_{m_2} \ualpha^\rho_n
  =
  C_2\left( \ydiagram{2,1} \right)
  \,
  s_{ (\ydiagram{2,1})  \mu \nu | \rho }%\YTxxy }
  \,
  \ualpha^\mu_{m_1} \ualpha^\nu_{m_2} \ualpha^\rho_n
.
\end{equation}
Let us start writing the states for $i\ne j$ and $N=2$ 
\begin{align}
  \uA^i_{-2} | \uk_T\rangle
  &=
  \left[
    \left( \Pi^i \ualpha_{-2} \right)
    +
    \left( \Pi^i \ualpha_{-1} \right)
    \ualpha^+_{-1} \frac{N}{\sdap \uk^+} \frac{1}{1!}
    \right]
  |\uk\rangle
  ,
  \nonumber\\
  \uA^i_{-1} \uA^j_{-1}| \uk_T\rangle
  =&
    \left( \Pi^i \ualpha_{-1} \right)
    \left( \Pi^j \ualpha_{-1} \right)
  |\uk\rangle
  .
\end{align}

Then we introduce the following transverse tensors
\begin{align}
  \Pi^i_\mu
  &=
  \delta^i_\mu - \frac{k^i}{k^+} \delta^+_\mu
  ,
  \nonumber\\
  P^\rho_\mu
  &=
  \eta^{\perp\, \rho}_{\,\,\,\, \mu}
  =
  \delta^\rho_\mu + \frac{k^\rho k_\nu}{ M^2 }
  =
  \delta^\rho_\mu - \frac{k^\rho k_\nu}{ k^2 }
  ,
\end{align}
which have the following properties
\begin{align}
  \Pi^i_\mu k^\mu
  =&
  P^\rho_\mu k^\mu
  =
  0
  ,
  \nonumber\\
  \Pi^i \cdot \Pi^j
  &=
  \delta^{ i j}
  ,
  \nonumber\\
  P^\rho_\mu\, \eta^{\mu \nu}\, P^\sigma_\nu
  &=
  P^{\rho \sigma}
  ,
  \nonumber\\
  \Pi^i_\mu\, \eta^{\mu \nu}\, P^\sigma_\nu
  &=
  \eta^{i \sigma}
  .
\end{align}

Using these  objects we can rewrite the covariant states associated
with the DDF states  using the irreps as
\begin{align}
  \uA^i_{-2} | \uk_T\rangle
  &=
  \left[
    \left(
    \Pi^i_\mu
    \right)_{\ydiagram{1}}
      \alpha_{-2}^\mu
    +
    \left(
    \frac{N}{\sdap \uk^+} \Pi^i_{[\mu} P_{\nu]}^+
    \right)_{\ydiagram{1,1}}
    \alpha_{-1}^\mu \ualpha_{-1}^\nu
    +
    \left(
    \frac{N}{\sdap \uk^+} \Pi^i_{(\mu} P_{\nu)}^+
    \right)_{\ydiagram{2}}
    \alpha_{-1}^\mu \ualpha_{-1}^\nu
    \right]
  |k\rangle
  ,
  \nonumber\\
  \uA^i_{-1} \uA^j_{-1}| \uk_T\rangle
  =&
  \left(
  \Pi^i_{(\mu}  \Pi^j_{\nu)}
    \right)_{\ydiagram{2}}
    \alpha_{-1}^\mu \ualpha_{-1}^\nu
  |k\rangle
  ,  
\end{align}
where the last equation is true only because we have chosen $i\ne j$
and the antisymmetric $\ydiagram{1,1}$ is obviously zero.

Now we can check the splitting into irreps by computing the states norm.
We expect the norm to be the sum of the squares of the different
irreps up to coefficients associated with the norms of the states.

If we compute the norm of the generic sum of irreps
\begin{align}
  |v \rangle
  =&
  \left[
    v_{\ydiagram{1}}^\mu
    \alpha_{-2 \mu}
    +
    s_{\ydiagram{2}}^{\mu\nu}
    \alpha_{-1 \mu} \alpha_{-1 \nu}
    +
    \frac{1}{M^2} k^\mu s_{\ydiagram{1}}^{\nu}
    \alpha_{-1 \mu} \alpha_{-1 \nu}
    \right] |0\rangle
  ,
\end{align}
we get
\begin{align}
\langle v  |v \rangle
=&
2\, v_{\ydiagram{1}}^\mu v_{\ydiagram{1} \mu}
+
2\, s_{\ydiagram{2}}^{\mu\nu} s_{\ydiagram{2} \mu\nu}
-
\frac{1}{M^2} s_{\ydiagram{1}}^{\nu} s_{\ydiagram{1} \nu}
,
\end{align}
where in our case $\ap M_N^2 = N-1 = 1$ since $N=2$.

Then we can easily compare with the direct computation of the norms as
\begin{align}
    \langle \uk_T | \uA^i_2\,\uA^i_{-2} | \uk_T \rangle
    =&
    2\, \Pi^i \cdot \Pi^i
    +
    2 \left( \frac{N}{\sdap \uk^+} \right)^2
    \oh \Pi^i \cdot \Pi^i\, P^+ \cdot P^+
    -
    \frac{1}{M^2} \left( \frac{N}{\sdap \uk^+} \right)^2
    \oh \Pi^i \cdot \Pi^i
    \nonumber\\
    &=
    2
    +
    \frac{N^2}{2 \ap M^2}
    -
    \frac{N^2}{2 \ap M^2}
    =2
    ,
    \nonumber\\
    \langle \ul_T | \uA^i_{1} \uA^j_{1}\, \uA^i_{-1} \uA^j_{-1} |
    \uk_T
    =&
    2\, \oh  \Pi^i \cdot \Pi^i \Pi^j \cdot \Pi^j
    =
    1
    .
\end{align}
Finally we can compute the Casimirs
\begin{align}
  C_2\left( \uA^i_{-2} | \uk_T\rangle  \right)
  =&
  \oh
  \times
  \left[
    C_2(\ydiagram{1})\,
    2
    +
    C_2(\ydiagram{2})\,
    \frac{N^2}{2 \ap M^2}
    -
    C_2(\ydiagram{1})\,
    \frac{N^2}{2 \ap M^2}
    \right]
  ,
  \nonumber\\
    C_2\left( \uA^i_{-1} \uA^j_{-1}| \uk_T\rangle  \right)
    =&
    1
    \times
    \left[
    C_2(\ydiagram{2})
    \right]
    ,
\end{align}
where the first factor is the normalization.
When we use
\begin{equation}
  N=2,~~
  \ap M_N^2 =1,~~
  C_2(\ydiagram{2}, so(D-1) )= 2 (D-1),~~
  C_2(\ydiagram{1}, so(D-1) )= D-2
  ,
\end{equation}
where $D-1$ is the space dimension, i.e. $D-1=25$
and $C_2(*,d)$ is the Casimir for the $so(D-1)$ irrep $*$
we get
\begin{equation}
  C_2\left( \uA^i_{-2} | \uk_T\rangle  \right)
  =
  (D-2) + 2(D-1) - (D-2)
  =
  C_2\left( \uA^i_{-1} \uA^j_{-1}| \uk_T\rangle  \right)
  = 2 (D-1)
  ,
  \label{eq:C2 N2 covariant}
\end{equation}
which is valid in all dimension as long as the mass shell condition is
true.

%%%%%%%%%%%%%%%%%%%%%%%%%%%%%%%%%%%%%%%%%%%%%%%%%%%%%%%%%%%%%%%%%%%%%%
%%%%%%%%%%%%%%%%%%%%%%%%%%%%%%%%%%%%%%%%%%%%%%%%%%%%%%%%%%%%%%%%%%%%%%
%%%%%%%%%%%%%%%%%%%%%%%%%%%%%%%%%%%%%%%%%%%%%%%%%%%%%%%%%%%%%%%%%%%%%%

%_______________________________________%
\section{Conclusions}

In this paper, we have presented a reformulation of DDF operators which
makes the underlying structure more clear and repackages all data in a
nice simple structure, the local frame $E$.

Moreover, in the proposed formulation the framed DDFs are true
conformal operators and do not have a cut, which is present in the usual
formulation as shown in Section \ref{sec:Conformal properties}.

This allows us to go off-shell easily, while maintaining all the Virasoro
constraints but the $L_0$ one as discussed in section \ref{sec:going minimally offshell}.

Another important point worth stressing in this off-shell formulation
is that the proper states to use to describe the gauge transformations
are not the ones obtained using $L_n$ as in the usual string field
theory, but the states generated by using the improved Brower operators.
This means that we can cover the whole momentum space except the
$k_\mu=0$ point.

Using the framed DDFs in Section
\ref{sec:DDF_off_shell_and_general_solution}
we have written the general solution of the Virasoro
constraints in terms of \lc polarizations albeit in an unconventional gauge.
In Section \ref{sec:mapping lc to DDFs}
we have also described how they can be used to make concrete the
natural idea that DDFs are the embedding of \lc states into the
covariant formulation.
It is worth stressing again that the improved Brower operators are the
proper operators to consider.

Finally, in Section \ref{sec:casimirs}
we check that the identification is correct by comparing
the Lorentz algebras and the expectation values of the second Casimir
for some \lc states and the corresponding covariant states obtained by
the stated embedding.

Using this formalism, we can easily get the $N$ Reggeon vertex,
i.e. the generating function of $N$ point open string framed DDF
correlators \cite{BiswasMarottaPesando}.

It would be interesting to investigate how and whether it is possible
to extend these ideas and findings to the supersymmetric case and to
other open string backgrounds, for example with a constant magnetic field.

%%%%%%%%%%%%%%%%%%%%%%%%%%%%%%%%%%%%%%%%%%%%%%%%%%%%%%%%%%%%%%%%%%%%%%
%%%%%%%%%%%%%%%%%%%%%%%%%%%%%%%%%%%%%%%%%%%%%%%%%%%%%%%%%%%%%%%%%%%%%%
%%%%%%%%%%%%%%%%%%%%%%%%%%%%%%%%%%%%%%%%%%%%%%%%%%%%%%%%%%%%%%%%%%%%%%

\section*{Acknowledgments}
We would like to thank Raffaele Marotta for sharing his insight into the existence of the projector. This research is partially supported by the MUR PRIN contract 2020KR4KN2 “String
Theory as a bridge between Gauge Theories and Quantum Gravity” and by the INFN project
ST\&FI “String Theory \& Fundamental Interactions”.

\appendix

%%%%%%%%%%%%%%%%%%%%%%%%%%%%%%%%
\section{$\uA^-$ and $\tilde \uA^-$ operators}
\label{app:uA}
If we proceeded to define the $\uA^-_m$ operators in direct analogy with the framed DDF operators $\uA^i_{m}$, we run into problems with cubic poles. To differentiate with the proper definition derived at the end of this section, we refer to this naive version as,
\begin{equation}
 {\widehat    \uA}^-_m(E)
  =
  i \sqrt{\frac{2}{\ap}}
  \oint_{z=0} \frac{d z}{ 2\pi i}
  : \partial_z \uL^-(z) e^{i m \frac{ \uL^+(z) }{\ap \up^+_0} } :
.
\label{eq:Ahat}
\end{equation}
To see the non-vanishing cubic pole, we use $L_n = \oint \frac{dz}{2\pi i} z^{n+1}\left(-\frac{1}{\alpha'}:\del \uL(z)\del \uL(z):\right)$ to obtain,
\eq{
[L_n,\widehat    \uA^-_m] &= \left[\oint_z z^{n+1}\left(-\frac{2}{\alpha'}\right)e^{\delta \cdot \del \uL(z)}, i\stap \oint_w e^{\eps \cdot \del \uL(w) + i \uk \cdot \uL(w)}\right]\Biggm|_{\delta^2,\eps_-,k_+} \nonumber \\
= -i \frac{2}{\alpha'}\stap& \oint_w \oint_{z=w} z^{n+1} :e^{\delta\cdot \del \uL(z)} e^{\eps \cdot \del \uL(w) + i\uk \cdot \uL(w)}: e^{-\frac{\alpha'}{2}\frac{\delta \cdot \eps}{(z-w)^2}}e^{-i\frac{\alpha'}{2}\frac{\delta\cdot k}{(z-w)}}\Biggm|_{\delta^2,\eps_-,k_+} \nonumber \\
= i \stap \oint_w \oint_{z=w} z^{n+1}&\left[-i\frac{\alpha'}{2} \frac{(\delta\cdot \eps)(\delta \cdot k)}{(z-w)^3} + \frac{\delta\cdot\eps}{(z-w)^2}\delta\cdot\del\uL(z) + \delta \cdot i\del\uL(z) \eps\cdot \del \uL(w) \frac{\delta \cdot k}{(z-w)}\right]e^{i\uk \cdot \uL(w)} \nonumber\\
}
Using $\delta_\mu \delta_\nu = \eta_{\mu\nu}$ and $\eps\cdot k = -\frac{m}{\alpha' p_0^+}$ we get,
\eq{
[L_n,\widehat    \uA^-_m] = i\stap \oint_{w=0}&:\left[\eps \cdot \del(w^{n+1}\del\uL(w)) + i w^{n+1}\eps\cdot\del\uL(w) \uk\cdot\del\uL(w)\right]e^{i\uk\cdot \uL(w)}: \nonumber \\
+ &i\stap \oint_{w}\oint_{z=w}z^{n+1}\left(-i \frac{\alpha'}{2}\frac{\eps \cdot \uk}{(z-w)^3}\right)e^{i\uk\cdot \uL(w)} \nonumber \\
= &i\stap \oint_{z=0} \frac{im}{4p_0^+}\del^2(z^{n+1})e^{\frac{im\uL^+(z)}{\alpha'p_0^+}} + ...,
\label{eq:cubic_term}
}
where, "..." denotes the usual total derivative terms which vanish. To remove this cubic pole contribution, we first compute,
\eq{
\left[\del_w \uL^-, \del^2_z\uL^+/\del_z \uL^+\right]&= -\del_z \int_0^\infty \frac{d\xi}{\xi} 
  [\partial_w \uL^-,
  e^{- \xi \partial_z \uL^+}]
  \nonumber\\
=&\hap \del_z \int_0^\infty d\xi \frac{1}{(w-z)^2}e^{-\xi \del_z \uL^+} + :..:\nonumber \\
=&
  \hap\partial_z
  \left[
    \frac{1}{(w-z)^2}     \frac{1}{\partial_z \uL^+}
    \right]
  +
  :
  \partial_w \uL^-\,
  \frac{\partial^2_z \uL^+}{\partial_z \uL^+}
  :,
\label{eq:Ln_corr1}
}
where we used the OPE for $\del_w \uL^- \del_z \uL^+$ directly in the second step. Using \eqref{eq:Ln_corr1}, we get
\eq{
[L_n,&\del^2_z\uL^+/\del_z \uL^+] = -\frac{2}{\alpha'}\oint_w\left[\del_w \uL^- , \frac{\del^2_z\uL^+}{\del_z \uL^+} \right]w^{n+1}\del_w \uL^+ \nonumber \\
&= \del_z \left[\del_w\left(w^{n+1}\del_w \uL^+\right)\Biggm|_{w=z} \frac{1}{\del_z \uL^+}\right] = \del^2_z(w^{n+1}) + \del_z\left(z^{n+1} \frac{\del^2_z\uL^+}{\del_z \uL^+}\right)
\label{eq:Ln_corr2}
}
Finally we observe that
\eq{
\Bigg[L_n,&
    i \ishap
    \oint_{z=0} \frac{d z}{ 2\pi i}
    \frac{\partial^2_z \uL^+}{\partial_z \uL^+}
    e^{i m \frac{ \uL^+(z) }{\ap \up^+_0} }
  \Bigg] = i\ishap \oint_z \left(\frac{\del^2_z \uL^+}{\del_z \uL^+}\Big[L_n,e^{i\frac{m \uL^+(z)}{\ap p_0^+}}\Big] + \left[L_n,\frac{\del^2_z \uL^+}{\del_z \uL^+}\right]e^{i\frac{m \uL^+(z)}{\ap p_0^+}}\right) \nonumber \\
&= i \ishap \oint_z \left(\del^2_z(z^{n+1})e^{i\frac{m\uL^+(z)}{\ap p_0^+}} + \del_z\left(z^{n+1}\frac{\del^2_z \uL^+}{\del_z \uL^+}\right) + \frac{\del^2_z \uL^+}{\del_z \uL^+} z^{n+1}\del_z\left(e^{i \frac{m\uL^+}{\ap p_0^+}}\right)\right) \nonumber \\
&= i\ishap\oint_z \del^2_z(z^{n+1})e^{i\frac{m\uL^+(z)}{\ap p_0^+}} + i\ishap \oint_z \del_z\left(z^{n+1}\frac{\del^2_z \uL^+}{\del_z \uL^+}e^{i\frac{m\uL^+(z)}{\ap p_0^+}}\right),
}
where the second term is zero since it does not contain any branch cuts and is a total derivative. Therefore, we can define the `good' conformal operator (free of cubic pole contributions) as,
\begin{equation}
\uA^-_m(E)
  =
  i \sqrt{\frac{2}{\ap}}
  \oint_{z=0} \frac{d z}{ 2\pi i}
  :
  \left[
    \partial_z \uL^-(z)
    -
    i \frac{m}{4 \up_0^+}
        \frac{\partial^2_z \uL^+}{\partial_z \uL^+}
    \right]
    e^{i m \frac{ \uL^+(z) }{\ap \up^+_0} } :
    ,    
\label{eq:A-_deriv}
\end{equation}
which satisfies $[L_n, \uA^-_m] = 0$ by construction.
%%%%%%%%%%%%%%%%%%%%%%%%%%%%%%%%%%%%
\section{Derivation of the algebra}
\label{app:Derivation of algebra}
The $\uA^i_m$ algebra can be obtained via the usual means. Let us  write
\begin{align}
\uA^i_m =
  \cN
  \oint_{z=0} \frac{d z}{ 2\pi i}
  : \partial_z \uL^i(z) e^{i m \delta_+ \uL^+(z)} :
  ,
  \label{eq:DDFflat}
\end{align}
with
\begin{equation}
\cN = i \sqrt{\frac{2}{\ap}},~~~~
\delta_+ =  \frac{1}{\ap \up_0^+},
~~~~
(\mbox{when } \up^+_0\ne0)
\label{eq:N_delta_vals}
,
\end{equation}
then  we get,
\begin{align}
[\uA^i_m, \uA^j_n]
&=
\cN^2
\left[
\oint_{z=0, |z|>|w|} \frac{d z}{ 2\pi i}
\oint_{w=0} \frac{d w}{ 2\pi i}
-
\oint_{w=0} \frac{d w}{ 2\pi i}
\oint_{z=0, |z|<|w|} \frac{d z}{ 2\pi i}
\right]
\nonumber\\
&
R\left[
: \left( \partial_z \uL^i e^{i m \delta_+ \uL^+} \right)(z) :
: \left( \partial_w \uL^j e^{i n \delta_+ \uL^+} \right)(w) :
\right]
%% ,\label{eq:start_AA_commutation_derivation}
%
%
\nonumber\\    
&=
\cN^2
\oint_{w=0} \frac{d w}{ 2\pi i}
\oint_{z=w} \frac{d z}{ 2\pi i}
\Biggl[
-\hap \frac{ \delta^{i j} }{(z-w)^2} e^{i (m+n) \delta_+ \uL^+(w) }
\nonumber\\
&~~~~
-
\hap \frac{ \delta^{i j} }{(z-w)}
i \delta_+ m \partial_w \uL^+ e^{i (m+n) \delta_+ \uL^+(w) }
+\dots
\Biggr]
\nonumber\\
&=
\left( -\oh (\ap\cN)^2 \delta_+ \up^+_0 \right)
m \delta_{m+n,0} \delta^{i j}
.
\label{eq:AA_comm_norm}
\end{align}
Using the definitions above we can compute,
\eq{
[\uA^i_m,\alpha_0^+ \uA^-_n] &= -\frac{2}{\alpha'} \alpha_0^+\oint_{w=0}\oint_{z=w} :e^{\delta\cdot\del_z\uL + iq\cdot\uL(z)}::e^{\gamma\cdot\del_w\uL + ik\cdot \uL(w)}:\Bigm|_{\delta_i,\gamma_-} \nonumber \\
= -\frac{2}{\alpha'} &\alpha_0^+\oint_{w=0}\oint_{z=w} :e^{\delta\cdot\del_z\uL + iq\cdot\uL(z)} e^{\gamma\cdot\del_w\uL + ik\cdot \uL(w)}: e^{i\hap\frac{q\cdot\gamma}{(z-w)}}\Bigm|_{\delta_i,\gamma_-}\nonumber\\
=&-\frac{2}{\ap}\alpha_0^+\oint_{w=0}\oint_{z=w}:\del_z \uL^i e^{i\frac{(m \uL(z) + n \uL(w))}{\ap p_0^+}}: \left(-\frac{im}{\ap p_0^+}\right)\hap\frac{1}{(z-w)} \nonumber \\
&\implies [\uA^i_m,\alpha_0^+ \uA^-_n] = m \uA^i_{m+n},
\label{eq:AiA-}
}
where, $\delta_i = 1, q_+ = m/(\ap p_0^+), \gamma_- = 1, k_+ = n/(\ap p_0^+)$ are the only non-zero components.

We shifted to the exponential trick for evaluating commutators since the second term in the definition of $\uA^-_n$ obviously commutes with $\uA^i_m$.

We now calculate the commutator,
\eq{
[\azp \uA^-_m. \azp \uA^-_n] = (\azp)^2([\widehat \uA^-_m,\widehat    \uA^-_n] - [\widehat \uA^-_m,C_n] - [C_m,\widehat    \uA^-_n] + [C_m,C_n]),
\label{eq:AmAm_full}
}
where, $C_m =i\stap \left(\frac{im}{4p_0^+}\right)\oint_{z=0} :\frac{\del^2_z \uL^+}{\del_z\uL^+} e^{\frac{im\uL^+}{\ap p_0^+}}:$. We see that,
\eq{
[\widehat \uA^-_m,\widehat    \uA^-_n] =& -\frac{2}{\ap}\oint_{w=0}\oint_{z=w}:e^{\delta\cdot \del_z\uL + iq\cdot\uL(z)}e^{\eps\cdot\del_w\uL+ik\cdot\uL(w)}:e^{-i\frac{\delta\cdot k}{(z-w)}\hap}e^{i \frac{\eps\cdot q}{(z-w)}\hap}\Bigm|_{\delta_-,\eps_-}\nonumber\\
=-\frac{2}{\ap}\oint_{w=0}&\oint_{z=w}\left[-\frac{im}{\ap p_0^+}\frac{\del_z\uL^-}{(z-w)}\hap + \frac{in}{\ap p_0^+}\frac{\del_w \uL^-}{(z-w)}+\left(\hap\right)^2\frac{mn}{(z-w)^2(\ap p_0^+)^2}\right] \nonumber \\
&\times e^{i\frac{(m \uL(z) + n \uL(w))}{\ap p_0^+}} \nonumber \\
=&~i\frac{(m-n)}{\ap p_0^+} \oint_z :\del_z\uL^- e^{i \frac{(m+n)\uL^+(z)}{\ap p_0^+}}: - \hap \frac{inm^2}{(\ap p_0^+)^3}:\del_z \uL^+ e^{i \frac{(m+n)\uL^+(z)}{\ap p_0^+}}:, \nonumber \\
\label{eq:Ahat_Ahat}
}
and,
\eq{
[\widehat \uA^-_m,C_n] = &\left(\frac{-in}{2\ap p_0^+}\right)\left(\left[\oint_w \del_w \uL^-,\oint_z \frac{\del^2_z \uL^+}{\del_z \uL^+}\right]e^{i\frac{(m\uL^+(w) + n\uL^+(z))}{\ap p_0^+}} \right.\nonumber \\
&\left.+ \left[\oint_w \del_w L^-,\oint_z e^{i\frac{n \uL^+(z)}{\ap p_0^+}}\right]e^{i\frac{m \uL^+(w)}{\ap p_0^+}}\frac{\del^2_z \uL^+}{\del_z \uL^+}\right) \nonumber \\
= &\frac{inm^2}{4(\ap)^2 (p_0^+)^3}\oint_z :\del_z \uL^+ e^{i\frac{(m+n)\uL^+(z)}{\ap p_0^+}}: + \frac{n^2}{4\ap (p_0^+)^2}\oint_z:\frac{\del^2_z \uL^+}{\del_z \uL^+}e^{i\frac{(m+n)\uL^+(z)}{\ap p_0^+}}:.
\label{eq:Ahat_Cn}
}
Using \eqref{eq:Ahat_Ahat} and \eqref{eq:Ahat_Cn} in \eqref{eq:AmAm_full}, we get,
\eq{
[\azp \uA^-_m, &\azp \uA^-_n] = \oint_z \Bigg[i\frac{(m-n)}{\ap p_0^+} \oint_z :\del_z\uL^- e^{i \frac{(m+n)\uL^+(z)}{\ap p_0^+}}: - \hap \frac{inm^2}{(\ap p_0^+)^3}:\del_z \uL^+ e^{i \frac{(m+n)\uL^+(z)}{\ap p_0^+}}: \nonumber \\
&- \frac{inm^2}{4(\ap)^2 (p_0^+)^3}\oint_z :\del_z \uL^+ e^{i\frac{(m+n)\uL^+(z)}{\ap p_0^+}}: - \frac{n^2}{4\ap (p_0^+)^2}\oint_z:\frac{\del^2_z \uL^+}{\del_z \uL^+}e^{i\frac{(m+n)\uL^+(z)}{\ap p_0^+}}: \nonumber \\ 
 &+ \frac{imn^2}{4(\ap)^2 (p_0^+)^3}\oint_z :\del_z \uL^+ e^{i\frac{(m+n)\uL^+(z)}{\ap p_0^+}}: + \frac{m^2}{4\ap (p_0^+)^2}\oint_z:\frac{\del^2_z \uL^+}{\del_z \uL^+}e^{i\frac{(m+n)\uL^+(z)}{\ap p_0^+}}:\Bigg] \nonumber \\
&\implies [\azp \uA^-_m, \azp \uA^-_n] = (m-n)\azp \uA^-_{m+n} + 2m^3 \delta_{m+n,0}
\label{eq:AmAm_result}
}
%%%%%%%
We now define the following operator,
\begin{equation}
   {\widetilde    \uA}^-_m(E)
   =
   {\uA}^-_m(E)
   -
   \frac{1}{\ualpha_0^+} \cL_m(E)
   -
   \frac{D-2}{24}
   \frac{1}{\ualpha_0^+}\,
   \delta_{m,0} 
   ,
\end{equation}
where we have defined the Virasoro generators as,
\begin{align}
\cL_m(E)
=
&
\frac{1}{2}
\sum_{j=2}^{D-1} \sum_{l\in \Z}: \uA^j_l(E)\, \uA^j_{m-l}(E) :
,
\end{align}
and they satisfy the standard Virasoro algebra for a theory with $D-2$ bosons, namely,
\begin{equation}
  [\cL_m,\, \cL_n]
    = 
    (m-n) \cL_{m+n}
    +
    (D-2)
    \frac{1}{12} m(m^2-1)
    \delta_{m+n,0}
    .
\end{equation}
This is easy to check since the $\uA^i_m$ and $\ualpha^i_m$ satisfy the exact same algebra.
We then observe that,
\eq{
[\uA^i_n,\widetilde \uA^-_m] &= [\uA^i_n,\uA^-_m] - \frac{1}{\azp}[\uA^i_n, \cL_m] \nonumber \\
= \frac{n}{\azp}&A^i_{m+n} - \frac{1}{\azp}\frac{1}{2}\sum_{j=2}^{D-1}\sum_{l \in \mathbb{Z}}[\uA^i_n, :\uA^j_{l}\uA^j_{m-l}:] \nonumber \\
= \frac{n}{\azp}A^i_{m+n}& - \frac{1}{\azp}\frac{1}{2}(n \uA^i_{m+n} + n\uA^i_{m+n}) = 0,
\label{eq:AiAtil_comm}
}
where in reaching the last line, we have used the commutator $[A^i_m,A^j_n] = m \delta_{m+n,0}\delta^{ij}$.

We also have,
\eq{
[\azp &\uA^-_m,\cL_n] = \frac{1}{2}\sum_{j=1}^{D-2} \left(\sum_{p=-\infty}^0[\azp \uA^-_m, \uA^j_p \uA^j_{n-p}] + \sum_{p=1}^\infty[\azp \uA^-_m, \uA^j_{n-p}\uA^j_p]\right) \nonumber\\
= -\frac{1}{2}&\sum_j \left[\sum_{p=-\infty}^0\left(p \uA^j_{m+p}\uA^j_{n-p} + (n-p)\uA^j_p\uA^j_{m+n-p}\right) + \sum_{p=1}^\infty\left((n-p)\uA^j_{m+n-p}\uA^j_p + p \uA^j_{n-p}\uA^j_{m+p}\right)\right] \nonumber \\
&= -\frac{1}{2}\sum_j\left[\sum_{p=-\infty}^0(n-p)\uA^j_{p}\uA^j_{m+n-p} + \sum_{q=-\infty}^m (q-m)\uA^j_{q}\uA^j_{m+n-q} \right. \nonumber \\
&~~~\left. + \sum_{p=1}^\infty (n-p)\uA^j_{m+n+p}\uA^j_{p} + \sum_{q=m+1}^\infty(q-m)\uA^j_{m+n-q}\uA^j_{q}\right] \nonumber \\
&= -\frac{1}{2}\sum_j\left[\sum_{p=-\infty}^0(n-m)\uA^j_{p}\uA^j_{m+n-p} + \sum_{q=1}^m (q-m)\uA^j_{q}\uA^j_{m+n-q} \right. \nonumber \\
&~~~\left.+ \sum_{p=m+1}^\infty (n-m)\uA^j_{m+n+p}\uA^j_{p} + \sum_{q=1}^m(q-m)\uA^j_{m+n-q}\uA^j_{q}\right] \nonumber \\
=& (m-n)\frac{1}{2}\sum_j\sum_{l\in \mathbb{Z}}:\uA^j_{p}\uA^j_{m+n-p}: - \frac{1}{2}\sum_j\left(\sum_{q=1}^m q(q-m) \delta_{m+n,0}\right),
}
where, the last summation of the last line in obtained from normal ordering the second term in the penultimate line (all others are already normal ordered!). Finally, we can evaluate the summation using,
\eq{
\sum_{q=1}^m q^2 = \frac{1}{6}m(m+1)(2m+1), ~~ \sum_{q=1}^m q = \frac{1}{2}m(m+1),
}
to obtain,
\eq{
[\azp \uA^-_m,\cL_n]=(m-n) \cL_{m+n} + \frac{D-2}{12}m(m^2-1)\delta_{m+n,0}.
\label{eq:Am_cL_comm}
}
Using \eqref{eq:AiAtil_comm} and \eqref{eq:Am_cL_comm}, we can finally calculate,
\eq{
[\widetilde \uA^-_m, \widetilde \uA^-_n] 
&= 
[\uA^-_m,\uA^-_n] 
- \frac{1}{\azp}[\cL_m,\uA^-_n] 
- \frac{1}{\azp}[\uA^-_m,\cL_n] 
+ \frac{1}{(\azp)^2}[\cL_m,\cL_n] 
\nonumber \\
=& 
\frac{(m-n)}{\azp}\uA^-_{m+n} 
+ \frac{2m^3}{(\azp)^2}\delta_{m+n,0} 
\nonumber\\
&+ 
\frac{1}{(\azp)^2}
\left[(n-m)\cL_{m+n} 
+ \frac{D-2}{12}n(n^2-1)\delta_{m+n,0}\right] \nonumber \\
&- 
\frac{1}{(\azp)^2}
\Bigg[(m-n)\cL_{m+n} 
+ \frac{D-2}{12}m(m^2-1)\delta_{m+n,0}
\Bigg] 
\nonumber\\
&+ \frac{1}{(\azp)^2}
\left[
(m-n)\cL_{m+n} 
+ \frac{D-2}{12}m(m^2-1)\delta_{m+n,0}
\right] 
\nonumber \\
=& \frac{(m-n)}{\azp}
\left[
\uA^-_{m+n} 
- \frac{1}{\azp}\cL_{m+n}
\right] 
- \frac{D-2}{12(\azp)^2}n\delta_{m+n,0} 
+ \frac{D-26}{12}\frac{n^3}{(\azp)^2}\delta_{m+n,0} 
\nonumber \\
=&\frac{(m-n)}{\azp}
\left[\uA^-_{m+n} 
- \frac{1}{\azp}\cL_{m+n} 
+\frac{D-2}{24(\azp)^2}\delta_{m+n,0}\right] 
+ \frac{26-D}{12}\frac{m^3}{(\azp)^2}\delta_{m+n,0} 
\nonumber \\
\implies
&
\left[
\azp\widetilde \uA^-_m, 
\azp\widetilde \uA^-_n\right] 
= 
(m-n)\azp \widetilde \uA^-_{m+n} 
+  \frac{26-D}{12}m^3\delta_{m+n,0}.
}
%%%%%%%%%%%%%%%%%%%%%%%%%%%%%%%%%
\section{Derivation of hermiticity properties}
\label{app:Derivation of hermiticity properties}
We calculate the Hermitian conjugation of the $\widetilde{A}^-_m$ operators using explicitly, the mode expansion of the string solution. Thereby, we note that,
\eq{
\left[e^{i\frac{m \uL^+(z)}{\ap p_0^+}}\right]^\dag &= e^{\left[i\frac{m \uL^+(z)}{\ap p_0^+}\right]^\dag} =  \exp\left[\frac{im}{\ap p_0^+}\left(\oh x^+_0 - i \ap p^+_0 \ln(z) + i \shap \sum_{n\ne 0} \frac{\alpha^+_n}{n} z^{-n}\right)\right]^\dag \nonumber \\
&=\exp\left[-\frac{im}{\ap p_0^+}\left(\oh x^+_0 + i \ap p^+_0 \ln(\zbar) - i \shap \sum_{n\ne 0} \frac{\alpha^+_{-n}}{n} \zbar^{-n}\right)\right] \nonumber \\
&= \exp\left[-\frac{im}{\ap p_0^+}\left(\oh x^+_0 - i \ap p^+_0 \ln(\frac{1}{\zbar}) + i \shap \sum_{m\ne 0} \frac{\alpha^+_{m}}{m} \left(\frac{1}{\zbar}\right)^{-m}\right)\right] \nonumber \\
\implies &\left[e^{i\frac{m \uL^+(z)}{\ap p_0^+}}\right]^\dag = e^{-\frac{im \uL^+\left(\frac{1}{\zbar}\right)}{\ap p_0^+}},
}
where we have used that $x_0^\mu$ and $p_0^\mu$ are Hermitian and $(\alpha^+_n)^\dag =\alpha^+_{-n}$. We also compute,
\eq{
[\del \uL^\mu(z)]^\dag &= \left[-i\shap\sum_{n\in \mathbb{Z}}\alpha^\mu_n z^{-n-1}\right]^\dag = i\shap \sum_{n\in \mathbb{Z}} \alpha^\mu_{-n}\zbar^{-n-1}\nonumber \\
&= i\shap \frac{1}{\zbar^2}\sum_{m\in \mathbb{Z}}\alpha^\mu_m \left(\frac{1}{\zbar}\right)^{-m-1} = -\frac{1}{\zbar^2}\left[-i\shap\sum_{m\in \mathbb{Z}}\alpha^\mu_m \left(\frac{1}{\zbar}\right)^{-m-1}\right] \nonumber \\
\implies& [\del \uL^\mu(z)]^\dag = -\frac{1}{\zbar^2} \del \uL^\mu\left(\frac{1}{\zbar}\right),
}
and the Hermitian conjugate of the second derivative,
\eq{
[\del^2 \uL^\mu(z)]^\dag &= -i\shap \sum_{n}(n+1)\alpha^\mu_{-n}\zbar^{-n-2} \nonumber\\
=& -i\shap \sum_m (1-m)\alpha^\mu_m \left(\frac{1}{\zbar}\right)^{-m-2} \frac{1}{\zbar^4} \nonumber \\
&= \frac{1}{\zbar^4}i\shap \sum_m (m+1) \alpha^\mu_m \left(\frac{1}{\zbar}\right)^{-m-2} - 2 i\shap \sum_m \alpha^\mu_m \left(\frac{1}{\zbar}\right)^{-m-1}\frac{1}{\zbar^{3}} \nonumber \\
\implies& [\del^2 \uL^\mu(z)]^\dag  = \frac{1}{\zbar^4}\del^2\uL^\mu\left(\frac{1}{\zbar}\right) + 2 \frac{1}{\zbar^{3}} \del \uL^\mu\left(\frac{1}{\zbar}\right).
}
Finally we compute the last piece,
\eq{
\left[\hat Q_{l; m}(E) \right]^\dagger = -\oint \frac{d\zbar}{2\pi i}\frac{1}{\zbar^{l+1}}e^{-\frac{im\uL^+\left(\frac{1}{\zbar}\right)}{\ap p_0^+}},
}
where the $'-'$ outside fixes the direction of the loop to anti-clockwise after taking the complex conjugate. Then,
\eq{
\left[\hat Q_{l; m}(E) \right]^\dagger &= (\zbar)^2 \oint \frac{d\left(\frac{1}{\zbar}\right)}{2\pi i}\frac{1}{\left(\frac{1}{\zbar}\right)^{-l+1} \left(\frac{1}{\zbar}\right)^{-2}}e^{-\frac{im\uL^+\left(\frac{1}{\zbar}\right)}{\ap p_0^+}} \nonumber \\
&= \oint \frac{d\left(\frac{1}{\zbar}\right)}{2\pi i}\frac{1}{\left(\frac{1}{\zbar}\right)^{-l+1}}e^{-\frac{im\uL^+\left(\frac{1}{\zbar}\right)}{\ap p_0^+}} \nonumber \\
\implies & \left[\hat Q_{l; m}(E) \right]^\dag = \hat Q_{-l;-m}(E).
\label{eq:Q_conj}
}
To compute the Hermitian conjugate of $\widehat \uA^-_m$, we first note that the action of H.C on a normal ordered product of two operators $A := \del \uL^-$ and $B:= e^{\frac{im \uL^+}{\ap p_0^+}}$ is given by,
\eq{
(:AB:)^\dag = B^\dag A^\dag - \langle{AB}\rangle^\dag \nonumber \\
=~:A^\dag B^\dag: + [B^\dag,A^\dag].
}
Using the above results, we then observe that,
\eq{
:\left(\del \uL^-(z) e^{\frac{im\uL^+(z)}{\ap p_0^+}}\right):~&=~:\left(-\frac{1}{\zbar^2}\right)\del \uL^-\left(\frac{1}{\zbar}\right)e^{-\frac{im\uL^+\left(\frac{1}{\zbar}\right)}{\ap p_0^+}}: + \frac{im}{\ap p_0^+}\left(-\frac{1}{\zbar^2}\right)(-i\ap)\frac{\zbar}{2}i e^{-\frac{im\uL^+\left(\frac{1}{\zbar}\right)}{\ap p_0^+}} \nonumber \\
\implies& [\widehat \uA^-_m]^\dag = \widehat \uA^-_{-m} + \frac{m}{\alpha^+_0}\widehat Q_{0;-m}
}
Finally, we can see that,
\eq{
[\uA^-_m(E)]^\dag
=&
[\widehat \uA^-_m(E)]^\dag
+ i\ishap \oint \frac{d\zbar}{2\pi
  i}:\left[\frac{im}{4p_0^+}\left(-\frac{1}{\zbar^2}\frac{\del^2
    \uL^+}{\del \uL^+}
  -
  \frac{2}{\zbar}\right)\right]e^{-\frac{im\uL^+\left(\frac{1}{\zbar}\right)}{\ap
    p_0^+}}:
\nonumber \\
=
\widehat \uA^-_{-m}
+ &
\frac{m}{\alpha^+_0}\widehat Q_{0;-m}
+ i\ishap \oint \frac{d\left(\frac{1}{\zbar}\right)}{2\pi i}:\left(-
\frac{i(-m)}{4 p_0^+}\frac{\del^2 \uL^+}{\del
  \uL^+}\right)e^{-\frac{im\uL^+\left(\frac{1}{\zbar}\right)}{\ap
    p_0^+}}:
\nonumber \\
&+
i\ishap \oint \frac{d\left(\frac{1}{\zbar}\right)}{2\pi
  i}:\frac{2im}{4 p_0^+}\frac{1}{\left(\frac{1}{\zbar}\right)^1}
e^{-\frac{im\uL^+\left(\frac{1}{\zbar}\right)}{\ap p_0^+}}:
\nonumber \\
=&
\uA^-_{-m}(E)
+ \frac{m}{\alpha^+_0}\widehat Q_{0;-m} - \frac{m}{\alpha^+_0}\widehat Q_{0;-m} \\
\implies& [\uA^-_m(E)]^\dag = \uA^-_{-m}(E) \\
\implies& [\uA^-_{m}(E)]^{\dag\dag} = \uA^-_m(E).
}

We also note that the $\widetilde \uA^-_m$ follows the same algebra
under Hermitian conjugation as $\uA^-_m$. This is evident from its
definition and the properties of the involved quantities derived in
the preceding sections of the Appendix.

%%%%%%%%%%%%%%%%%%%%%%%%%%%%%%%%%%%%%%%%%%%%%%%%%%%%%%%%%%%%%%%%%%%%%%
\section{Details on Lorentz algebra}
\subsection{$M^{+ 2}$ Infinitesimal rotation}
To better understand the discussion above, we explicitly compute the
variation of the flat space DDF operators using an infinitesimal 
rotation such that,
\begin{equation}
  E^{1'} = E^1 + \sqrt{2} \epsilon E^2
  ,~~~~
  E^{2'} = E^2 - \sqrt{2} \epsilon E^1  ,
\end{equation}
where the factor $\sqrt{2}$ has been chosen so that,
\begin{equation}
  E^{\pm'} = E^\pm \pm \epsilon E^2
  ,~~~~
  E^{2'} = E^2 - \epsilon E^+ + \epsilon E^-
  .
\end{equation}
Under this condition we can write,
\begin{align}
    e^{i m \frac{ E^{+'} \cdot L(z) }{\ap E^{+'} \cdot p_0} } 
  =&
  e^{i m \frac{ E^{+} \cdot L(z) }{\ap E^{+} \cdot p_0} }
  \left[
    1
    +
    \epsilon
     \frac{ i m  }{\ap (\up^+_0)^2}
    \left(
    \up_0^+ \uL^2 - \up_0^2 \uL^+
    \right)
    + O(\epsilon^2)
    \right]
  \nonumber\\
  =
  e^{i m \frac{ E^{+} \cdot L(z) }{\ap E^{+} \cdot p_0} }
 & \left[
    1
    +
    \epsilon
     \frac{ i m  }{\ap (\up^+_0)^2}
    \left(
    \up_0^+ \uL_{(\ne)}^2 - \up_0^2 \uL_{(\ne)}^+
    \right)
    + O(\epsilon^2)
    \right]
  ,
\end{align}
where we have defined,
\begin{equation}
  \uL_{(\ne)}(z) = \uL(z) - i \ap \up_0 \ln(z)
  ,
\end{equation}
i.e. the part of $\uL$ without the logarithmic cut.
Note the normal ordering relations,
\begin{align}
  \uL_{(\ne)}(z)\, \uL_{(\ne)}(w)
  =&
  :   \uL_{(\ne)}(z)\, \uL_{(\ne)}(w) :
  -\frac{\ap}{2} \ln\left( 1 - \frac{z}{w} \right)
  ,~~~~
  |z|>|w|
  \nonumber\\  
  \uL(z)\, \uL_{(\ne)}(w)
  =&
  :   \uL(z)\, \uL_{(\ne)}(w) :
  -\frac{\ap}{2} \ln\left( z -w  \right)
  ,
  \label{eq:normal_orderings_wo_p0}
\end{align}
where the first one has no cuts.

It is important to use $\uL_{(\ne)}$ if we want to use the usual formula
\begin{equation}
  \left[\oint_{z=0} \cA(z), \oint_{w=0} \cE(w)\right]
  =
  \oint_{w=0} \oint_{z=w} R[\cA(z) \cE(w)]
  ,
\end{equation}
since its derivation requires the absence of cuts.

From the previous expression we get,
\begin{align}
  \delta_\epsilon A^i_m(E)
  =&~
  A^i_m(E') - A^i_m(E)
  \nonumber\\
  =
  \epsilon
     &\frac{ i m  }{\ap (\up^+_0)^2}
    i \sqrt{\frac{2}{\ap}}
    \oint_{z=0} \frac{d z}{ 2\pi i}
    :
    \partial_z \uL^i(z)
    \left( \up_0^+\uL_{(\ne)}^2(z) - \up_0^2 \uL_{(\ne)}^+(z) \right)
    e^{i m \frac{ \uL^+(z) }{\ap \up^+_0} }
    :
    \nonumber\\
    &+
    \delta^{i 2}
    (-\epsilon)
    i \sqrt{\frac{2}{\ap}}
    \oint_{z=0} \frac{d z}{ 2\pi i}
    : \left(
    \partial_z \uL^+(z)
    -
    \partial_z \uL^-(z)
    \right)
    e^{i m \frac{ \uL^+(z) }{\ap \up^+_0} }
    : 
    ,
    \label{eq:infinitesimal_DDF_rotation}
\end{align}
where the contribution in the last line is only for $i=2$ and does
not use $\uL_{(\ne)}$, but $\uL$ since it comes from the rotation of
$\partial \uL^2$. We shall study the variation above separately for two cases ($i\neq 2$ and $i=2$).
%___________________________________________%
\subsubsection{Rewriting the variation $\delta_\epsilon A^i_m(E)$ for $i\neq2$}
If the mapping $A^i_n \leftrightarrow \alpha^i_{n (\lc)}$ is indeed true, we should have that the previous variation $  \delta_\epsilon A^i_m(E)= \delta_\epsilon \alpha^i_{m (\lc)}$, up to terms which vanish in the
\lc gauge.

To check this, we evaluate the commutator,
\begin{equation}
  [A^i_m(E),   \delta_\epsilon A^j_n(E)]
  ,
\end{equation}
and then determine the coefficients $c_l$ and $\cN$ in
\begin{equation}
\delta_\epsilon \uA^j_n(E)
=\sum_l  c_l\, \uA^j_l(E)\, \uA^2_{n-l}(E)
+
\cN
{\tilde \uA}^-_n(E)
,
\label{eq:ansatz_infinitesimal_change_of_frame}
\end{equation}
in order to reproduce the previous commutators.
The expression in \eqref{eq:ansatz_infinitesimal_change_of_frame} is an ansatz based on the operator's level and
$SO(D-2)$ rotations.

We evaluate \eqref{eq:infinitesimal_DDF_rotation} for $j\ne2$ as follows,
\begin{align}
  [&\uA^i_m(E),
    ~
    \delta_\epsilon \uA^j_n(E)]
  =
  \epsilon \frac{i n}{\ap (\up^+_0)^2}
  \cdot
  \oint_{w=0} \frac{d w}{2\pi i}
  \oint_{z=w} \frac{d z}{2\pi i}
  \Biggl[
    \frac{ \delta^{i j} }{ (z-w)^2 }
    \left( \up_0^+\uL_{(\ne)}^2(w) - \up_0^2 \uL_{(\ne)}^+(w) \right)
  \nonumber\\
  &
  +
  \frac{ \delta^{i 2} }{ z-w }
  \up_0^+
  \partial \uL^j(w)
  +
  \frac{-2}{\ap}
  :
  \partial  \uL^i(z)
  \partial \uL^j(w)
  \left( \up_0^+\uL_{(\ne)}^2(w) - \up_0^2 \uL_{(\ne)}^+(w) \right)
  :  
    \Biggr]
    \,
    e^{ i m \frac{\uL^+(z)}{\ap p^+_0} }
    e^{i n \frac{ \uL^+(w) }{\ap \up^+_0} }
    %% =
    %% %
    %% %
    \nonumber\\
  &~=
    \epsilon \frac{i n}{\ap (\up^+_0)^2}
  \oint_{w=0} \frac{d w}{2\pi i}
  \Biggl[
    \delta^{i j}
    \,
    \left( \up_0^+ \uL_{(\ne)}^2(w) - \up_0^2 \uL_{(\ne)}^+(w) \right)
    \,
    \frac{i m}{\ap \up^+_0}
    \,
    \partial \uL^+(w)
    \,
    e^{i (m+n) \frac{ \uL^+(w) }{\ap \up^+_0} }
\nonumber\\
&
\phantom{
    \epsilon \frac{i n}{\ap (\up^+_0)^2}
}
+
  \delta^{i 2}
  \,
  \up_0^+
  \,
  \partial \uL^j(w)
  \,
  e^{i (m+n) \frac{ \uL^+(w) }{\ap \up^+_0} }
  \Biggr]
    .
\end{align}
There are now two cases: $m+n\ne0$ and $m+n=0$.
For the first case, we can integrate by parts
since,
$    \left( \up_0^+ \uL_{(\ne)}^2(w) - \up_0^2 \uL_{(\ne)}^+(w) \right)$
has no cut and
we can write,
\begin{align}
    [\uA^i_m,
    ~
    \delta_\epsilon \uA^j_n]=
  \epsilon \frac{i n}{\sdap (\up^+_0)^2 }
  \Biggl[
    \left(
    -
    \up_0^+ \uA^2_{m+n} \frac{m}{m+n}   \delta^{i j}
    +
    \up_0^+ \uA^j_{m+n}   \delta^{i 2}
    \right)
    +
    \up_0^2 \uA^+_{m+n} \frac{m}{m+n}   \delta^{i j}
    \Biggr]
  ,
\end{align}
where we have defined,
\begin{equation}
  \uA^+_m(E)
  =
  i \sqrt{\frac{2}{\ap}}
  \oint \frac{d z}{2 \pi i}
  \partial \uL^+(z)
  e^{ i m \frac{\uL^+(z)}{\ap \up^+_0} }
  =
  \delta_{m,0} \ualpha^+_0
,
\end{equation}
which is trivial for $m\ne0$.

%% \igor{Details to be inserted...}
In order to determine the $c_l$ coefficients in
eq. \eqref{eq:ansatz_infinitesimal_change_of_frame}, we insert the
ansatz into the commutator and obtain,
\begin{align}
  [A^i_m(E),&   \delta_\epsilon A^j_n(E)]
  =
  m \delta^{i j}\,  c_{-m}\, \uA^2_{m+n}(E)
  +
  m \delta^{i 2}\,  c_{n+m}\, \uA^j_{m+n}(E)
  , \nonumber \\
  \implies&
  m\, c_{-m}
  =
  -\epsilon \frac{i}{\ap \up^+_0} \frac{m n}{m+n} \left(-i \sqrt{\hap} \right)
  ;~
  m\, c_{n+m}
  =
  +\epsilon \frac{i}{\ap \up^+_0} n \left(-i \sqrt{\hap} \right)
,
\end{align}
when compared with the actual result.
Taking the ratio we get
\begin{equation}
  \frac{ c_{-m} }{ c_{n+m} }
  =
  \frac{-m}{ n+m}
  ,
\end{equation}
which suggests two possible classes of solutions and a combination of the two.
The two basic possible solutions are,
\begin{equation}
  c_l = \tilde c_1\, l
  ,~~~~
  c_l = \hat c_1 \frac{1}{n - l}
  .
  \label{eq:basic_rec_soln}
\end{equation}
Inserting them into the original equations we get,
\begin{align}
  c_l = \tilde c_1\, l
  &\Rightarrow
  \tilde c_1 = \epsilon \frac{1}{\sdap \up_0^+} \frac{n}{m}
  \frac{1}{m+n}
  ,
  \nonumber\\
  c_l = \hat c_1 \frac{1}{n - l}
  &\Rightarrow
  \hat c_1 = - \epsilon \frac{1}{\sdap \up_0^+} n
  .  
\end{align}
Since $c_l$ may depend on $n$, but must not depend on $m$,
only the second solution is viable.

Now, for the second case ($n+m=0$), we see that we get two contributions.
The first one corresponds to the term $l=0$ in \eqref{eq:ansatz_infinitesimal_change_of_frame}.
The second one is related to the $\underline{M}^{+ 2}$ generator. 
Explicitly we get
\begin{align}
  [\uA^i_{-n}(E),  \delta_\epsilon \uA^j_n(E) ]
  =&
  +
  \epsilon
  \frac{n}{\sdap \up_0^+}
  \delta^{i 2}
  \uA^j_0(E) \nonumber \\
  +&~
  \epsilon
  \frac{n^2}{(\ap \up_0^+)^2}
  \delta^{i j}
  \oint \frac{d z}{2 \pi i}
  (\up_0^+ \uL^2_{(\ne)}(z) - \up_0^2 \uL^+_{(\ne)}(z) )
  \partial \uL^+(z)
  .
\end{align}    
From the first term above, we see that,
$c_0=- \epsilon   \frac{1}{\sdap \up^+_0}$ as expected for the $l=0$ term.

We are left with the term $\epsilon\frac{n^2}{\sdap\up_0^+}\delta^{ij}I^{(2)}$, where,
\begin{equation}
  I^{(2)}=  \oint \frac{d z}{2 \pi i}
  (\up_0^+ \uL^2_{(\ne)}(z) - \up_0^2 \uL^+_{(\ne)}(z) )
  \partial \uL^+(z)
  \equiv
  \oh \up_0^+  \underline{M}^{2 +}(E)
  ,
\end{equation}
and $\underline{M}^{+ 2}(E)$ is the Lorentz generator in the frame $E$.
To see why this is so, we observe that,
\begin{align}
  \oint \frac{d z}{2 \pi i}
  \uL^+_{(\ne)}(z) \partial \uL^+(z)
  =&
  \oint \frac{d z}{2 \pi i}
  \uL^+_{(\ne)}(z) \partial \left(\uL_{(\ne)}^+(z) -i \ap \up_0^+ \ln(z) \right)\nonumber\\
  =&~
  -i \ap \up_0^+   \oint \frac{d z}{2 \pi i z} \uL^+_{(\ne)}(z) 
  =
  -\oh i \ap \up_0^+ \ux_0^+
  .
\end{align}
Then the first term can be rewritten as,
\eq{
  \oint \frac{d z}{2 \pi i}
  \uL^2_{(\ne)}(z) \partial \uL^+(z)
  =
  +
  \oint \frac{d z}{2 \pi i}
  \uL^2_{(\ne)}(z) \partial \uL_{(\ne)}^+(z)
  -
  \oh i \ap \up_0^+ \ux_0^2 \nonumber\\
  =
  -
  \oint \frac{d z}{2 \pi i}
  \uL^+_{(\ne)}(z) \partial \uL_{(\ne)}^2(z)
  -
  \oh i \ap \up_0^+ \ux_0^2
  ,
}
or taking the semi-sum as,
\begin{equation}
  \oint \frac{d z}{2 \pi i}
  \uL^2_{(\ne)}(z) \partial \uL^+(z)
  =
  \oh
  \oint \frac{d z}{2 \pi i}
  (
  \uL^2_{(\ne)}(z) \partial \uL_{(\ne)}^+(z)
  -
  \uL^+_{(\ne)}(z) \partial \uL_{(\ne)}^2(z)
  )
  -
  \oh i \ap \up_0^+ \ux_0^2
  .
\end{equation}
Assembling all the contributions and using $\partial \uL$ instead of
$\partial \uL_{(\ne)}$
we get,
\begin{align}
  I^{(2)}
  =
  \oh
  \up_0^+
  &\left[
  \oint \frac{d z}{2 \pi i}
  (
  \uL^2_{(\ne)}(z) \partial \uL^+(z)
  -
  \uL^+_{(\ne)}(z) \partial \uL^2(z)
  )
  +
  \oh i \ap
  ( \up_0^2 \ux_0^+ - \up_0^+ \ux_0^2 )  
  \right] \nonumber \\
  &=
  \oh
  \up_0^+
  \underline{M}^{2 +}(E)  
  ,
\end{align}
which matches \eqref{eq:Lorentz_generator} up to an overall factor in the frame $E$.

Noting that $M^{+2}$ rotates $x^2 \rightarrow x^+$
and $x^- \rightarrow x^2$ but keeps $x^+$ fixed,
the term $I^{(2)}$ commutes with all DDFs $A^i_m$ with $i\ne 2$.
Explicitly we get,
\begin{equation}
  [\uA^i_{-m}(E), I^{(2)}]
  =
  -\oh \sqrt{\hap} \delta^{i 2}
  (\ualpha^+_0)^2
  \delta_{m, 0}
  .
\end{equation}

Assembling all contributions for $j\ne 2$ we get,
\begin{align}
  \delta_\epsilon \uA^j_n(E)
  &
  =
  -
  \epsilon
  \frac{n}{\sdap \up^+_0}\left[
  \sum_{l\ne n} \frac{1}{n-l} \uA^j_l(E)\, \uA^2_{n-l}(E)
  +
  \frac{1}{\sdap \up^+_0}
  \underline{M}^{+ 2}(E)
  \uA^j_n(E)\right]
  ,
  \label{eq:jneq2_res}
\end{align}
where the contribution from $\uA^-_n$ is absent since in
$  \delta_\epsilon \uA^{j\ne 2}_n(E)$
there are no $\ualpha^-$.
There are also no normal ordering issues in \eqref{eq:jneq2_res} since,
$\underline{M}^{+ 2}(E)$ commutes with all DDFs as noted above.
%____________________________________________%
\subsubsection{Normal ordering subtlety in the $i=2$ case}
In \eqref{eq:infinitesimal_DDF_rotation} we would like to consider the details of the variation
$\delta_\epsilon A^2_m(E)$ since some normal ordering is required and
because the integrand $\partial \uL^- e^{i k^. \uL^+}$ is not a good
conformal operator - it has a cubic pole in the OPE with the
stress energy tensor.

The key observation is that,
$\lim_{z \rightarrow w} \partial \uL^2(z) e^{i k^. \uL^+(w)}$ is a
perfectly well defined conformal operator and therefore its
infinitesimal Lorentz rotation must be as well.
We expect it to be given by the normal ordering of the single
components rotations but other quantum i.e. vanishing classically
terms may in principle appear.

We start by computing the infinitesimal Lorentz rotation,
\begin{align}
  [
    \sqrt{2} \epsilon M^{1 2}
  &
    ,
    \partial \uL^2(z) e^{i \frac{m}{\ap \up_0^+} \uL^+(w)}
  ]
  =
  \delta_\epsilon (\partial \uL^2(z))
  e^{i \frac{m}{\ap \up_0^+} \uL^+(w)}
  +
  \partial \uL^2(z)
  \delta_\epsilon \left( e^{i \frac{m}{\ap \up_0^+} \uL^+(w)} \right)
  \nonumber\\
  =&
  -
  \epsilon ( \partial \uL^+(z) -\partial \uL^-(z) )
  \,
  e^{i \frac{m}{\ap \up_0^+} \uL^+(w)}
  \nonumber\\
  &+
  \partial \uL^2(z)
  \,
  \left(
  \epsilon
  \frac{ i m  }{\ap (\up^+_0)^2}
  \left(
  \up_0^+ \uL_{(\ne)}^2(w) - \up_0^2 \uL_{(\ne)}^+(w)
  \right)
  e^{i \frac{m}{\ap \up_0^+} \uL^+(w)}
  \right)
.
\end{align}
We can then
evaluate the singular part of the OPE to get,
\begin{align}
  \left.\left[
    \sqrt{2} \epsilon M^{1 2}
    ,
    \partial \uL^2(z) e^{i \frac{m}{\ap \up_0^+} \uL^+(w)}
  \right] \right\vert_{\mbox{sing}}
  =
  \frac{-\epsilon \ap}{2(z-w)}
  \left(
  i \frac{m}{\ap \up_0^+}
  \eta^{+-}
  +
  \frac{ i m  }{\ap (\up^+_0)^2}  \up_0^+
  \right)
  e^{i \frac{m}{\ap \up_0^+} \uL^+(w)}
  ,
\end{align}
which vanishes and therefore the infinitesimal Lorentz rotation is
given by the normal ordering of the rotations of the single components.

Finally the infinitesimal Lorentz rotation is,
\begin{align}
  \delta_\epsilon
  \left(
      \partial \uL^2 e^{i \frac{m}{\ap \up_0^+} \uL^+}
  \right)
  =&
  \epsilon
  \,
  :
  \left[
  \frac{ i m  }{\ap (\up^+_0)^2}
  \partial \uL^2
  \left(
  \up_0^+ \uL_{(\ne)}^2 - \up_0^2 \uL_{(\ne)}^+
  \right)
  -
  ( \partial \uL^+ -\partial \uL^- )
  \right]
  e^{i \frac{m}{\ap \up_0^+} \uL^+}
  :
  \nonumber\\
  =&
  \epsilon
  \,
  :
  \left[
    \partial \uL^-
    +
    \frac{ i m  }{\ap (\up^+_0)^2}
    ( \up_0^+ \uL_{(\ne)}^2 -\up_0^2  \uL_{(\ne)}^+ )
    \partial \uL^2 
    \right]
  e^{i \frac{m}{\ap \up_0^+} \uL^+}
  :
  \nonumber\\
  &
  %% -
  %% \epsilon
  %% \frac{ i m  \up_0^2}{\ap (\up^+_0)^2}
  %% \partial \uL^2
  %% \uL_{(\ne)}^+ 
  %% e^{i \frac{m}{\ap \up_0^+} \uL^+}
  -
  \epsilon
  \partial
  \left(
  \frac{\ap \up_0^+}{i m}
  e^{i \frac{m}{\ap \up_0^+} \uL^+}
  \right)
  ,
\end{align}
which contains pieces which are not good conformal operators.
However the sum of first two operators is free from cubic poles,

$T= - \frac{2}{\ap} : (\partial L)^2 :$ to be
\begin{align}
  -\frac{\ap}{2}  T(z)
  \,
  &
  \left(
  \partial \uL^-
  e^{i \frac{m}{\ap \up_0^+} \uL^+}
  \right)(w)
  =
  \left.
  2 e^{ \delta \cdot \partial L(z) }
  \right|_{\delta_\mu \delta_\nu   \rightarrow  \eta_{\mu\nu} }
  \,
  \left.
  e^{ \epsilon \cdot \partial L + i \frac{m}{\ap \up_0^+} \uL^+ }
  \right|_{\epsilon_-}
  \nonumber\\
  %
  % 2nd line
  =&
  2
  \left.
  :
  e^{
    \delta \cdot \uL(z)
    +
    \epsilon \cdot \partial L(w)
    +
    i \frac{m}{\ap \up_0^+} \uL^+(w) }
  :
  e^{
    -
    \hap \frac{\epsilon \cdot \delta}{ (z-w)^2}
    -
    \hap \frac{i \delta \cdot k}{ z-w}    
  }
  \right|_{\delta_\mu \delta_\nu   \rightarrow  \eta_{\mu\nu},~
    \epsilon_-,
    k_+ \rightarrow \frac{m}{\ap \up_0^+}}
  \nonumber\\
  =&
  2 \left(\hap\right)^2
  \frac{
    i \epsilon \cdot k
  }{
    (z-w)^3
  }
  + \dots
  \nonumber\\
  =&
  2 \left(\hap\right)^2
  \frac{
     i \frac{m}{\ap \up_0^+} \eta^{+-}
  }{
    (z-w)^3
  }
  + \dots
.  
\end{align}
In a similar way we get
\begin{align}
  -\frac{\ap}{2}  T(z)
  \,
  &
  \left(
  \partial \uL^2
  L^2_{(\ne)}
  e^{i \frac{m}{\ap \up_0^+} \uL^+}
  \right)(w)
  =
  \left.
  2 e^{ \delta \cdot \partial L(z) }
  \right|_{\delta_\mu \delta_\nu   \rightarrow  \eta_{\mu\nu} }
  \,
  \left.
  e^{ \epsilon_{(1)} \cdot \partial L
    + \epsilon_{(2)} \cdot \partial L_{(\ne)}
    + i \frac{m}{\ap \up_0^+} \uL^+ }
  \right|_{\epsilon_{(1) 2}, \epsilon_{(2) 2}} 
  \nonumber\\
  %
  % 2nd line
  =&
  2 \left(\hap\right)^2
  \frac{
    \epsilon_{(1)} \cdot \epsilon_{(2)} 
  }{
    (z-w)^3
  }
  + \dots
  \nonumber\\
  =&
  2 \left(\hap\right)^2
  \frac{
    1
  }{
    (z-w)^3
  }
  + \dots
.  
\end{align}
Now we can examine the cubic pole contributions and see that,
\begin{align}
    -\frac{\ap}{2}  T(z)
  \,
  &
  \left(
  \dots
  \right)
  =
  2 \left(\hap\right)^2
  \frac{
     -i \frac{m}{\ap \up_0^+}
  }{
    (z-w)^3
  }
  +
  \frac{ i m  }{\ap (\up^+_0)^2}
  \,
  \up_0^+
  \times
  2 \left(\hap\right)^2
  \frac{
    1
  }{
    (z-w)^3
  }
  +
  \dots
  ,
\end{align}
which in fact vanishes.

%__________________________________%
\subsubsection{Rewriting the variation $\delta_\epsilon A^j_m(E)$ for $j=2$}

%% \igor{
  Finally we consider the $j=2$ case where there is a further contribution in
eq. \eqref{eq:infinitesimal_DDF_rotation}, when $i=2$.
If we proceed as before, we can evaluate the full commutator as,
\begin{align}
  [A^i_m(E),   \delta_\epsilon A^{j=2}_n]
  =& \epsilon \frac{i n}{\ualpha^+_0}
    \left(
    -
    \uA^2_{m+n} \frac{m}{m+n}  
    +
    \uA^2_{m+n} 
    \right)\delta^{i 2}
  -m \delta^{j 2}
  \frac{\epsilon}{\ualpha^+_0}
  \uA^i_{m+n}
  .
\label{eq:Ai_dAj2_comm}
\end{align}

We observe that the ansatz in \eqref{eq:ansatz_infinitesimal_change_of_frame} can generate the first term (in parentheses), wherein we get explicitly,
\eq{
{c}_{-m} + {c}_{m+n} = -\frac{i \epsilon}{\ualpha^+_0} \frac{n^2}{m(m+n)}.
}
Once again, there are two basic possible linear homogeneous solutions to this recurrence relation, given by \eqref{eq:basic_rec_soln}. Furthermore, we must choose the second possibility for the exact same reason as before, to obtain,
\eq{
c_{l} = -\frac{i \epsilon}{\ualpha^+_0} \frac{n}{n-l}.
\label{eq:rec_soln_j2}
}
We see that this term is neatly absorbed in the expansion of $\delta_\epsilon A^{j}$, which now reads,
\begin{align}
  \delta_\epsilon \uA^j_n(E)
  &
  =
  -
  i\epsilon
  \frac{n}{\sdap \up^+_0}\left[
  \sum_{l\ne n} \frac{1}{n-l} \uA^j_l(E)\, \uA^2_{n-l}(E)
  +
  \frac{1}{\sdap \up^+_0}
  \underline{M}^{+ 2}(E)
  \uA^j_n(E)\right]
  ,
  \label{eq:jeq2_res1}
\end{align}
and is no longer limited to $j\neq 2$.

Finally, we consider the last piece in \eqref{eq:Ai_dAj2_comm} - since there are no more terms like $\delta^{i \dots}$,
we assume an expression like,
\begin{equation}
\delta_\epsilon \uA^{j=2}_n(E)|_{\mbox{extra}}
=
\delta^{j 2}
\left[
\sum_l  d_l\, \uA^k_l(E)\, \uA^k_{n-l}(E)
+
\bar \cN
{\tilde \uA}^-_n(E)
\right]
,
\label{eq:ansatz_infinitesinal_change_of_frame_J=2}
\end{equation}
where, 'extra' labels the last term of \eqref{eq:Ai_dAj2_comm}.
Comparison with the true expression gives,
\begin{equation}
  d_{m+n} + d_{-m}
  =
  \frac{- i \epsilon}{\sdap \up_0^+}
  ,
\end{equation}
whose simplest solution is,
\begin{equation}
  d_l
  =
  \oh
  \frac{- i \epsilon}{\sdap \up_0^+}
  .
\end{equation}
Putting everything together, the complete variation is given by,
\eq{
  \delta_\epsilon \uA^j_n(E)
  =
  -
  i\epsilon
  \frac{n}{\sdap \up^+_0}&\left[
  \sum_{l\ne n} \frac{1}{n-l} \uA^j_l(E)\, \uA^2_{n-l}(E)
  +
  \frac{1}{\sdap \up^+_0}
  \underline{M}^{+ 2}(E)
  \uA^j_n(E)\right]\nonumber\\
  - &\delta^{j 2} 
\left[\oh\frac{i \epsilon}{\sdap \up_0^+} 
\sum_l \uA^k_l(E)\, \uA^k_{n-l}(E)
+
\bar \cN
{\tilde \uA}^-_n(E)
\right].
\label{eq:app varDDF_full}
}

\section{Details on Lorentz generators expressed DDF and Brower operators}
\label{app:Lorentz_algebra_using_DDF}
In this appendix we report the details of how to write the action of
Lorentz generators on DDF and Brower operators using DDF and Brower
operators.
This is not the same as expressing the Lorentz generators using DDF
and Brower operators since they span a subset of all covariant operators.

\subsection{Preliminaries}
The covariant Lorentz generator
\begin{equation}
  M^{\mu\nu}
  =
  x_0^\mu p_0^\nu - x_0^\nu p_0^\nu
  +
  i \sum_{n\ne 0} \frac{ \alpha^\mu_n \alpha^\nu_{-n} }{n}
  ,
\end{equation}
satisfies
\begin{align}
  [M^{\mu\nu}, x_0^\lambda]
  &=
  2 i \delta^{\lambda [\mu} x_0^{\nu]}
  ,
  \nonumber\\
  [M^{\mu\nu}, \alpha_m^\lambda]
  &=
  2 i \delta^{\lambda [\mu} \alpha_m^{\nu]}
  ,
\end{align}
or for the chiral field
\begin{align}
  [M^{\mu\nu}, L(z)^\lambda]
  &=
  2 i \delta^{\lambda [\mu} L(z)^{\nu]}
  .
\end{align}

%%%%%%%%%%%%%%%%%%%%%%%%%%%%%%%%%%%%%%%%%%%%%%%%%%
\subsubsection{A comment on commutators involving the not derived
  $L(z)$}

Naively one would expect that the operator
\begin{align}
{\widehat {\hat  M}}^{\mu\nu}
  =&
  \cN
  \oint_{z=0} \frac{d z}{2\pi i}
  \left(
  L^\mu(z) \partial L^\nu(z) - L^\nu(z) \partial L^\mu(z)
  \right)
  ,
\end{align}
be the Lorentz generator. But it is not so because of the term $-i \ap
p_0^\mu\, \ln z$ which implies a cut.
Then one could think of
\begin{align}
\hat  M^{\mu\nu}
  &=
  \cN
  \oint_{z=0} \frac{d z}{2\pi i}
  \left(
  L^\mu_{(\ne)}(z) \partial L^\nu(z) - L^\nu_{(\ne)}(z) \partial L^\mu(z)
  \right)
  ,
  ~~~
  \cN = \frac{i}{\ap}
\end{align}
be the Lorentz generator. But it is not so.
We can evaluate directly the operator in terms of the oscillators as
\begin{align}
\frac{1}{\cN}
  \hat M^{\mu\nu}
  =&
  -i \hap \left( x_0^\mu p_0^\nu - x_0^\nu p_0^\mu \right)
  +
  \ap \sum_{m \ne 0} \frac{1}{m} \alpha_m^\mu \alpha_{-m}^\nu
  ,
\end{align}
and then the commutators
\begin{align}
  \frac{1}{\cN}
  [\hat M^{\mu\nu} ,x^\lambda]
  =&
  \ap \delta^{ [\mu} x_0^{\nu]}
  ,
  \nonumber\\
  \frac{1}{\cN}
  [\hat M^{\mu\nu} ,\alpha_l^\lambda]
  =&
  2 \ap \delta^{ [\mu} \alpha_l^{\nu]}
  ,
\end{align}
which show explicitly what stated since the two variations differ by a
factor $2$.

We can evaluate the same using the fields.
To this aim one has to go back to the original definitions and compute
\begin{align}
  \frac{1}{\cN}
  [\hat M^{\mu\nu}  &,L^\lambda(w)]
  =
  \nonumber\\
  =&
  \oint_{z=0, |z|>|w| } \frac{d z}{2\pi i}
  \left(
  L^\mu_{(\ne)}(z) \partial L^\nu(z) - L^\nu_{(\ne)}(z) \partial L^\mu(z)
  \right)
  \, L^\lambda(w)
  \nonumber\\
  &-
  L^\lambda(w)
  \,
  \oint_{z=0, |z|<|w| } \frac{d z}{2\pi i}
  \left(
  L^\mu_{(\ne)}(z) \partial L^\nu(z) - L^\nu_{(\ne)}(z) \partial
  L^\mu(z)
  \right)
  ,
\end{align}
and carefully use eq.s \eqref{eq:normal_orderings_wo_p0}.

The proper Lorentz generator is then
\begin{equation}
  M^{\mu\nu}
  =
    \hat M^{\mu\nu}
  +
  \oh ( x^\mu_0 p^\nu_0 - x_0^\nu p_0^\mu)
  \label{eq:Lorentz_generator_M}
  .
\end{equation}

%%%%%%%%%%%%%%%%%%%%%%%%%%%%%%%%%%%%%%%%%%%%%%%%%%
\subsection{Generators $\underline{M}^{i j}$}
It is immediate to compute
\begin{align}
  [\underline{M}^{i j}, \, \uL^k]
  &=
  2 i \eta^{k [i} \uL^{j]}
  ,
  \nonumber\\
 [\underline{M}^{i j}, \, \uL^+]
  &=
 0
  ,
 \nonumber\\
 [\underline{M}^{i j}, \, \uL^-]
  &=
 0
 ,
 \end{align}
so that
\begin{align}
  [\underline{M}^{i j}, \, \uA^k_n]
  &=
  2 i \eta^{k [i} \uA_n^{j]}
  ,
 \nonumber\\
 [\underline{M}^{i j}, \, \uA_n^-]
 &=
 [\underline{M}^{i j}, \, \tilde \uA_n^-]
 =
 0
 .
 \end{align}
Then we are led to the expression for the physical states
\begin{equation}
  \underline{M}^{i j}|_{DDF}
  =
  i
  \sum_{n\ne 0} \frac{1}{n} \uA_n^{i} \uA^j_{-n}
  .
\end{equation}
This expression reproduces the desired commutation relations for
$\uA_n$ with $n\ne 0$ but fails for $\uA_0$.
Let us see whether it is not possible to add some zero modes contributions to
fix this issue.
For example the natural guess $2\ux_0^{[i}  \up_0^{j]}$ fails.
In facts
\begin{equation}
  [ 2\ux_0^{[i}  \up_0^{j]},\, \uA^k_m]
  = 2 i \eta^{k [i} \uA_0^{j]}
  ,
\end{equation}
for all $n$ because of the $\ux_0$ zero mode interacts with $\partial \uL$.
But this failure can be compensated as
\begin{align}
  [ -2 i \sum_{n\ne 0} \frac{1}{n} \uA_n^{[i} \uA_0^{j]},\,
    \uA^k_m]
  =& -2 i \eta^{k [i} \uA_0^{j]}
  ,~~~m\ne 0
  .
\end{align}
The final expression which is valid also for $\uA_0^i$ is then
\begin{align}
  \underline{M}^{i j}
  =&
  i
  \sum_{n\ne 0} \frac{1}{n} \uA_n^{i} \uA^j_{-n}
  \nonumber\\
  &
  - 2 i
  \sum_{n\ne 0} \frac{1}{n} \uA_n^{[i} \uA_0^{j]}
  +
  \frac{2}{\sdap} \ux_0^{[i} \uA_0^{j]}
  .
\end{align}

%%%%%%%%%%%%%%%%%%%%%%%%%%%%%%%%%%%%%%%%%%%%%%%%%%
\subsection{Generators $\underline{M}^{+ i}$}
It again is immediate to compute
\begin{align}
  [\underline{M}^{+ i}, \, \uL^k]
  &=
  - i \delta^{k i} \uL^{+}
  ,
  \nonumber\\
 [\underline{M}^{+ i}, \, \uL^+]
  &=
 0
  ,
 \nonumber\\
 [\underline{M}^{+ i}, \, \uL^-]
  &=
 -i \uL^i
 ,
 \end{align}
so that
\begin{align}
  [\underline{M}^{+ i}, \, \uA^k_n]
  &=
  - i \delta^{k i} \uA_0^{+}\, \delta_{n, 0}
  ,
 \nonumber\\
 [\underline{M}^{+ i}, \, \uA_n^-]
 &=
 -i \uA^i_n
 ,
\nonumber\\
 [\underline{M}^{+ i}, \, \tilde \uA_n^-]
 &=
 0
 .
\label{eq:M+i}
\end{align}
Then we are led to the expression for the physical states which
do not require the use of DDF zero modes $\uA_0$ 
\begin{equation}
  \underline{M}^{+ i}|_{DDF}
  =
  0
  .
\end{equation}
Notice that this expression is valid {\sl only when}
using $\uA^i$ and the improved
Brower operators $\tilde \uA^-$.
This means that when we have an expression with $\uA^-$ it has to be
reexpressed using $\tilde \uA^-$ and $\cL$: this observation is
important for the computation of the expectation values of the Casimir operators
of the different physical states obtained by the DDF operators.

%%%%%%%%%%%%%%%%%%%%%%%%%%%%%%%%%%%%%%%%%%%%%%%%%%
\subsection{Generator $\underline{M}^{+ -}$}
It again is immediate to compute
\begin{align}
  [\underline{M}^{+ -}, \, \uL^k]
  &=
  0
  ,
  \nonumber\\
 [\underline{M}^{+ -}, \, \uL^+]
  &=
 +i \uL^-
  ,
 \nonumber\\
 [\underline{M}^{+ -}, \, \uL^-]
  &=
 -i \uL^+
 ,
 \end{align}
so that
\begin{align}
  [\underline{M}^{+ -}, \, \uA^k_n]
  &=
  0
  ,
 \nonumber\\
 [\underline{M}^{+ -}, \, \uA_n^-]
 &=
 -i \uA^-_n
 ,
\nonumber\\
 [\underline{M}^{+ -}, \, \tilde \uA_n^-]
 &=
 -i \tilde \uA_n^-
 .
\end{align}
Notice that these results rely on the invariance of
$\exp\left( i \frac{n}{\ap} \frac{\uL^+}{\up_0^+} \right)$.

Exactly as for the generators $\underline{M}^{+ i}$ it cannot be
represented using $\uA^i$ and $\tilde \uA^-$, not even in the physical sector.
The reason is that it can be only a functional of $\tilde \uA_n^-$
since $\underline{M}^{+ -}$ commutes with all $\uA^i$
but $\tilde \uA_n^-$  satisfy a Virasoro algebra and therefore 

%%%%%%%%%%%%%%%%%%%%%%%%%%%%%%%%%%%%%%%%%%%%%%%%%%
\subsection{Generators $\underline{M}^{- i}$}
These are the most interesting generators since they combine a true
rotation and a compensating
conformal transformation in the transverse directions.

It again is immediate to compute
\begin{align}
  [\underline{M}^{- i}, \, \uL^k]
  &=
  -i \delta^{k i} \uL^-
  ,
  \nonumber\\
 [\underline{M}^{- i}, \, \uL^+]
  &=
 -i \uL^i
  ,
 \nonumber\\
 [\underline{M}^{- i}, \, \uL^-]
  &=
 0
 .
 \end{align}
Because of the non trivial rotation of $\uL^+$ is not immediate to
compute the effect  on DDF and Brower operators.

The final result is that
\begin{align}
  [\underline{M}^{- i}(E), \, \uA^k_n(E)]
  =&
%%  \delta \uA^k_n(E)
%%  =&
  i \frac{n}{\ualpha^+_0}
  \sum_{l\ne n} \frac{1}{n-l} \uA^k_l(E)\, \uA^i_{n-l}(E)
  +
  \frac{n}{(\ualpha^+_0)^2}
  \underline{M}^{+ i}(E)
  \uA^k_n(E)
  \nonumber\\
  &+
  \delta^{k i} 
  \left[
    - \frac{i}{2 \ualpha_0^+}     
    \sum_l \uA^j_l(E)\, \uA^j_{n-l}(E)
    -
    i
    {\tilde \uA}^-_n(E)
    \right].
  \nonumber\\
  =&
  i \frac{n}{\ualpha^+_0}
  \sum_{l\ne n} \frac{1}{n-l} \uA^k_l(E)\, \uA^i_{n-l}(E)
  +
  \frac{n}{(\ualpha^+_0)^2}
  \underline{M}^{+ i}(E)
  \uA^k_n(E)
  -
  i \delta^{k i} 
  {\uA}^-_n(E)
  ,
  \label{eq:M-i on DDF}
%%   \\
%%   %
%%   %
%%  [\underline{M}^{- i}(E), \, \uA_n^-(E)]
%%  &=
%% %
%%  ,
%% \nonumber\\
%%  [\underline{M}^{- i}(E), \, \tilde \uA_n^-(E)]
%%  &=
%% %
 .
\end{align}
The expression for the variation of $\uA^k$
under $  \underline{M}^{- i}(E)$  may seem quite strange
since it  involves $  \underline{M}^{+ i}(E)$
but a simple consistency check is immediate  to perform:
the scaling \wrt $\underline{M}^{+ -}(E)$
works because of $1/ (\ualpha_0^+)^2$.
The same is true for the first term.

%% \framebox{%
%%   \begin{minipage}{\textwidth}
%% Because of this we remark that only the term proportional to $\uA^-$
%% matters when computing the expectation values for the Casimirs
%% operators for the DDF states.
%% \end{minipage}
%% }

The previous result is obtained first computing the variation of $\uA^k$
then inferring the expression of the generator $\underline{M}^{- i}$
and then using this expression we can compute the variation of $\uA^-$
and $\tilde \uA^-$.
The expression of the generator $\underline{M}^{- i}$ restricted to
DDF operators is
\begin{align}
  \underline{M}^{- i}|_{DDF}
  =&
  i
  \sum_{m\ne 0} \frac{1}{m} \uA_m^-(E) \uA^i_{-m}(E)
  -
  \frac{1}{ (\ualpha_0^+)^2} \underline{M}^{+ i}(E) \cL_0(E)
  \nonumber\\
  =&
  i
  \sum_{m\ne 0} \frac{1}{m} {\tilde \uA}_m^-(E) \uA^i_{-m}(E)
  +
  \frac{i}{ 2 \ualpha_0^+}
  \sum_{m\ne 0} \frac{1}{m} \cL_m(E) \uA^i_{-m}(E)
  -
  \frac{1}{ (\ualpha_0^+)^2} \underline{M}^{+ i} \cL_0(E)
  ,
\end{align}
where $\cL_n$ is defined in \eqref{eq:cL virasoro transverse}.
The first term in the \rhs of the first line gives to the first
and third terms of the final form of the $\uA^k(E)$ variation. 

Notice that there are apparently normal ordering problems
since $\uA_m^-(E)$ and $\uA^i_{-m}(E)$ do not commute, however the
ambiguity is proportional to the zero modes which are not DDF operators.

%%%%%%%%%%%%%%%%%%%%%%%%%%%%%%%%%%%%%%%%%%%%%%%%%%%%%%%%%%%%%%%%%%%%%%

\section{Details on extracting Poincar\'e irreps from Lorentz irreps}
In this appendix we would like to give some details on how to perform
the splitting of Lorentz polarizations into Poincar\'e irreps.
This is useful if we want to compute the Casimirs in the covariant formalism.

We consider massive states with mass $M$ and perform the decomposition
in rest frame and then we covariantize the result.

We use $I,J, \dots=1,2, \dots D-1$.

%%%%%%%%%%%%%%%%%%%%%%%%%%%%%%%%%%%%%%%%
\subsection{Lorentz vector splits}
Let us consider the Lorentz vector $T_\mu$.
In the rest frame from the $SO(D-1)$ point of view we have actually two irreps
\begin{equation}
  T_\mu
  \rightarrow
  T_I\, \oplus\, T_0
  ,
\end{equation}
which have two different Casimirs.

This split seems to break Lorentz invariance but it is nor so.
We have in fact the following covariant characterization of the two
irreps\footnote{
Remember that in rest frame $k^0=-k_0= M$.}
\begin{align}
  T_0:&
  ~~
  t=\frac{k^\mu}{M} T_\mu \ne0
  ,~~~
  &T&_\mu^{(\cdot)} = \frac{k^\mu}{M} t
  ~~~~
  &\rightarrow&
  ~~~~
  \mbox{not transverse vector and proportional to } k_\mu
  \nonumber\\
  T_I:&
  ~~
  k^\mu t_{(\ydiagram{1}) \mu} =0
  ,~~~
  &t&_{(\ydiagram{1}) \mu} = T_\mu + \frac{k^\mu}{M} t
  &\rightarrow&
  ~~~~
  \mbox{transverse vector}
  .
\end{align}
Finally we can write the decomposition of the Lorentz vector into
Poincar\'e irreps as
\begin{equation}
  T_\mu
  =
  t_{(\ydiagram{1}) \mu} + \frac{k^\mu}{M} t
.
\end{equation}

%%%%%%%%%%%%%%%%%%%%%%%%%%%%%%%%%%%%%%%%

\subsection{Lorentz antisymmetric two tensor splits}
Let us consider the Lorentz antisymmetric tensor $A_{[\mu \nu]}$.
From the $SO(D-1)$ point of view
in the rest frame we have actually two irreps
\begin{equation}
  A_{[\mu \nu]}
  \rightarrow
  A_{[I J]}\, \oplus\, A_{[I 0]}
  ,
\end{equation}
which again have two different Casimirs.

The covariant characterization is
\begin{align}
  A_{[I 0]}:&
  ~~
  a_{(\ydiagram{1})\mu}=\frac{k^\mu}{M} A_{\mu \nu}
  ,~~~
  k^\mu a_{(\ydiagram{1})\mu} =0
  ,~~~
  &A&_{(\ydiagram{1}) \mu \nu}=  -2 \frac{1}{M}k_{[\mu}   a_{(\ydiagram{1})\nu]}
  %% ~~~~
  %% &\rightarrow&
  %% ~~~~
  %% \mbox{not transvese tensor and proportional to } k_\mu
  \nonumber\\
  A_{I J}:&
  ~~
  k^\mu a_{(\ydiagram{1,1})\mu \nu} =0
  ,~~~
  &a&_{(\ydiagram{1,1}) \mu \nu}= A_{ \mu \nu} - A_{(\ydiagram{1}) \mu \nu}
  %% &\rightarrow&
  %% ~~~~
  %% \mbox{transverse tensor}
  .
\end{align}
All these relations can be verified either directly or in rest frame.

Finally we can write the decomposition of the Lorentz antisymmetric
two tensor into Poincar\'e irreps as
\begin{equation}
  A_{ \mu \nu}
  =
  a_{(\ydiagram{1,1}) \mu \nu}
  - 
  2 \frac{1}{M}k_{[\mu}   a_{ (\ydiagram{1}) \nu]}
  .
\end{equation}

%%%%%%%%%%%%%%%%%%%%%%%%%%%%%%%%%%%%%%%%
\subsubsection{Lorentz symmetric traceless two tensor splits}
Let us consider the Lorentz symmetric traceless tensor $S_{\mu \nu}$.
The traceless condition means that
\begin{equation}
  S_{00} = S_{K K}
  .
\end{equation}
From the $SO(D-1)$ point of view
in the rest frame we have actually three irreps
\begin{equation}
  S_{\mu \nu}
  \rightarrow
  \left(
  S_{I J}- \frac{1}{D-1}\delta_{I J} S_{K K}
  \right)
  \,
  \oplus\,
  S_{I 0}\,
  \oplus\,
  S_{0 0} \equiv S_{K K}
  ,
\end{equation}
which have once again three different Casimirs.

The covariant characterization of the lowest irreps is easy and reads
\begin{align}
  S_{00}:&
  ~~
  s_{(\cdot)}
  =
  \frac{k^\mu}{M} \frac{k^\nu}{M} S_{\mu \nu}
  ,~~~
  &S&_{(\cdot) \mu \nu}=  \frac{k^\mu}{M} \frac{k^\nu}{M} s_{(\cdot)}
  \nonumber\\
  S_{0 I}:&
  ~~
  s_{ (\ydiagram{1}) \mu}
  =
  \frac{k^\nu}{M} S_{\mu \nu} + \frac{k_\mu}{M} s_{(\cdot)}
  ,~~~
  k^\mu s_{ (\ydiagram{1}) \mu} = 0
  ,~~~
  &S&_{(\ydiagram{1}) \mu \nu}=  \frac{k_{(\mu} s_{ (\ydiagram{1}) \nu)} }{M}
  .
\end{align}
Notice that not all tensors are traceless, i.e.
\begin{equation}
  S_{(\cdot) \mu \mu}
  \ne 0,
  ~~~
  S_{(\ydiagram{1}) \mu \mu}
  =
  0
  .
\end{equation}

While the highest irrep is a little trickier since one would wrongly
guess that the last irrep is the left over
\begin{equation}
  S_{I J}:~~~
  s_{(\ydiagram{2}) \mu \nu}
  =
  S_{ \mu \nu} - S_{(\ydiagram{1}) \mu \nu} - S_{(\cdot) \mu \nu}
%%  ~~~ WRONG!
  ,
\end{equation}
but this is wrong since it misses
$ \frac{1}{D-1}\delta_{I J} S_{K K}$ in the irrep definition.
The proper definition is
\begin{equation}
  S_{I J} - \frac{1}{D-1}\delta_{I J} S_{00}:
  ~~
  s_{(\ydiagram{2}) \mu \nu}
  =
  S_{ \mu \nu} - S_{(\ydiagram{1}) \mu \nu} - S_{(\cdot) \mu \nu}
  -
  \frac{1}{D-1} P_{\mu \nu} s_{(\cdot)}
  ,
\end{equation}
with $ P_{\mu \nu} = \eta_{\mu\nu} - \frac{k_\mu k_\nu}{ k^2} $
so that $ P_{\mu \nu} k^\nu=0$ and $P^\lambda_\lambda = D-1$.
This expression can be verified in the rest frame
and making use of $S_{00} = S_{K  K}$.
It has the wanted properties
\begin{align}
  k^\mu s_{(\ydiagram{2})\mu \nu} =0
  ,~~~
  s_{(\ydiagram{2}) \rho \rho}
  =
  0
  .
\end{align}
%% This can be verified either directly or simply going to the rest frame.
%% Transversality follows because
%% \begin{equation}
%%   k^\mu S_{(\ydiagram{2})\mu \nu}
%%   =
%% ,  
%% \end{equation}
%% and traceless-ness from
%% \begin{equation}
%% \end{equation}

Finally we can write the decomposition of the Lorentz symmetric traceless
two tensor into Poincar\'e irreps as
\begin{equation}
  S_{ \mu \nu}
  =
  s_{(\ydiagram{2}) \mu \nu}
  -
  2 \frac{1}{M}k_{(\mu}   s_{ (\ydiagram{1}) \nu)}
  +
  \left(
  \frac{1}{D-1} P_{\mu\nu} + \frac{k_\mu k_\nu}{ M^2}
  \right)
  s_{(\cdot)}
  .
\end{equation}

%%%%%%%%%%%%%%%%%%%%%%%%%%%%%%%%%%%%%%%%%%%%%%%%%%%%%%%%%%%%%%%%%%%%%%
%%%%%%%%%%%%%%%%%%%%%%%%%%%%%%%%%%%%%%%%%%%%%%%%%%%%%%%%%%%%%%%%%%%%%%
%%%%%%%%%%%%%%%%%%%%%%%%%%%%%%%%%%%%%%%%%%%%%%%%%%%%%%%%%%%%%%%%%%%%%%
%%%%%%%%%%%%%%%%%%%%%%%%%%%%%%%%%%%%%%%%%%%%%%%%%%%%%%%%%%%%%%%%%%%%%%

\printbibliography[heading=bibintoc]

\end{document}